%% file: main.tex
\documentclass[11pt, letterpaper]{article}
\input{preamble}

\usepackage{amsmath}
\usepackage{graphicx}
\usepackage{enumerate}

\usepackage[resetlabels]{multibib}
\newcites{appendix}{Appendix References}

\title{Distilling heterogeneous treatment effects: \\Stable subgroup estimation in causal inference\acknowledgements}
\author{
  Melody Huang\textsuperscript{\affiliationA,\equalcontributor},
  Tiffany M. Tang\textsuperscript{\affiliationB,\equalcontributor},
  Ana M. Kenney\textsuperscript{\affiliationC}
}
\date{\today}

\begin{document}

\renewcommand{\thefootnote}{\myfnsymbol{footnote}}
\maketitle
\footnotetext[1]{The authors would like to thank Naoki Egami, Erin Hartman, Kosuke Imai, Sam Pimentel, Josh Clinton, Yan Shuo Tan, and participants at the American Causal Inference Conference, Johns Hopkins Biostatistics Causal Inference Seminar, UNC Chapel Hill Biostatistics Seminar, NYU Quantitative Methods Seminar, Yale Econometrics Lunch Seminar, WUSTL Political Science Seminar, UW Madison Statistics Seminar, Vanderbilt Research on Behavior and Institutions Workshop, Penn State Statistics Seminar, ENAR Spring Meeting, ICSA Applied Statistics Symposium, and ND RISE AI Conference for their feedback and comments.}%
\footnotetext[2]{Denotes equal contribution.}
\footnotetext[3]{Yale University, CT, USA. Email: \texttt{melody.huang@yale.edu}, URL: \texttt{www.melodyyhuang.com}}%
\footnotetext[4]{University of Notre Dame, IN, USA. Email: \texttt{ttang4@nd.edu}, URL: \texttt{https://tiffanymtang.github.io/}}%
\footnotetext[5]{University of California Irvine, CA, USA. Email: \texttt{anamaria.kenney@uci.edu}}%

\setcounter{footnote}{0}
\renewcommand{\thefootnote}{\arabic{footnote}}
\maketitle 

\begin{abstract}
Recent methodological developments have introduced new black-box approaches to better estimate heterogeneous treatment effects; however, these methods fall short of providing interpretable characterizations of the underlying individuals who may be most at risk or benefit most from receiving the treatment, thereby limiting their practical utility. In this work, we introduce \textit{causal distillation trees} (CDT) to estimate interpretable subgroups. CDT allows researchers to fit \textit{any} machine learning model to estimate heterogeneous treatment effects, and then leverages a simple, second-stage tree-based model to ``distill’’ the estimated treatment effects into meaningful subgroups. As a result, CDT inherits the improvements in predictive performance from black-box machine learning models while preserving the interpretability of a simple decision tree. We derive theoretical guarantees for the consistency of the estimated subgroups using CDT, and introduce stability-driven diagnostics for researchers to evaluate the quality of the estimated subgroups. We illustrate our proposed method on a randomized controlled trial of antiretroviral treatment for HIV from the AIDS Clinical Trials Group Study 175 and show that CDT out-performs state-of-the-art approaches in constructing stable, clinically relevant subgroups.  
\end{abstract}

\noindent%

\newpage
\allowdisplaybreaks
\doublespacing
\section{Introduction}
While much of the causal inference literature has focused on estimating an average treatment effect for a specific intervention, researchers are often interested in understanding the underlying treatment effect heterogeneity for a given intervention. For example, in medical settings, researchers are often concerned about potential subsets of units who may be harmed by a particular drug or medical intervention. Similarly, in the social sciences, being able to understand who will benefit or could be potentially harmed by a specific intervention is crucial for evaluating the cost-benefit associated with a specific intervention. Though different approaches have been proposed to consider how to better estimate treatment effect heterogeneity across individuals, being able to \textit{characterize} subgroups of individuals in a study in an interpretable way remains challenging.

Many existing subgroup analyses in causal inference focus on pre-specified subgroups based upon prior substantive knowledge. 
However, it can be challenging to reason about higher-order interactions between covariates.
In the absence of prior knowledge, various tree-based approaches have been proposed to discover subgroups with heterogeneous treatment effects \citep[e.g.,][]{su2009subgroup, lipkovich2011subgroup, loh2015regression, lechner2024comprehensive}, of which causal trees \citep{athey2016recursive} are arguably the most popular.\footnote{Citation count was used as a proxy for popularity. In comparing the citation count of the different papers, as of 2025, the original causal tree paper by \citet{athey2016recursive} has over 2,000 citations, while the alternative methods have anywhere from 100-500 citations.} Similar to decision trees \citep{breiman1984classification}, causal trees partition the underlying covariate space into groups by finding clusters of individuals who have similar treatment effects. Moreover, unlike linear regression approaches, causal trees output a readily-interpretable partition of subgroups while also accounting for potential high-order interactions. 

Unfortunately, a drawback to causal trees is that they can be highly unstable and overfit in the presence of even small amounts of noise \citep[e.g.,][]{cattaneo2022pointwise}. Recent literature has proposed alternatives to better model heterogeneous treatment effects with `metalearners'---e.g., $T$-learners \citep[e.g.,][]{foster2011subgroup}, $S$-learners \citep[e.g.,][]{hill2011bayesian}, $X$-learner \citep{kunzel2019metalearners}, and $R$-learner \citep{nie2021quasi}, to name a few. These black-box metalearners provide theoretical guarantees in the form of quasi-oracle properties in recovering the conditional average treatment effect; however, they no longer produce a tree-like output and thus cannot be directly used to identify interpretable subgroups. 

In this work, we propose \textit{causal distillation trees} (CDT) to stably estimate interpretable subgroups. CDT allows researchers to leverage the power of black-box meta learners, while preserving the interpretability of a simple tree output. CDT is a two-stage learner. The first stage estimates a heterogeneous treatment effect model using a flexible metalearner to generate a predicted heterogeneous treatment effect for each unit. The second stage then \textit{distills} the information from the predicted heterogeneous treatment effect through a decision tree to construct interpretable subgroups. Intuitively, the first step serves as a de-noising step to help mitigate the instabilities that decision trees traditionally suffer from. The second step of `distillation' draws upon existing ideas in machine learning \citep[e.g.,][]{hinton2015distilling, menon2021statistical, dao2021knowledge}. However, unlike typical machine learning settings, where the primary goal is to improve the second stage learner's predictive ability, we leverage distillation to improve our ability to estimate interpretable subgroups using the second stage decision tree learner.

Our work provides several key theoretical and methodological contributions. First, while recent papers have similarly proposed the use of a two-stage learner to estimate subgroups \citep[e.g.,][]{foster2011subgroup, zhang2021subgroup, rehill2024distilling}, CDT is agnostic to what first-stage learner researchers use, allowing for a large degree of flexibility in choosing an informative teacher model. Second, unlike previous work, which has largely relied on simulated evidence, we derive theoretical guarantees, proving that CDT consistently recovers the optimal subgroups. To our knowledge, we are the first to prove the theoretical properties in recovering subgroups from distillation. We highlight that distillation can offset the instabilities incurred from using causal trees (or other tree-based approaches) in low signal-to-noise regimes, while preserving the interpretability of these methods. 

Finally, we propose a novel model selection procedure to help researchers select a teacher model in CDT. Unlike existing model selection procedures, which compare the goodness-of-fit of different models, our selection procedure is explicitly tailored for estimating subgroups. We introduce a new measure, which we call the \textit{Jaccard Subgroup Similarity Index} (SSI), that quantifies the similarity of subgroups. Researchers can use the subgroup similarity index to select the teacher model that recovers the most stable subgroups. 

The paper is structured as follows. Section~\ref{sec:setup} presents notation and formalizes the notion of a subgroup in causal inference. In Section~\ref{sec:cdt}, we introduce causal distillation trees and derive its theoretical properties. In Section~\ref{sec:model_selection}, we propose a stability-driven model selection procedure to select a teacher model in CDT. We demonstrate the effectiveness of CDT across extensive simulation settings (Section~\ref{sec:sims}), and an AIDS clinical trial case study (Section~\ref{sec:case_study}). Section~\ref{sec:discussion} concludes. 

\section{Setup and Notation}\label{sec:setup}
Throughout the paper, we will use the potential outcomes framework, where $Y(1)$, $Y(0)\in \R$ denote the potential outcome under treatment and control, respectively. Let $Z \in \{0, \;1\}$ denote the treatment assignment indicator. We assume consistency of treatment assignment across the study and no interference (i.e., the stable unit treatment value assumption). Let $Y := Y(1) Z + Y(0) (1-Z)$ be the observed outcomes. Finally, we assume that researchers have access to a set of pre-treatment covariates $X \in \R^p$ for each unit in their study, and the tuple $\{Y_i(0),\; Y_i(1),\; X_i,\; Z_i\}$ are drawn independently and identically distributed from an arbitrary joint distribution for $i = 1, ..., n$.

A common estimand of interest is the \textit{average treatment effect} (ATE) for a given study: 
$$\tau_{\textsc{ATE}} = \E\left \{ Y(1) - Y(0)\right\}.$$

In practice, a more policy-relevant quantity is the conditional average treatment effect (CATE), which considers the ATE across subsets of individuals in the study population with covariate value $X = x$: 
$$\tau(x) = \E \left\{Y(1) - Y(0) \mid X = x \right\} $$
Existing literature has recommended researchers leverage their substantive expertise to posit what characteristics could potentially moderate the treatment effect. 

The primary focus of this paper considers how to construct interpretable subgroups. We define a \textit{subgroup} as a subset of individuals, defined by a discrete partition in the space of $X$, where the partitions in the space correspond to substantively different treatment effects. In particular, we are interested in characterizing different groups of individuals in a study that have different effects from receiving the same treatment. 

We formally define a \textit{subgroup} $\mathcal{G}'_g: \mathcal{X} \to \{0,1\}$ as
\begin{align}\label{eq:subgroup}
    \cG'_g(X) = \prod_{j=1}^p \1\left\{X^{(j)} \in R^{(j)}_g\right\},
\end{align}
where $X^{(j)}$ denotes the $j^{th}$ covariate, and $R^{(j)}_g \subseteq \text{supp}(X^{(j)})$. Here, $\cG'_g(X)$ represents a collection of binary decision \textit{rules} that characterize a subset of individuals in the study \citep{lipkovich2017tutorial}. For example, if $\cG'_g(X) = \1\{X^{\text{gender}} = \text{Female}\} \cdot \1\{X^{\text{age}} < 35\}$, then $\cG'_g(X)$ corresponds to the subset of female-identifying individuals who are under the age of 35. Throughout the paper, we focus on subgroups that are a discrete partition of the covariate space--i.e., $\sum_{g=1}^G \cG'_g(X_i) = 1$ for all units $i = 1, \ldots, n$ (i.e., each unit belongs to exactly one subgroup). 

Given a subgroup $\cG'_g(X)$, we define $\tau^{(g)}$ as the \textit{subgroup average treatment effect}:
\begin{align}\label{eq:pop_subgroup_ate}
    \tau^{(g)} :=\E \left[ Y(1) -Y(0) \mid \cG'_g(X) = 1 \right].
\end{align}
The primary aim in this paper is to estimate an \textit{optimal partition}, which maximizes treatment effect heterogeneity across subgroups using the most parsimonious set of rules. Specifically, for a fixed cardinality $G$, we focus on constructing the most parsimonious partition that minimizes the squared loss between the conditional average treatment effects and the associated subgroup ATEs: 
\begin{align}\label{eq:optimal_partition}
    \{\mathcal{G}_g(X)\}_{g=1}^G =
    \argmin_{\{\cG'_g(X)\}_g \in \Omega_G} \E\left[ \left\{\tau(X) - \sum_{g} \cG'_g(X) \cdot  \tau^{(g)} \right\}^2 \right],
\end{align}
where $\Omega_G$ represents a class of partitions in the covariate space of $X$ with cardinality $G$. Under squared loss, the optimal partition corresponds to a subset in the covariate space of $X$ that maximizes the difference in treatment effect heterogeneity across subgroups. In other words, \textit{within} a partitioned group, the treatment effect heterogeneity is minimized, thereby \textit{maximizing} the variance in treatment effects \textit{across} groups. In practice, researchers can choose alternative loss functions that correspond to what are substantively meaningful partitions (or subgroups) in the covariate space. However, the theoretical guarantees we derive utilize the squared loss, and generalizing the results for other loss functions is an open avenue of future research.

Our work is distinct from existing literature in considering subgroup ATEs. For example, alternative quantities in the literature include the sorted group average treatment effect (GATE)\footnote{Following \citet{chernozhukov2018double, imai2022statistical, dwivedi2020stable}, we use GATE as an acronym for the sorted group average treatment effect, though we note other papers, such as \citet{heiler2025heterogeneity} and \citet{lechner2024comprehensive} use the acronym to generally refer to any group average treatment effect. One can view the subgroup average treatment effect as a special case of the generic group average treatment effect, with the goal of the paper to estimate the group membership.}, which considers the ATE across subsets formed based on the magnitude of the estimated effect
(i.e., the ATE across individuals who would most benefit) \citep[e.g.,][]{imai2022statistical, chernozhukov2018generic, dwivedi2020stable}. GATE relies on using a machine learning model to generate predictions of the CATE, and then groups the predicted CATEs by highest to lowest to estimate the ATE across each group. While quantities like GATE provide a way for researchers to evaluate the range of possible treatment effects within a study, it falls short of providing interpretability for the \textit{characteristics} of the units within each group.

\section{Causal distillation trees} \label{sec:cdt}
\subsection{Overview}\label{subsec:cdt}
\begin{figure}
    \centering
    \includegraphics[width=0.95\linewidth]{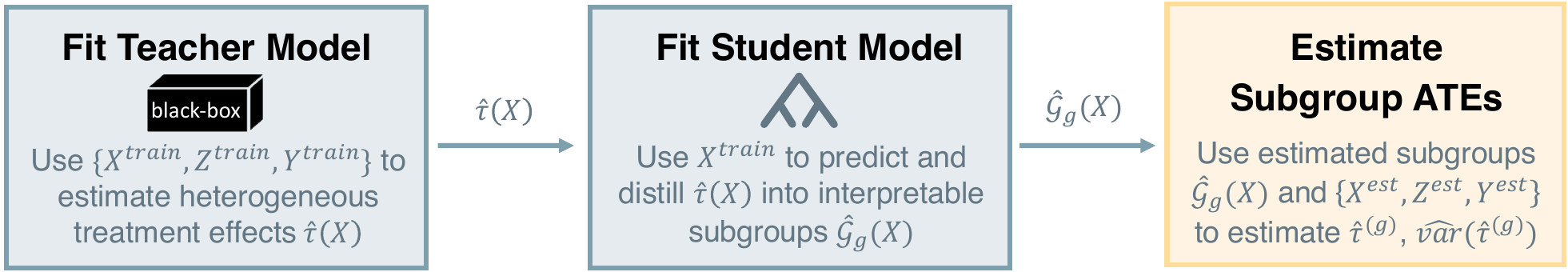}
    \caption{Overview of Causal Distillation Trees (CDT). CDT leverages a two-stage procedure, which first fits a teacher model to estimate heterogeneous treatment effects and then fits a student model (e.g., a decision tree) to distill the estimated heterogeneous treatment effects and produce interpretable subgroups. The two-stage learner is learned using the training data (blue-gray boxes). Using the estimated subgroups, the subgroup ATEs are honestly estimated with a held-out estimation set (yellow box).}
    \label{fig:cdt}
\end{figure}

We propose \textit{causal distillation trees} (CDT) to stably estimate interpretable subgroups. CDT is a two-stage learner (Figure~\ref{fig:cdt} and Algorithm~\ref{alg:cdt}). The first stage learner, referred to as the \textit{teacher} model, learns an informative model for the heterogeneous treatment effect using the observed data $\{Y, \; Z,\; X\}$. Using this teacher model, we generate a prediction $\hat \tau(X_i)$ for the heterogeneous treatment effect for all units.\footnote{With slight abuse of notation, $\hat \tau(X_i)$ corresponds to the predicted conditional average treatment effect for unit $i$ (i.e., for unit $i$, with covariates $X_i = x$, $\hat \tau(X_i) := \hat \tau(x)$)} In the second stage, we \textit{distill} the predicted $\hat \tau(X_i)$'s into interpretable subgroups by fitting a decision tree, called the \textit{student} model, to predict the $\hat \tau(X_i)$'s from the covariates $X$. The estimated decision rules from the decision tree define a discrete partition, which naturally maps to the collection of subgroups $\left \{ \hat \cG_1(X),\; \ldots,\; \hat \cG_G(X)\right\}$ (see Figure~\ref{fig:tree_to_subgroup}). Finally, using the estimated partitions, we estimate the subgroup ATE for each subgroup.

Informally, the first stage learner smooths the heterogeneous treatment effects by projecting the observed data into the basis space of the covariates $X$. By using the projected version of the heterogeneous treatment effect instead of the original outcomes, the second stage learner will be able to more stably estimate the subgroups, as the teacher model will have de-noised the outcomes. 

Throughout the paper, we use a decision tree---specifically, a classification and regression tree (CART) \citep{breiman1984classification}---as the second-stage learner. While researchers can use alternative interpretable models as the second-stage learner \citep[e.g.,][]{friedman2008predictive, bargagli2020causal, singh2021imodels, wang2022causal, wan2023rule, tan2025fast}, we focus on CART because of its popular use in practice and because the output maps intuitively to a standard interpretation of a subgroup as a discrete partition of the covariate space. In settings when researchers are interested in a \textit{linear} CATE, it would be useful to consider alternative second-stage learners. 

We also incorporate careful sample splitting in CDT to avoid post-selection bias. Specifically, we randomly split the data into a training set to estimate the subgroups $\left \{ \hat \cG_1(X), \ldots, \hat \cG_G(X)\right\}$ and a hold-out estimation set to estimate the subgroup ATEs $\hat{\tau}^{(1)}, \ldots, \hat{\tau}^{(G)}$ (see Figure~\ref{fig:cdt}). This provides an honest estimate of the subgroup ATE \citep{athey2016recursive}. In addition, to avoid overfitting to the teacher model, we estimate the $\hat \tau(X)$'s using out-of-sample procedures (e.g., using out-of-bag samples \citep{wager2018estimation} or cross-fitting \citep{chernozhukov2018double}) in the first stage (details in Algorithm~\ref{alg:cdt}).

\subsection{Consistency of subgroup estimation: Single rule setting} \label{subsec:single}

We now investigate the theoretical properties of CDT. We show that under a set of regularity assumptions, CDT consistently recovers the optimal subgroups, and can improve the rate of convergence in estimating subgroups over standard decision trees. The rate of convergence depends on (1) the smoothness in the underlying CATE, and (2) how much noise there is in the underlying process. We focus our discussion on standard greedy tree algorithms, such as CART. In settings when researchers employ alternative algorithms to construct decision trees, some of the assumptions for consistency can be potentially relaxed. 

Throughout, we denote $\E_n(A)$ as the expectation of a variable $A$ over the observed sample (i.e., $\E_n(A) := \frac{1}{n} \sum_{i=1}^n A_i$), while $\E(A)$ refers to the population-level expectation. 
Furthermore, we define $f_\tau(X;k)$ as the population-level prediction generated from a decision tree estimated with the oracle CATEs $\tau(X)$ and a single split at threshold $k$ (i.e., $f_\tau(X;k) := \1\{X \leq k \} \E(\tau(X) \mid X \leq k) + \1\{X > k\} \E(\tau(X) \mid X > k)$). Similarly, we define $f^d_\tau(X;k)$ as the population-level prediction generated from a decision tree using the estimated CATEs $\hat \tau(X)$ (i.e., $f^d_\tau(X;k) := \1\{X \leq k \} \E(\hat \tau(X) \mid X \leq k) + \1\{X > k\} \E(\hat \tau(X) \mid X > k)$).

In this section, we will consider the simplest univariate setting, in which we observe a single covariate and the underlying subgroups are defined by a single rule. Section~\ref{subsec:multi} will then generalize the results to show consistency of the estimated subgroups, comprised of multiple rules in the multivariate setting. More formally, let $X \in \mathcal{X} \subseteq \R$, with the underlying subgroups defined as $\cG_1(X) = \1\{X \leq s\}$ and $\cG_2(X) = \1\{X > s\}$ for some optimal split threshold $s \in \R$. We define an \textit{optimal split} under the squared loss as follows.

\begin{definition}[Optimal Split] \label{def:optim_split}
An optimal split $s \in \mathcal{X}$ satisfies
    $s := \argmin_{s' \in \mathcal{X}} \E \left[\ell(\tau(X); X, s') \right],$
where $\ell$ is defined as the squared loss function: 
$\ell(\tau(X); X, s') := ( \tau(X) - f_\tau(X;s') )^2.$
\end{definition}

To ensure that there exists an optimal split, we must assume the following. 

\begin{assumption} \label{assum:min} 
Let $F_X$ be the cumulative distribution function of the covariate $X$. Define the points $s_l$ and $s_u$ such that $0 < F_X(s_l) < F_X(s_u) < 1$. Let $s$ be an optimal split as defined in Definition \ref{def:optim_split}. Assume that (i) $F_X$ is absolutely continuous with a density $dF_X$ that is bounded away from zero and is continuous in a neighborhood of $s$, and (ii) $s$ is unique.
\end{assumption} 
Assumption \eqref{assum:min} guarantees that there exists subgroups of interest to estimate. Intuitively, the constraint on $F_X$ ensures that the squared loss function will attain a minimum value for $s \in [s_l, s_u]$. In particular, we are implicitly ruling out the setting considered in \citet{cattaneo2022pointwise}, in which an optimal split does not exist. Furthermore, Assumption \eqref{assum:min} rules out the possibility that there could be more than one potential split point. 

Given that the subgroups exist, we next show that the estimated subgroups $\hat \cG(X)$ from CDT converge to the true subgroups $\cG(X)$. Informally, to establish the consistency of CDT, we must consider two gaps. First, CDT uses the predicted CATEs $\hat \tau(X)$ in lieu of the true CATEs $\tau(X)$, which are inherently latent. Second, we must show that the within-sample optimal partitioning will asymptotically recover the population-level optimal partitioning.

To start, we assume that the amount that the true CATE $\tau(X)$ changes as the split point $s'$ changes within a neighborhood of $s$ is bounded. 

\begin{assumption} \label{assum:smooth}
Let $\tau_0 = \frac{1}{2} \left\{\E(\tau(X) \mid X \leq s) + \E(\tau(X) \mid X > s) \right\}$. Define $H(s) = \E \left[(\tau(X) - \tau_0) 1\{X \leq s\}\right]$. Let $\mathcal{N}$ be a neighborhood around $s$, where $\mathcal{N} := (l,u) \subset [s_l, s_u]$. Assume the following holds: 
\begin{enumerate}[left=0em]
\item[(a)] $\tau(X)$ is continuous in $(l, s)$ and $(s, u)$, where
$\displaystyle \lim_{x \to s^-} \tau(x) = \tau(s^-) \text{ and } \lim_{x \to s^+} \tau(x) = \tau(s^+)$
\item[(b)] For $\epsilon > 0$, there exists an $1 \leq \alpha \leq 2$ and constants $0 < c_h \leq C_h < \infty$ such that when $|s-s'| < \epsilon$, $\displaystyle c_h  \leq \frac{\lvert H(s) - H(s') \rvert}{\lvert s-s'\rvert^\alpha}  \leq C_h.$
\item[(c)] There exists $0 < \eta \leq 1$ such that $\forall ~\epsilon > 0$, 
$$\E \left[ (\tau(X) - \tau_0)^2 \1\{s - \epsilon \leq X \leq s+\epsilon \} \right]\leq C \epsilon^{2\eta}.$$
\end{enumerate} 
\end{assumption}
Assumption \eqref{assum:smooth} is a standard assumption in establishing consistency of $M$-estimators \citep[e.g.,][Theorem 5.52]{van2000asymptotic}. Informally, $\alpha$ and $\eta$ are parameters that control the smoothness of the underlying function around the optimal split point $s$. The assumption rules out settings where $\tau(X)$ can take on values of infinity close to the optimal split point $s$. In practice, we expect Assumption \eqref{assum:smooth} to generally hold across a large class of data generating processes. For example, when $\tau(X)$ is continuously differentiable in the neighborhood $\mathcal{N}$ and bounded in $s$, then $\alpha = 2$ and $\eta = 1/2$. Assumption \eqref{assum:smooth} also allows for discontinuities at the split point $s$, so long as $\tau(X)$ is continuous within the neighborhood around the split point. 

We assume the conditional average treatment effects have bounded moments.
\begin{assumption}[Bounded Moments] \label{assum:bounded_moments}
Assume there exists constants $M_2 < \infty$ and $M_4 < \infty$ such that
$\E\left\{ \tau(X) ^2 \right\} \leq M_2$ and $\E\left \{ \tau(X)^4 \right\} \leq M_4$.
\end{assumption} 

Finally, we assume $L_2$ consistent estimation of the CATE. 
\begin{assumption}[$L_2$ consistent estimation of CATE] \label{assum:consistent_teacher}
Assume there exists a sequence $\delta_n$ converging to zero as $n$ grows to infinity such that
$$\norm{\hat \tau(X) - \tau(X)}_{2} \leq \delta_n.$$
Furthermore, assume $\norm{\hat \tau(X) - \tau(X)}_\infty \leq c_1 < \infty$.
\end{assumption}
This constrains the amount of error that occurs in recovering the CATE model. Existing literature has established the different convergence rates for different CATE estimators under different function classes and conditions \citep[see, e.g.,][for a few examples]{nie2021quasi, kunzel2019metalearners, kennedy2022minimax}.

With Assumptions \eqref{assum:min}-\eqref{assum:consistent_teacher}, we can establish convergence in recovering the optimal subgroups. 

\begin{proposition}[Convergence in Subgroup Recovery, for Single Covariate Setting] \label{prop:converge_one}
Let $\hat \cG_1(X) := \1\{X \leq \hat s_n\}$ denote the subgroup estimated by CDT. Then, under Assumptions \eqref{assum:min}-\eqref{assum:consistent_teacher}, CDT will consistently recover the optimal subgroups: 
\begin{align*} 
\E\left[ \left \lvert \hat \cG_1(X) - \cG_1(X) \right \rvert \right] \lesssim
 \left \lvert \hat s_{n} - s \right \rvert = O_p \left(n^{-\frac{1}{2(\alpha - \eta)}} + \delta_n^{\frac{2}{2 \alpha - 1}} \right).
\end{align*} 
\end{proposition}
Proposition \ref{prop:converge_one} formalizes that the rate of convergence in subgroup recovery depends on two factors: (1) how smooth the underlying data generating process is (represented by $n^{-\frac{1}{2(\alpha - \eta)}}$); and (2) the convergence of the teacher model's CATE predictions to the true CATEs (represented by $\delta_n^{\frac{2}{2\alpha - 1}}$). 

To build intuition, consider the two terms separately. This first term represents how quickly a decision tree, estimated on the \textit{oracle} CATEs $\tau(X)$, will recover the optimal splits. The rate of convergence here inherits a characteristic highlighted in \citet{escanciano2020estimation} and \citet{song2015asymptotics}, which bridges the gap in convergence rate of an estimated split in a decision tree $\hat s_n$ between a fully continuous setting (considered in \citealp{banerjee2007confidence} and \citealp{buhlmann2002analyzing}) and discontinuous settings (considered in \citealp{chan1993consistency, kosorok2008introduction}). Consider the setting in which $\tau(X_i)$ is continuously differentiable (i.e., $\alpha = 2, \eta = 1/2$). This results in a convergence rate of $O_p(n^{-1/3})$, which recovers the cube root convergence from \citet{buhlmann2002analyzing}. On the other hand, when $\tau(X)$ is discontinuous at $X=s$, then $\hat s_n - s = O_p(n^{-1})$. Intuitively, as the underlying data generating process becomes more discontinuous at the split point $s$, the faster the estimated split $\hat s_n$ will converge to the true split $s$. While these results have been studied in the decision tree and change point estimation literature (see, for example, \citealp{banerjee2007confidence, buhlmann2002analyzing, escanciano2020estimation}), we are, to the best of our knowledge, the first to formally connect the notion of split consistency to subgroup recovery. 

The second term represents how quickly the teacher model's predictions converge to the true CATEs. This formalizes the intuition that in settings when the teacher model outputs noisier predictions of the CATE, the rate of convergence to recovering the optimal subgroups will be slower. In contrast, when the teacher model has less estimation error, then CDT will converge to recovering the optimal subgroups faster. In practice, because researchers cannot use the true CATEs to estimate a decision tree, the second term can also be thought of as a penalty we pay for having to estimate the CATEs.

In addition to consistently recovering the subgroups, CDT improves the stability of the split estimation by de-noising the underlying data and increasing the signal strength. To illustrate, in the following example, we consider a setting in which researchers have access to the true individual-level treatment effect $\tau_i$ and estimate a decision tree using the covariates $X_i$. In practice, this is infeasible, as researchers can never observe $\tau_i$, but this comparison serves as a helpful benchmark to compare the potential improvements from distillation. 

\begin{example}[Improving stability of splits with distillation] \label{ex:split_stability} 
Consider a setting where $\tau(X)$ is continuously differentiable and bounded in a neighborhood around $s$. Furthermore, assume we can write the individual-level treatment effect and the predicted treatment effects as a function of $\tau(X_i)$ and a noise term (i.e.,  $\tau_i = \tau(X_i) + v_i$, and $\hat \tau(X_i) = \tau(X_i) + v_i^{d}$), where $\E(v_i) = \E(v_i^d) = 0$, both $\var(v_i)$ and $\var(v_i^d)$ are finite, and $\var(\tau(X_i)) > 0$. Then, under squared loss and mild regularity assumptions (Assumption \eqref{assum:regularity}), the relative asymptotic variance of the estimated splits without distillation $\hat{s}^{orig}_n$ and with distillation $\hat{s}_n$ is
$$\frac{\asyvar(\hat s_n)}{\asyvar(\hat s^{orig}_n)} = \left(\frac{\text{SNR}_{original}}{\text{SNR}_{distil}} \right)^{2+4/3},$$
where $\text{SNR}_{distil} = \var(\tau(X))/\var(v^{d})$ and  $\text{SNR}_{original} = \var(\tau(X))/\var(v)$. 
\end{example}
The results from Example \ref{ex:split_stability} follow from noting that the asymptotic distribution of the estimated splits in a linear setting follow Chernoff's distribution \citep{buhlmann2002analyzing, groeneboom1989brownian}. While the variance itself is not straightforward to interpret and depends on an Airy function, the \textit{ratio} of variances simplifies into a function of the signal-to-noise ratios (derivation in Appendix \ref{app:proofs}). Example \ref{ex:split_stability} formalizes the intuition that as the first-stage learner more closely approximates $\tau(X)$ (i.e., the conditional expectation function of $\tau_i$, given the pre-treatment covariates), the stability of the estimated splits after distillation also improves. In particular, so long as the variance in $\hat \tau(X)-\tau(X)$ is smaller than the variance in $\tau_i - \tau(X)$, then $\asyvar(\hat s_n) \leq \asyvar(\hat s_n^{orig})$. We can hence view the first-stage learner as a de-noising, or smoothing, step.

\subsection{General setting with multiple rules} \label{subsec:multi}
We next extend the consistency results to the multivariate setting, in which we observe $p$ total covariates (i.e., $X \in \R^p$) and the optimal subgroups, satisfying \eqref{eq:optimal_partition}, consist of an arbitrary number of rules with different covariates and different split points. Throughout, for each subgroup $\cG_g(X)$, we denote the total number of binary decision rules as $r_g$. 

Under this multivariate setting, estimating consistent subgroups requires not only recovering the correct split points, but also selecting the right covariates to split on. We thus must introduce an additional separability condition, which states that conditional on the previously selected binary decision rules, the differences in loss from constructing a relevant and irrelevant rule must be greater than zero (i.e., separable).
\begin{assumption}[Separability Condition]
\label{assum:sep}
For each subgroup index $g \in \{1,\ldots,G\}$ and rule index
$r \in \{1,\ldots,r_g\}$, write the subgroup defined by the $r - 1$ previously selected binary decision rules as $\cG_g^{(r-1)}(X) = \prod_{k=1}^{r-1} \1\left\{ X^{(j_k)} \lesseqgtr s^{(j_k)} \right\}$, where $\cG^{(0)}_g(X) \equiv 1$.
Also, conditioned on the first $r-1$ previously selected rules, let $A_r = \{j_r,\ldots,j_{r_g}\}$ denote the set of remaining relevant subgroup feature indices while $B_r = \{1,\ldots,p\} \setminus A_r$ denotes the set of irrelevant subgroup feature indices. We assume that there is separability, conditional on the previously selected binary decision rules, i.e., there exists a constant $\Delta > 0$ such that
\begin{align*}
\min_{b\in B_r}\E\left[ \ell\!\left(\tau(X);X^{(b)},s^{(b)}\right)\,\middle|\,\cG_g^{(r-1)}(X)=1\right]
-
\min_{a\in A_r}\E\left[\ell\!\left(\tau(X);X^{(a)},s^{(a)}\right)\,\middle|\,\cG_g^{(r-1)}(X)=1\right]
\geq \Delta,
\end{align*}
where $s^{(j)}$ denotes the population-optimal split point for feature
$X^{(j)}$ conditional on $\cG_g^{(r-1)}(X)=1$.
\end{assumption}

Informally, Assumption \eqref{assum:sep} states that given the previously constructed rules, whether a rule is relevant or not must be distinguishable through the loss function. The constant $\Delta$ represents the minimum difference of the population-level loss associated with splitting on a relevant and irrelevant covariate, conditional on the previous splits. 
By assuming $\Delta > 0$, Assumption \eqref{assum:sep} ensures that the underlying subgroups have an optimal substructure so that a greedy second-stage decision tree learner is valid. 
If Assumption \eqref{assum:sep} is not met, standard greedy tree algorithms like CART can fail to recover the optimal partitions even with infinite amounts of data \citep{tan2026statistical}. 

In fact, Assumption~\eqref{assum:sep} is closely related to both the sufficient impurity decrease (SID) \citep{chi2022asymptotic,mazumder2023convergence} and merged staircase property (MSP) \citep{abbe2022merged,tan2026statistical} assumptions, which have played key roles in guaranteeing the consistency of greedy tree algorithms in the previous literature.
Informally speaking, we show in Propositions~\ref{prop:sid} and \ref{prop:msp} that as long as splitting on irrelevant features does not substantially change the population loss, Assumption~\eqref{assum:sep} (a) is equivalent to the SID condition, and (b) in the context of uniform-binary covariates, is equivalent to MSP up to a minor genericity constraint.
By building these connections, we can directly characterize the hierarchical structure that $\tau(X)$ must satisfy in order for Assumption~\eqref{assum:sep} to hold.
In particular, an important implication is that Assumption~\eqref{assum:sep} can be satisfied by a number of common nonparametric function classes, including additive models with components that are strongly increasing functions, smooth and strongly convex functions, and/or polynomial functions \citep{mazumder2023convergence}. A more rigorous and formal discussion of these connections is provided in Appendix~\ref{app:assum5}.

Beyond greedy tree models, recent work has introduced alternative algorithms to construct optimal trees under weaker assumptions, but at a much higher computational cost. These optimal trees are also often restricted to specific settings such as binary features, with limited out-of-sample performance gains relative to greedy trees \citep[e.g.,][]{hu2019optimal, van2024optimal}. Given the popularity of greedy search approaches and the computational burden of many optimal tree construction approaches, we focus our discussion on settings in which greedy search algorithms can feasibly construct optimal subgroups. 

With separability, we can now extend the results from Proposition \ref{prop:converge_one}.

\begin{theorem}[Consistency of subgroup estimation with decision trees] \label{thm:subgroup_consistency}
Let Assumptions \eqref{assum:min}-\eqref{assum:smooth} hold for all relevant covariates, conditional on the previously selected decision rules. Additionally, assume bounded moments (Assumption \eqref{assum:bounded_moments}), consistent teacher model (Assumption \eqref{assum:consistent_teacher}), and separability (Assumption \eqref{assum:sep}). Then, for all groups $g \in \{1, \ldots, G\}$,
\begin{align*} 
\E \left( \left \lvert \hat \cG_g(X) - \cG_g(X) \right \rvert \right) 
\leq& \frac{( 2C_\tau + K \Delta)r_g}{\Delta} \left \lvert \hat s^{(k)}_{n} - s^{(k)} \right \rvert+ \\
&\qquad \frac{2r_g}{\Delta} \left\{ \E_n\left[ \ell(\hat \tau(X); X^{(k)}, s^{(k)})\right] - \E\left[ \ell(\tau(X); X^{(k)}, s^{(k)})\right] \right\} ,
\end{align*} 
where 
$k = \argmax_{k'} \left\{ \left \lvert \hat s_n^{(k')} - s^{(k')} \right \rvert + \E_n \left [ \ell(\hat \tau(X); X^{(k')}, s^{(k')}) \right] - \E\left[\ell( \tau(X); X^{(k')}, s^{(k')}) \right] \right\}$, $K$ is a finite positive constant, and $\Delta$ corresponds to the constant in the separability condition. Furthermore, applying Proposition \ref{prop:converge_one}, $\hat \cG_g(X) \cip \cG_g(X).$
\end{theorem} 
The results of Theorem \ref{thm:subgroup_consistency} follow from first noting that the estimated splits will converge in probability to the optimal split (see Proposition \ref{prop:converge_one}), and then showing that the within-sample loss will converge in probability to the true population-level loss (see Appendix \ref{app:proofs}). In particular, the rate of convergence of $\hat \cG_g(X)$ to $\cG_g(X)$ is dependent on (1) the degree of separability in the losses between relevant and irrelevant rules (i.e., $\Delta$), and (2) the underlying convergence rate of the sample splits to the true optimal splits (i.e., the rate at which $\left \lvert \hat s_n^{(k)} - s^{(k)} \right \rvert \to 0$). From Theorem \ref{thm:subgroup_consistency}, we see that the tree will not only  consistently split at the correct cutoff points, but also correctly select the important features that should be used in the splitting criteria. An immediate implication of Theorem \ref{thm:subgroup_consistency} is that for a fixed $G$, as $n \to \infty$, the true positive rate of the subgroups obtained under CDT will converge to 1, and similarly, the false positive rate will converge to 0. 

In practice, the total number of subgroups present is unknown, and can be treated as a tuning parameter. We show empirically that using standard pruning approaches results in consistent subgroup estimation with CDT (Appendix \ref{app:sim_results}). Furthermore, if researchers want to constrain the maximum number of rules in a subgroup, Theorem \ref{thm:subgroup_consistency} guarantees that for a fixed number of rules, CDT gives the optimal partitioning of that cardinality.

To help illustrate the benefits of CDT, we provide a numerical example that compares the performance of CDT with a standard causal tree. 

\begin{figure}
    \centering
    \includegraphics[width=0.9\linewidth]{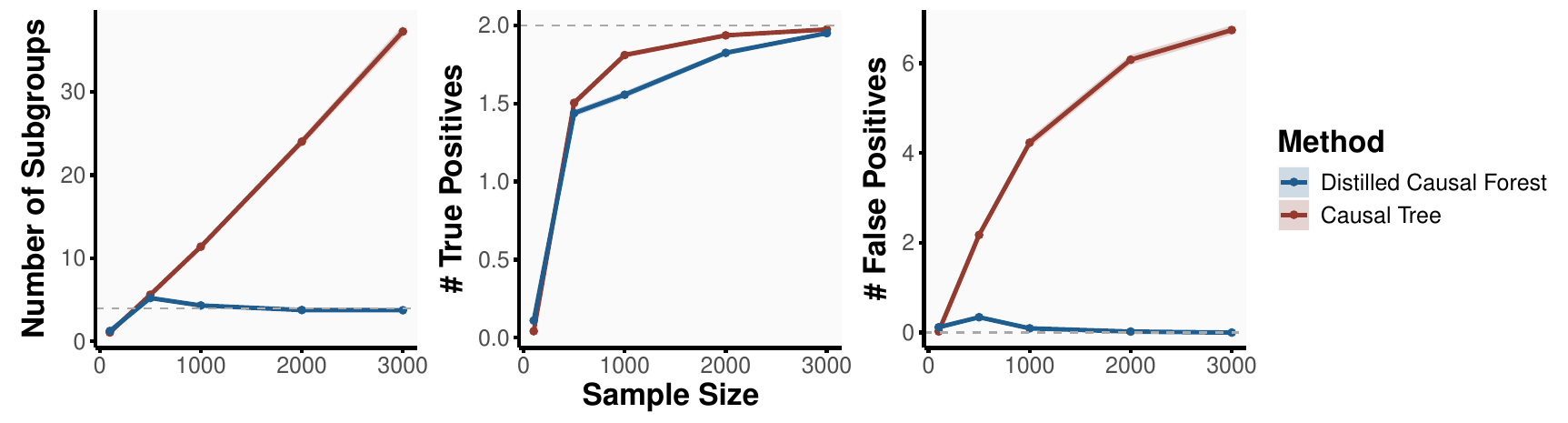}
    \caption{Comparing the performance of causal trees versus CDT (using causal forests as the teacher model), measured via (A) number of estimated subgroups as well as (B) number of true positive and (C) number of false positive features used in the estimated subgroups. The oracle number of subgroups, true positives, and false positives are shown as dashed gray lines. Results are averaged across 500 simulation replicates with ribbons denoting $\pm 1 SE$.}
    \label{fig:toy-example}
\end{figure}

\begin{example}[Numerical comparison of CDT with causal trees] \label{ex:numerical}
Consider a simple data generating process, in which $X_i \iid \text{MVN}(0, I)$ and $\tau_i = 2 \cdot \mathbbm{1} \left \{X_{i}^{(1)} > 0 \right\} - \mathbbm{1}\left \{X_{i}^{(2)} < -0.5 \right\} + \epsilon_i$, where $\epsilon_i \sim N(0, \sigma_{\tau}^2)$ (details in Section~\ref{sec:sims}). We see that the number of estimated subgroups using causal tree increases linearly as a function of the sample size, while CDT correctly recovers the true underlying subgroups. Figure \ref{fig:toy-example} visualizes the results.
\end{example} 

To summarize, we have shown that under a set of regularity assumptions and a well-specified teacher model, CDT is able to consistently recover the set of optimal subgroups. Furthermore, CDT improves the efficiency of subgroup recovery over existing tree-based approaches by first de-noising the underlying data using the teacher model. Distilling thus allows researchers to exploit the interpretable subgroup structure of decision trees, while offsetting the instabilities that usually adversarially impact tree estimation.

\subsection{Estimating subgroup average treatment effects}\label{subsec:subgroup-ate}
Next, we consider how to estimate the subgroup ATEs. For simplicity, we focus on an experimental setting, with extensions for estimating the subgroup ATE in observational settings in Appendix \ref{app:dr}. 
More formally, we assume there is random treatment assignment in the study, such that the treatment indicator $Z$ is independent of $Y(1), Y(0)$, and $X$. 
\begin{assumption}[Unconfounded Treatment Assignment] \label{assum:random} 
$\{Y(1), Y(0), X\} \ \indep \ Z$
\end{assumption} 
We also assume positivity, such that all units have a non-zero probability of receiving treatment. 
\begin{assumption}[Positivity] \label{assum:positivity}
There exists some constant $0 < \eta \leq 0.5$ such that
$\eta < \Pr(Z = 1) \leq 1-\eta.$
\end{assumption} 

Finally, we define a subgroup difference-in-means estimator $\hat \tau^{(g)}$ for $g = 1, \ldots, G$ as
\begin{equation} 
\hat \tau^{(g)} := \frac{1}{\sum_{i=1}^n Z_i \cdot \hat \cG_g(X_i)} \sum_{i=1}^n Z_i Y_i \hat \cG_g(X_i) - \frac{1}{\sum_{i=1}^n (1-Z_i) \cdot \hat \cG_g(X_i)} \sum_{i=1}^n (1-Z_i) Y_i \hat \cG_g(X_i).
\label{eqn:subgroup_dim} 
\end{equation}
Following existing literature, we adopt a design-based perspective, where the potential outcomes $Y(1), Y(0)$ are treated as fixed, such that randomness arises only from the treatment assignment process. Furthermore, we condition on a set of valid randomizations---i.e., the set of random treatment allocations, such that there are treatment and control units in each subgroup $g \in \{1, \ldots, G\}$.\footnote{Because of the sparsity induced by the causal distillation trees, we expect that the total number of subgroups $G$ should be small, relative to the sample size $n$. As such, we anticipate this restriction will have little impact on the results in practice. See \citet{schochet2024design} and \citet{miratrix2013adjusting} for more discussion.} With some abuse of notation, we suppress the explicit conditioning in the results presented next. Finally, we assume throughout that the observed sample is randomly drawn from the larger population.\footnote{For settings when the study sample is not a random sample from the population of interest, researchers must account for the confounding effects of selection. See e.g., \citet{huang2024towards, zhang2024minimax, huang2024sensitivity} for more discussion.}
Thus, the expectations and variances in the following subsection are taken over the treatment assignment process and the sampling process. We consider the setting in which researchers are interested in the finite-population subgroup ATE in Appendix \ref{app:discussion_pop}.

Because we are using sample splitting and honest estimation to estimate the subgroups, the estimated subgroups can be treated as \textit{a priori} defined strata within the study. As such, under the same assumptions as Theorem \ref{thm:subgroup_consistency}, the subgroup difference-in-means estimator, defined in Equation \eqref{eqn:subgroup_dim}, provides a consistent estimate of the true subgroup ATEs.

\begin{theorem}[Consistency and Asymptotic Sampling Distribution of Subgroup Difference-in-Means Estimator] \label{thm:cip_group_dim}
Under Assumptions \eqref{assum:min}-\eqref{assum:positivity}, the subgroup difference-in-means estimator will be a consistent estimator for the sample subgroup ATE. Furthermore, assume the proportion of units in each subgroup converges to a proportion $\pi^*_g \in (0,1)$ (i.e., $n_g/n \to \pi^*_g$ and $n \to \infty$, where $\sum_{g=1}^G \pi^*_g = 1$). Assume there exists some constant $c> 0$ such that $\min_{g} \pi^*_g (1-\pi^*_g) = c$. Then, 
$$\frac{\hat \tau^{(g)} - \tau^{(g)}}{\sqrt{\mathbbm{V}(\hat \tau^{(g)})}} \cid N(0,1),$$
where $\mathbbm{V}(\hat \tau^{(g)}) = \tilde \beta_g(1) \var_g\{Y(1)\} + \tilde \beta_g(0) \var_g\{Y(0)\}$, $\tilde \beta_g(z) = \E\{1/\sum_{i=1}^n \hat \cG_g(X_i)\1\{Z_i = z\}\}$, and $\var_g\left\{Y(z)\right\} := \E \left\{\frac{1}{n_g-1}\sum_{i=1}^n \hat \cG_g(X_i) (Y_i(z)- \right.$ $\left. \overline{Y_g(z)})^2\right\}$ for $z \in \{0,1\}$.
\end{theorem} 
The restriction $\min_g \pi^*_g (1-\pi^*_g) = c > 0$ rules out settings with near zero overlap. See \citet{khan2010irregular, hong2020inference, heiler2021valid} for discussion on how to deal with such settings.  

From Theorem \ref{thm:cip_group_dim}, we can also construct a variance estimator using the sample analogs for $\var_g(Y_i(z))$ for $z \in \{0,1\}$:
\begin{align}
    \widehat{\var}(\tau^{(g)}) = \frac{1}{n_g}  \left\{ \hat \beta_g(1) \widehat{\var}_g(Y_i(1)) + \hat \beta_g(0) \widehat{\var}_g(Y_i(0)) \right\} \label{eq:vhat}
\end{align}
where $\hat \beta_g(z) = \left( \frac{1}{n_g}\sum_{i=1}^n \hat \cG_g(X_i) \1\{Z_i = z\} \right)^{-1}$. \citet{miratrix2013adjusting} showed that the gap between $\hat \beta_g(z)$ and $\tilde \beta_g(z)$ can be upper bounded as a function of $1/n$. Asymptotically, this will be equivalent to the robust Huber–White variance estimator.

In the Appendix, we provide several extensions for other settings. We apply standard nonparametric tests of treatment effect heterogeneity for researchers to evaluate whether or not the different subgroup ATEs are statistically significantly different from one another (see Appendix \ref{subsec:nonparam-test}). Furthermore, we also discuss how to perform inference in settings when researchers are interested in performing inference for the largest treatment effect across the subgroups (see Appendix \ref{app:selection-inf}). 

\section{Stability-driven teacher model selection}\label{sec:model_selection}

In Section \ref{sec:cdt}, we showed that the benefits of distillation arise from the teacher model's ability to construct a meaningful basis representation of the individual-level treatment effect using the covariate data. While numerous metalearners have been developed and can be used as the teacher model, researchers may wonder how to choose an appropriate teacher model for CDT. Traditional model selection procedures cannot be used to evaluate the goodness-of-fit of our teacher models, which predict heterogeneous treatment effects --- quantities that are not observable in practice. Furthermore, checking whether the predicted outcomes match the observed outcomes is insufficient to guarantee that the \textit{treatment effects} are modeled well \citep{kunzel2019metalearners}. 

Given these challenges, we propose a novel stability-based model selection procedure, designed specifically for subgroup estimation. Our proposed procedure uses the training split to select a teacher model based on its ability to stably reconstruct subgroups. Stable reconstruction refers to being able to estimate similar subgroups given various perturbations in the underlying data generating process. We consider more stable teacher models as being more appropriate models for the data, as they demonstrate greater robustness to sampling error \citep[e.g.,][]{yu2013stability, yu2020veridical}. Moreover, as we will see through simulations in Section~\ref{sec:sims}, teacher models with higher stability generally correspond to more accurate estimation of the heterogeneous treatment effects and thus provides potential plausibility for the validity of the teacher model.

To begin, we must quantify what it means for two subgroups to be `similar' and develop a novel measure, referred to as the Jaccard Subgroup Similarity Index (SSI). Informally, SSI considers two different sets of estimated subgroups $\{\hat \cG^{(1)}, \hat \cG^{(2)}\}$ and computes the proportion of units that belong to the same subgroup across both partitions. If the units are similarly grouped in the two different partitions, then $\hat \cG^{(1)}$ and $\hat \cG^{(2)}$ are similar, and SSI is high. If the units are grouped differently, then $\hat \cG^{(1)}$ and $\hat \cG^{(2)}$ are less similar, and SSI will be low. 

Formally, let $\hat \cG^{(1)} = \{\hat \cG^{(1)}_1(X), \ldots, \hat \cG^{(1)}_G(X)\}$ and $\hat \cG^{(2)} = \{\hat \cG^{(2)}_1(X), \ldots, \hat \cG^{(2)}_G(X)\}$ denote two different candidate sets of subgroups. Then, for $k \in \{1,2\}$, construct the adjacency matrix $C^{(k)} \in \R^{n \times n}$ which denotes whether or not two distinct units belong in the same subgroup in $\hat \cG^{(k)}$. Specifically, for $i, j = 1, \ldots, n$, let $C_{ij}^{(k)} = 1$ if $X_i$ and $X_j$ belong to the same subgroup in $\hat \cG^{(k)}$ and $i \neq j$, and 0 otherwise. We define the \textit{Jaccard Subgroup Similarity Index} (SSI) as 
\begin{align}\label{eq:jaccard_subgroup}
    \text{SSI}(\hat \cG^{(1)}, \hat \cG^{(2)}) = \frac{1}{2 G} \sum_{\cH \in \hat \cG^{(1)} \cup \hat \cG^{(2)}}\frac{N_{11}(\cH)}{N_{01}(\cH) + N_{10}(\cH) + N_{11}(\cH)},
\end{align}
where $N_{qr}(\cH) = \lvert \{(i, j) \in [n] \times [n]: C^{(1)}_{ij} = q, \, C^{(2)}_{ij} = r, X_i \in \cH(X)\} \rvert$ for each subgroup $\cH \in \hat \cG^{(1)} \cup \hat \cG^{(2)}$ and $q, r = 0, 1$. In words, $N_{11}(\cH)$ is the number of sample pairs that belong to the same subgroup in both partitions ($\hat \cG^{(1)}$ and $\hat \cG^{(2)}$), conditioned on the sample pair belonging to subgroup $\cH(X)$; $N_{01}(\cH) + N_{10}(\cH)$ is the number of sample pairs that belong to subgroup $\cH(X)$ in one partition but two different subgroups in the other partition. 

The subgroup similarity index is bounded between 0 and 1. When SSI is close to 1 (i.e., when $N_{11}(\cH)$ is much larger than $N_{01}(\cH) + N_{10}(\cH)$), this implies the two sets of subgroups are similar. Figure~\ref{fig:ssi} illustrates the computation of SSI for an example subgroup $\cH \in \hat \cG^{(1)} \cup \hat \cG^{(2)}$. SSI is similar to the Jaccard index \citep{jaccard1901etude} used to compare similarities across clusters \citep{ben2001stability}. However, unlike the standard Jaccard index, which weights individual units equally, SSI re-weights units to give equal weight to each subgroup. This penalizes subgroups with very few units, which are typically highly unstable.\footnote{We can consider the extreme setting, in which a subgroup contains only a single individual. An estimated subgroup with only one unit would have very little impact on the standard Jaccard index, even though in practice, such a subgroup would not be substantively meaningful and would be undesirable.}

With SSI, we can then evaluate the similarity of estimated subgroups from each candidate teacher model across repeated data samples (e.g., via bootstrapping). 
Note however to ensure a fair comparison across the different data resamples and teacher models, SSI is always computed between subgroups with the same number of partitions by comparing (unpruned) trees that have grown to the same depth.
Additional details and the full proposed teacher model selection procedure are provided in Algorithm~\ref{alg:jaccard}.
Using this algorithm, the teacher model that produces the highest SSI across the different bootstrapped iterations corresponds to the most stable teacher model and should be selected.

In Section~\ref{sec:sims}, we will demonstrate that more stable teacher models often correspond to more accurate estimation of the underlying subgroups and treatment effects across a wide variety of simulations.
However, it is also important to caveat that this is not necessarily in all settings.
In general, the error (i.e., MSE) of the subgroup estimates can be decomposed into a bias and a variance component, and our stability-based procedure serves as a proxy for the variance component.
If the bias of the teacher model is large, then the most stable teacher model may not necessarily be the best teacher model for the data.
However, many commonly-used teacher models (e.g., based on random forests, boosting) are flexible and designed to minimize bias. 
In such cases, we anticipate the teacher model bias to be relatively low (or at least similar relative to each other\footnote{As long as the bias of the teacher models does not change the relative ordering of the overall subgroup estimation errors, selecting the teacher model based on stability will result in the same teacher model as if we had selected based on the overall subgroup estimation error.}); as a result, the variance of the subgroup estimates will dominate the subgroup estimation error, and the proposed stability-based teacher selection procedure will be most effective.

\section{Simulations}\label{sec:sims}
In this section, we demonstrate the effectiveness of CDT in accurately estimating subgroups through extensive simulations. We consider four different teacher models: causal forest \citep{wager2018estimation}
, Bayesian causal forest (BCF) \citep{hahn2020bayesian}
, R-learner with boosting (Rboost), and R-learner with natural cubic splines (Rspline) \citep{nie2021quasi}.\footnote{Rspline is a knowingly misspecified teacher model, but included to demonstrate the effectiveness of the teacher model selection procedure.} We benchmark CDT's performance, relative to popular subgroup estimation approaches --- namely, causal trees \citep{athey2016recursive}, virtual twins \citep{foster2011subgroup}, and interacted linear models, fitted without regularization (i.e., ordinary least squares) and with Lasso ($L_1$) regularization \citep[e.g.,][]{imai2013estimating}. Implementation and method details can be found in Appendix~\ref{app:sims}. Across the different simulation scenarios, we find that CDT consistently estimates the true subgroup features, thresholds, and CATEs with greater accuracy than existing methods. Moreover, our model selection procedure using the Jaccard SSI provides an effective data-driven approach to choosing an appropriate teacher model for CDT.

\paragraph{Simulation Setup.} We generate a random treatment assignment process, where $Z_i \iid \text{Bernoulli}(1/2)$, and generate $n = 500$ samples and $p = 10$ covariates from a standard multivariate normal distribution (i.e., $X_i \iid \text{MVN}(0, I)$, where $X_i \in \R^{p}$ and $i = 1, \ldots, n$). Furthermore, we simulate three different treatment effect heterogeneity models: 
\begin{enumerate}[noitemsep]
    \item `AND' Subgroup DGP: $\tau_i = 2 \mathbbm{1}\left\{X_{i}^{(1)} > 0 \right\} \cdot \mathbbm{1}\left\{X_{i}^{(2)} > 0.5\right\} + \epsilon_i$
    \item `Additive' Subgroup DGP: $\tau_i = 2 \mathbbm{1} \left \{X_{i}^{(1)} > 0 \right\} - \mathbbm{1}\left \{X_{i}^{(2)} < -0.5 \right\} + \epsilon_i$
    \item `OR' Subgroup DGP: $\tau_i = 2 \mathbbm{1}\left\{X_{i}^{(1)} > 0 \right\} - \mathbbm{1} \left\{(X_{i}^{(2)} > 0.5) \text{ or } (X_i^{(2)} < -0.5)\right\} + \epsilon_i$
\end{enumerate}
where $\epsilon_i \stackrel{iid}{\sim} N(0, \sigma_{\tau}^2)$, and we define the outcome process to be a function of this underlying treatment effect heterogeneity: 
$Y_i = Z_i \cdot \tau_i + X_i^{(3)} + X_i^{(4)} + \nu_i,$
where $\nu_i \stackrel{iid}{\sim} N(0, 0.1^2)$. 
We vary $\sigma_{\tau}$ such that the proportion of variance explained in $\tau(X)$ by the covariates $X$ vary across the values $\{0.2, 0.4, 0.6, 0.8, 1\}$.\footnote{The proportion of variance explained in $\tau(X)$ by $X$ is defined as $\var(\E[\tau_i \mid X]) / \var(\tau_i) \in [0, 1]$.} 
To evaluate the effectiveness of each subgroup estimation method, we consider how accurately each method recovers (1) selected subgroup features, (2) estimated subgroup thresholds, and (3) estimated subgroup ATEs (details in Appendix~\ref{app:sim_results}). We also consider additional data generating processes in Appendix \ref{app:sim_results}, and find that the results are largely consistent, even across more complex settings. 

\begin{figure}[tb]
    \centering
    \includegraphics[width=1\linewidth]{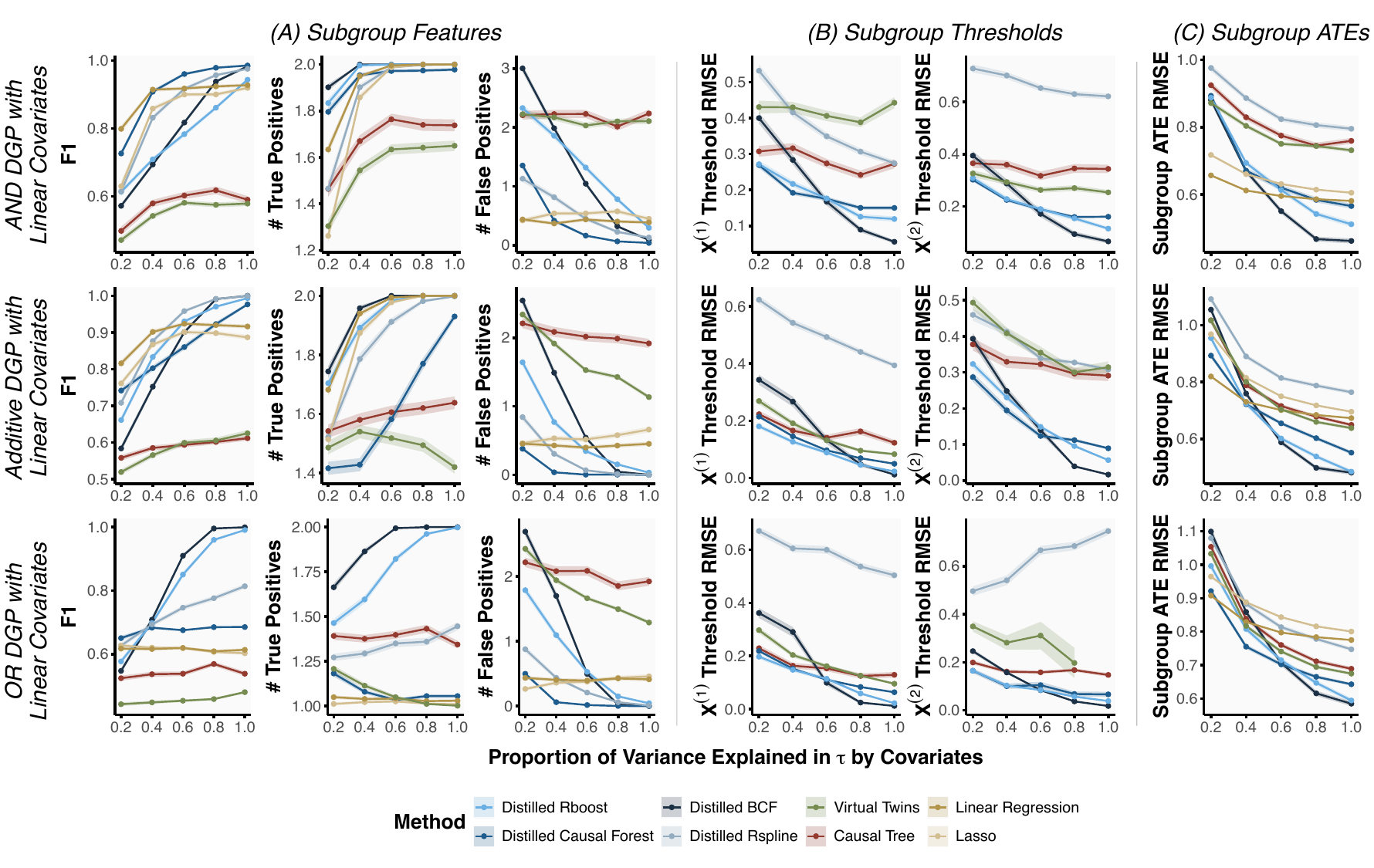}
    \caption{Performance of subgroup estimation methods for (A) identifying the true subgroup features, measured via $F_1$ score, number of true positives, and number of false positives, (B) estimating the true subgroup thresholds, measured via RMSE for each true subgroup feature, and (C) estimating the true subgroup ATE, measured via RMSE, across increasing treatment effect heterogeneity strengths (x-axis) and different subgroup data-generating processes with linear covariate effects (rows). Results are averaged across 500 simulation replicates with ribbons denoting $\pm1SE$.} 
    \label{fig:sim-subgroup-errors-cov}
\end{figure}

\paragraph{Subgroup Estimation Results.} Figure \ref{fig:sim-subgroup-errors-cov} visualizes the key simulation results, with more detailed summaries in Appendix \ref{app:sim_results}. Across the different simulation settings, we find that the CDT models with the tree-based teachers (i.e., Distilled Causal Forest, Distilled BCF, and Distilled Rboost) are able to accurately estimate the true subgroups in terms of finding the correct features (as measured by $F_1$ score in Figure \ref{fig:sim-subgroup-errors-cov}A), estimating the correct thresholds to construct the decision rules (Figure \ref{fig:sim-subgroup-errors-cov}B), and estimating the true subgroup ATEs (Figure \ref{fig:sim-subgroup-errors-cov}C). 
Meanwhile, Distilled Rspline struggles to estimate the correct subgroup thresholds, resulting in poor subgroup ATE estimation. This is consistent with the theoretical findings in Section \ref{sec:cdt}. The R-learner with natural cubic splines is a misspecified teacher model, which cannot capture the sharp discontinuities in the underlying treatment effect function. As a result, the corresponding misspecification error results in worse subgroup recovery.

In contrast to the distilled methods, standard subgroup estimation methods (i.e., causal trees and virtual twins) often result in many erroneously constructed subgroups, overfitting to noise features (shown by the high number of false positives). This reinforces our theoretical findings from Section \ref{sec:cdt} that distillation helps mitigate potential overfitting in the presence of noise. Furthermore, linear approaches (i.e., linear regression and Lasso) suffer from the opposite issue and often fail to pick up on existing subgroups (shown by the low number of true positives). This is expected given that these linear approaches are misspecified for the underlying data-generating processes, which are non-linear, thresholding functions.
We also find these linear methods to be highly sensitive to the form of the outcome model while CDT is relatively robust to the outcome model specification (see Figure~\ref{fig:sim-subgroup-errors-0} and Appendix~\ref{app:sim_results} for additional results).

\paragraph{Teacher Model Selection Simulations.} To assess our proposed teacher model selection procedure, we computed the Jaccard SSI for Distilled Rboost, Distilled Causal Forest, Distilled BCF, and Distilled Rspline using $B = 100$ bootstrap samples and four different choices of tree depths ($d = 1, 2, 3, 4$). As a baseline, we also computed the Jaccard SSI for the causal tree. 

\begin{figure}[tb]
    \centering
    \includegraphics[width=\linewidth]{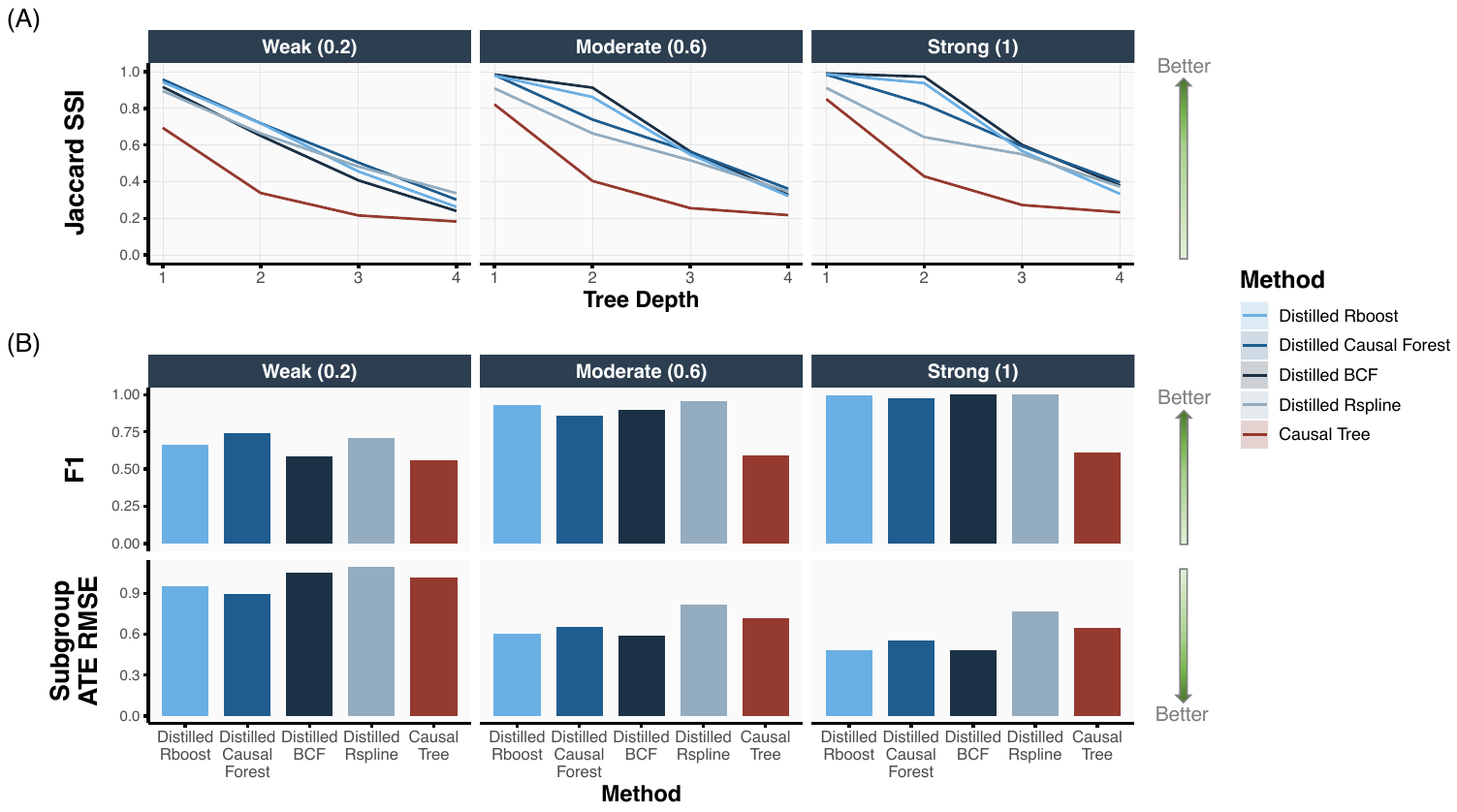}
    \caption{Under the `Additive' subgroup data-generating process with linear covariate effects, we examine (A) the Jaccard SSI across a range of tree depths and (B) the corresponding subgroup estimation accuracy for various subgroup estimation methods (colors) and treatment effect heterogeneity strengths, measured via the proportion of variance explained in $\tau$ by the covariates (columns). Choosing the teacher model in CDT which leads to the highest Jaccard SSI generally corresponds to more accurate subgroup estimation. Results are averaged across 500 simulation replicates.}
    \label{fig:jaccard-and-cov}
\end{figure}

Figure~\ref{fig:jaccard-and-cov}~summarizes the results under the `Additive' subgroup simulation.
Regardless of the noise level, teacher models which yield higher subgroup Jaccard stability scores, particularly at the tree depths of 1 and 2, tend to also yield more accurate subgroup estimation (i.e., lower RMSE in estimating the subgroup ATEs). 
Specifically, at weak treatment effect levels, Distilled Rboost and Distilled Causal Forest demonstrate the greatest stability and lowest subgroup ATE RMSEs. 
At moderate-to-strong treatment effect levels, Distilled BCF, followed by Distilled Rboost, exhibit the greatest stability and best subgroup ATE estimation.
Most importantly, the misspecified teacher model, Distilled Rspline, clearly yields the lowest Jaccard SSI among the candidate CDT teacher models and therefore would not be selected by our proposed model selection procedure.

It is also important to note here that our interpretations focus on the first two tree depths since the underlying treatment effect heterogeneity maps to a tree of depth 2.
For more realistic settings where the true oracle tree depth is unknown, we provide guidance and discussion on choosing an appropriate tree depth in Appendix~\ref{app:jaccard}.
We also provide the teacher model selection results for all other simulation scenarios in Appendix~\ref{app:jaccard}, which exhibit similar patterns and further demonstrate the efficacy of our stability-driven procedure to select the teacher model.

\section{Case Study: AIDS Clinical Trials Group Study 175}\label{sec:case_study}

To illustrate the stability and interpretability of CDT on real data, we turn to the AIDS Clinical Trials Group Study 175 (ACTG 175). ACTG 175 was a randomized controlled trial to determine the effectiveness of monotherapy compared to combination therapy on HIV-1 infected patients. To enter the study, participants' CD4 cell counts had to be between 200 to 500 cells per cubic millimeter upon screening. The main outcome of interest was whether a patient reached the primary end point ($Y_i = 0$), defined by at least a 50\% decline in CD4 cell count, the development of acquired immunodeficiency syndrome (AIDS), or death. 

Given that combination therapy is already known to slow the progression of HIV-1 on average, our goal is to further characterize subgroups of the population that may experience varying effectiveness of combination therapy using clinical and demographic variables as pre-treatment covariates. 
Moreover, we specifically focus on the impact of zidovudine and zalcitabine as our treatment compared to a monotherapy regime using zidovudine.
In total, there are $n = 1056$ participants, with an approximately balanced number of participants undergoing combination therapy and monotherapy. 
For each participant, the study measured $p = 12$ pre-treatment covariates (details in Table~\ref{tab:features})

\begin{figure}[tb]
    \centering
    \includegraphics[width=0.8\linewidth]{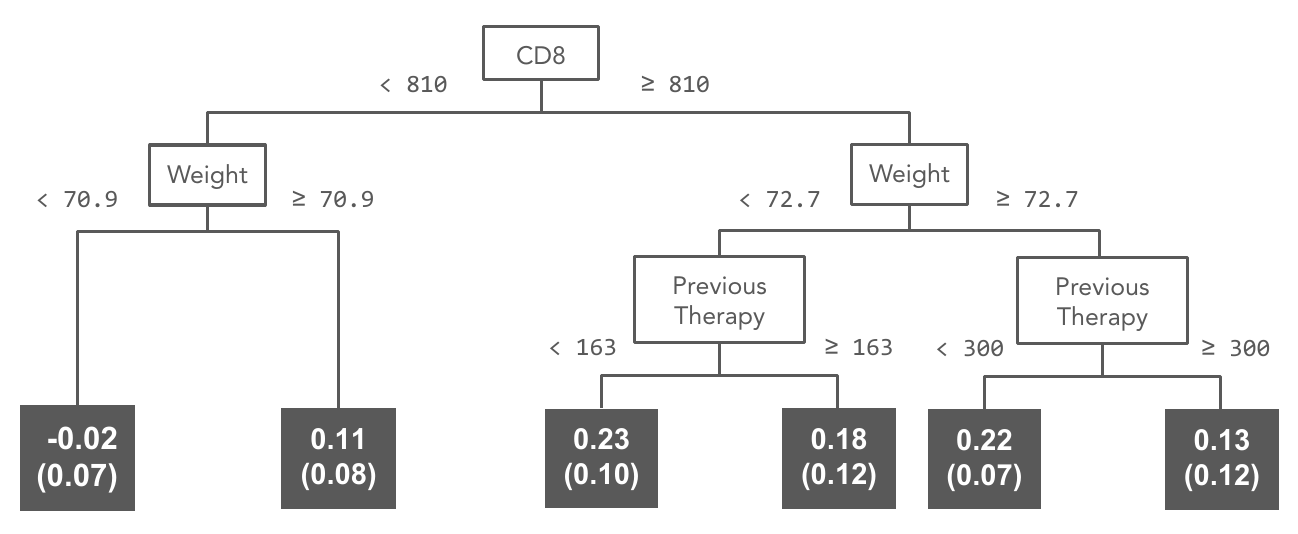}
    \caption{Decision tree produced by CDT using Rboost as the teacher model for AIDS case study. Estimated subgroup ATEs are shown, with standard errors in parentheses (computed using \eqref{eq:vhat}).}
    \label{fig:case-study-tree}
\end{figure}

To begin, we perform a 50-50\% split of the data into a training and a held-out estimation split. 
Using the training data, we apply CDT with the proposed teacher model selection procedure. 
We find that Distilled Rboost results in the highest Jaccard SSI across varying tree depths (Figure~\ref{fig:case-study-jaccard}). 
Thus, leveraging the estimated subgroup partitions from Distilled Rboost, we finally estimate the subgroup ATEs using the held-out estimation split and visualize the resulting tree in Figure~\ref{fig:case-study-tree}.

As seen in Figure~\ref{fig:case-study-tree}, the key covariates in our estimated subgroups are CD8 cell counts, weight (measured in kg) at the start of the study, and the amount of previous exposure to anti-retroviral therapy (measured in days). 
Importantly, these variables are well-known to influence the progression of HIV-1. 
CD8 cell counts in particular are commonly used in combination with CD4 counts \citep{mcbride2017imbalance} to characterize overall immune dysfunction. 
Lower baseline weight often serves as a proxy for HIV-associated wasting, nutritional depletion, or disease severity \citep{mangili2006nutrition}. 
Finally, while previous anti-retroviral therapy can initially help improve CD4 counts, over time, individuals can develop drug resistance and benefit from a different regime \citep{hammer1996trial}. 

These three covariates interact across our subgroups, producing varying treatment effects. 
Notably, it is only those with both low CD8 counts ($< 810$) and low starting weight ($< 70.9$ kg) who appear not to benefit from combination therapy compared to monotherapy.
These participants may be more likely to have advanced HIV-1 disease, which may limit the effectiveness of combination therapy.
In contrast, among participants with high CD8 counts ($> 810$), those with lower previous exposure to anti-retroviral therapy tend to experience greater benefit from combination therapy relative to others in their weight class and yield estimated subgroup ATEs between 0.22-0.23.
For reference, the estimated ATE in this study is 0.13 with a standard error of 0.027 based on a standard difference-in-means analysis.
This increase in benefit is consistent with the clinical understanding that prolonged exposure to anti-retroviral therapy can lead to drug resistance, which may reduce the effectiveness of combination therapy.

To summarize, CDT recovers clinically-relevant subgroups in the AIDS Clinical Trial data that we now know to be substantively important, thus providing a helpful validity check on the usefulness of the proposed method. This illustration also highlights that had researchers utilized CDT when the initial trial was conducted, they could have identified these potential drivers of treatment effect heterogeneity, without knowing them \textit{a priori}. 

We also evaluate the stability of the different methods across bootstrapped samples in Appendix~\ref{app:case-study}, and find distillation methods both consistently split on the same set of features and that estimated subgroups were overwhelmingly characterized by features with direct clinical relevance. In contrast, the other tree-based approaches split on a variety of different features, depending on the bootstrapped sample, while the linear approaches frequently failed to identify meaningful heterogeneity. For details on this stability analysis, we refer interested readers to Appendix \ref{app:case-study}.

\section{Discussion}\label{sec:discussion}
We introduced \textit{causal distillation trees} for interpretable subgroup estimation in causal inference. CDT allows researchers to leverage the flexibility and power of black-box metalearner approaches, while preserving the interpretability of simple decision trees. We prove that CDT allows researchers to consistently estimate the optimal set of subgroups, and we introduce a novel model selection procedure to help researchers use CDT in practice. Furthermore, we have developed an R package \texttt{causalDT} which not only fits CDT but also automatically conducts the subgroup stability and diagnostic checks, thus enabling researchers to interpret their estimated subgroups from CDT alongside important and informative diagnostics in a holistic manner.

There are several exciting avenues for future research. First, recent literature in external validity has emphasized the importance of recruiting a sufficiently heterogenous experimental sample that adequately captures the total treatment effect variation across the target population \citep[e.g.,][]{huang2024towards, huang2024overlap, zhang2024minimax}. CDT offers an opportunity for researchers to characterize the underlying treatment effect heterogeneity in a given study and to consider how to optimally recruit these subgroups of participants for an externally valid study.
Second, CDT provides a very general and flexible two-stage framework. While we focused on CART as the second-stage learner, there are numerous alternative tree-based models \citep[e.g.,][]{loh2002regression, quinlan2014c4, hu2019optimal} and many that have been developed specifically for subgroup estimation \citep[e.g.,][]{loh2015regression, su2009subgroup, lipkovich2011subgroup, seibold2016model, wang2024iterative}, which can be used as second-stage learners in CDT and may further improve performance. Similarly, efforts to further enhance the stability of distillation-based methods \citep{zhou2024approximation} can be easily incorporated into the broader CDT framework.
Third, given that the treatment effects are inherently unobservable, quantifying the bias of a metalearner continues to be a challenging, but important problem in causal inference. We introduced a stability-based procedure to select an appropriate teacher model for CDT and showed that it provides a useful starting point for model selection; however, accounting for the bias of the teacher model would help to further improve performance.
Lastly, while this work is focused on the causal inference setting and treatment effect heterogeneity, an extension of CDT could consider prediction models, where there may be variation in the underlying predictive performance of a specific model. In this prediction context, subgroup analyses could help diagnose critical disparities in the performance of predictive models. 

\bibliographystyle{agsm}
\bibliography{bibliography}
\clearpage

\appendix
\singlespacing
\renewcommand{\thefigure}{A\arabic{figure}} 
\setcounter{figure}{0} 

\renewcommand{\thetable}{A\arabic{table}} 
\setcounter{table}{0} 
\setcounter{page}{1}

\begin{center} 
\Large
\textbf{Appendix: ``Distilling heterogeneous treatment effects:
Stable subgroup estimation in causal inference''}
\end{center} 

\begin{table}[ht]
\centering 
\begin{tabular}{cl} \toprule 
Notation & Definition \\\midrule 
$\tau(X)$ & (Oracle) Conditional Average Treatment Effect \\
& (i.e.,  $\E[Y(1) - Y(0) \mid X]$)\\ \midrule 
$\hat \tau(X)$ & Predicted CATE from a teacher model\\\midrule 
$\tau^{(g)}$ & Subgroup average treatment effect, for subgroup $g$\\
&(i.e., $\E[Y(1) - Y(0) \mid \cG_g(X) = 1]$)\\ \midrule 
$\hat \tau^{(g)}$ & Estimate of the subgroup ATE for subgroup $g$ \\\midrule 
$f_\tau(X; k)$ & Population-level predicted output from a single split population-level decision tree \\
& using the \textit{oracle} CATEs\\
& (i.e., $\E[\tau(X) \mid X \leq k] \1\{X \leq k\} + \E[\tau(X) \mid X > k] \1\{X > k\}$)\\\midrule 
$f_\tau^d(X; k)$ & Population-level predicted output from a single split population-level decision tree \\ 
& using the \textit{distilled} CATEs\\
& (i.e., $\E[\hat \tau(X) \mid X \leq k] \1\{X \leq k\} + \E[\hat \tau(X) \mid X > k] \1\{X > k\}$) \\ \midrule 
$\hat f_\tau(X; k)$ & Sample-level predicted output from a single split population-level decision tree \\
& using the \textit{oracle} CATEs\\
& (i.e., $\E_n[\tau(X) \mid X \leq k] \1\{X \leq k\} + \E_n[\tau(X) \mid X > k] \1\{X > k\}$)\\\midrule 
$\hat f_\tau^d(X; k)$ & Sample-level predicted output from a single split population-level decision tree \\ 
& using the \textit{distilled} CATEs\\
& (i.e., $\E_n[\hat \tau(X) \mid X \leq k] \1\{X \leq k\} + \E_n[\hat \tau(X) \mid X > k] \1\{X > k\}$) \\ \bottomrule 
\end{tabular} 
\caption{Summary of notation for key terms used throughout the paper.}
\end{table} 
\section{Extended Discussion}\label{app:discussion}
\subsection{Pruning}\label{app:pruning}
As in the supervised learning setting, decision trees for causal effect estimation (including in the second stage of CDT) are often pruned to avoid overfitting. There are generally two types of pruning: pre-pruning and post-pruning. Pre-pruning refers to techniques or early stopping criteria that are used to stop growing the decision tree during the tree construction process. Common pre-pruning approaches include setting constraints on the maximum depth of the tree, the minimum number of samples per node, or the minimum information gain from making each split. Post-pruning refers to techniques used to remove uninformative nodes from the decision tree after it has already been constructed. For example, in the \texttt{rpart} R package \citep{therneau1997introduction}, cross-validation is performed to choose the best complexity parameter $\alpha$, where $\alpha$ measures the cost of adding another split to the tree, and all nodes with a cost greater than $\alpha$ are subsequently removed or pruned from the tree. 

Through simulations (see Section~\ref{sec:sims}), we demonstrate that standard pre-pruning and post-pruning approaches used in ordinary decision trees (e.g., CART \citepappendix{breiman1984classification}) also work well for the student decision tree model in \method~and are hence recommended. In particular, standard here means using the default pre-pruning criteria in the \texttt{rpart} R package and post-pruning based upon the cross-validated complexity parameter $\alpha$, as described previously. We provide additional simulation results for pruning in Appendix \ref{app:sim_results}.

\subsection{Inconsistent Teacher Model}
Throughout the paper, we have assumed that the teacher model is able to consistently recover the CATE $\tau(X)$. In settings when the teacher model is an inconsistent estimator for the CATE (i.e., $\hat \tau(X) \not \cip \tau(X)$), there are still settings when CDT could recover the optimal subgroups. More formally, let $\hat \tau(X) \cip \tilde \tau(X)$, where $\tilde \tau(X) \neq \tau(X)$. If $\tilde \tau(X)$ admits the same optimal partitioning as the true CATEs $\tau(X)$, the subgroups estimated using the misspecified teacher model will still converge in probability to the optimal subgroups under the true CATE. 

As a toy example of when this could occur, consider a simple setting in which $X \sim \text{Unif}([0, 1])$ and the true CATE $\tau(X)$ is a simple threshold model with a split at the midpoint 0.5, i.e., $\tau(X) = \1\{X \geq 0.5\}$. Suppose researchers misspecify the CATE and use a linear model, such that the estimated CATE model converges to $\tilde \tau(X) = \beta X$. In such a setting, the optimal partitioning is $\cG(X) = \{ \1\{X < 0.5\}, \1\{X \geq 0.5\}\}$, and in fact, using $\tilde \tau(X)$ will result in the same partitioning of the underlying space. We illustrate this phenomenon via simulation in Figure~\ref{fig:jaccard-example}. 

In this manner, a relaxed assumption in lieu of consistency is to consider the \textit{class} of possible teacher models that would result in the same optimal partitioning as the true CATE. However, this is challenging to reason about in practice, and for this reason, we focus on the assumption of consistent teacher models in the main manuscript. 

\subsection{Inference for observational settings}\label{app:dr}
While the main manuscript focused on experimental settings, we can similarly estimate the subgroup ATE in observational settings. However, we must account for potential confounding within a given subgroup. 

To begin, we must assume conditional ignorability of treatment and outcome. 
\begin{assumption}[Conditional ignorability of treatment and outcome]
$$Y(1), Y(0)\ \indep \ Z \mid X$$
\end{assumption} 
This assumption states that given a set of pre-treatment covariates $X$, the treatment assignment process is effectively `as-if' random. While the subgroups themselves are functions of the pre-treatment covariates, they represent a coarsened subset of units in the study. As such, we do not generally expect that all units within a given subgroup have identical covariate values $X$, especially in settings when $X$ is high dimensional. As such, we can no longer rely on just a subgroup difference-in-means estimator, and must adjust for potential confounding of treatment within a subgroup.

Instead of estimating the subgroup difference-in-means, researchers can estimate $\hat \tau^{(g)}_{\textsc{w-adj}}$: 
$$\left\{ \hat \tau^{(g)}_{\textsc{w-adj}}, \hat \alpha, \hat \beta \right\}= \argmin_{\tau, \alpha, \beta} \frac{1}{n_g} \sum_{i: \hat \cG_g(X_i) = 1} \hat w_i  \left\{Y_i - \left(\tau Z_i + \alpha + \beta^\top X_i \right) \right\}^2,$$
where $\hat w_i$ is defined as the inverse propensity score weights: 
$$\hat w_i = \begin{cases} 
1/\hat e(X_i) &\text{if } Z_i = 1 \\ 
1/\{1-\hat e(X_i)\} &\text{if } Z_i = 0
\end{cases},$$
and $\hat e(X_i)$ is the estimated probability of being assigned to treatment, given the covariates $X$. $\hat \tau^{(g)}_{\textsc{w-adj}}$ provides a doubly robust approach to estimating the subgroup ATE \citepappendix[e.g.,][]{kang2007demystifying}, with potential efficiency gains in finite samples.  Theorem \ref{thm:cid_group_dim} can be extended for $\hat \tau^{(g)}_{\textsc{w-adj}}$, allowing for valid inference and analogous hypothesis tests. 

There are two components to $\hat \tau^{(g)}_{\textsc{w-adj}}$: (1) the covariate adjustment taking place within the regression, and (2) a re-weighting adjustment to balance the treatment and control groups. The covariate adjustment within the weighted regression allows researchers to leverage variation across covariates within a specific subgroup. In particular, because the subgroups are defined using covariates that are explanatory of the treatment effect variation, we generally expect a sparser set of covariates that explain the treatment effect variation in contrast to covariates that can explain variation in the outcome. As such, within each subgroup, while we expect relatively small amounts of treatment effect variation, there could still be residual variation in the outcomes that can be explained by the other covariates. The re-weighting component of $\hat \tau^{(g)}_{\textsc{w-adj}}$ ensures that within each subgroup, researchers can also account for finite-sample imbalances that are present \citepappendix{xie2012estimating}. 

Notably, $\hat \tau_{\textsc{W-ADJ}}^{(g)}$ can similarly be used in experimental contexts to improve finite-sample performance. In practice, a drawback to subgroup analyses is the relatively small number of observations within each subgroup. In particular, even if the overall study contains a large number of observations, the number of observations available within a subgroup can be relatively limited. As a result, using the standard difference-in-means estimator can result in poor finite-sample performance. 

\subsection{Extension to finite population model}\label{app:discussion_pop}
In the main manuscript, we focus on the infinite target population setting for considering inference for the subgroup difference-in-means estimator. In settings when researchers are interested in estimating the finite population-level subgroup ATE, then we can show that the subgroup difference-in-means estimator will be unbiased for the sample subgroup ATE. 

To begin, we define the finite-sample subgroup ATE as: 
$$\tau^{(g)}(\mathcal{D}) = \frac{1}{n_g} \sum_{i=1}^n \cG_g(X) \{Y_i(1) - Y_i(0)\}.$$
Let $\mathcal{D}$ represent the finite-sample study data.
\begin{theorem}[Unbiasedness and Variance of Subgroup Difference-in-Means Estimator in Finite Populations] \label{thm:cip_group_dim_finite}
Under Assumptions \eqref{assum:min}-\eqref{assum:positivity}, the subgroup difference-in-means estimator will be an unbiased estimator for the sample subgroup ATE:
$$\E(\hat \tau^{(g)} \mid \mathcal{D}) = \tau^{(g)}(\mathcal{D}).$$
The variance of $\hat \tau^{(g)}$ is
$$\var(\hat \tau^{(g)}\mid \mathcal{D}) = \frac{1}{n_g} \left \{ \beta_g(1)\var_g(Y_i(1)) + \beta_g(0) \var_g(Y_i(0)) - \var_g(\tau_i) \right\},$$ 
where $\beta_g(z) = \E\{1/\sum_{i=1}^n \hat \cG_g(X_i)\1\{Z_i = z\}\}$ and $\var_g\left\{Y(z)\right\} := \E \left\{\frac{1}{n_g-1}\sum_{i=1}^n \hat \cG_g(X_i) (Y_i(z)- \right.$ $\left. \overline{Y_g(z)})^2\right\}$ for $z \in \{0,1\}$.
\end{theorem}

To consider the asymptotic distribution of the subgroup difference-in-means estimator, we must account for the fact that as the sample size increases, the underlying subgroup allocation will also change. We leverage recent work in developing central limit theorem results for finite populations (e.g., \citealp{schochet2024design, li2017general}) to show that under mild regularity conditions, the subgroup difference-in-means estimator will be asymptotically normal. 

\begin{theorem}[Asymptotic Normality of Subgroup Difference-in-Means Estimator] \label{thm:cid_group_dim}
Assume that as the sample size $n \to \infty$, the total number of subgroups $G$ remain fixed. Furthermore, assume the proportion of units in each subgroup converges to a proportion $\pi_g^* \in (0,1)$ (i.e., $n_g / n \to \pi_g^*$, where $0 < \pi_g^* < 1$ for all $g \in \{1, \ldots, G\}$ and $\sum_{g=1}^G \pi_g^* = 1$), and the proportion of treated units converges to $p_z$, where $p_z \in (0,1)$ (i.e., $\frac{1}{n} \sum_{i=1}^n Z_i \to p_z$). Furthermore, assume $\min_g \pi^*_g (1 - \pi^*_g) = c > 0$. Then, for $z \in \{0,1\}$ and $g \in \{1,\ldots, G\}$, if the following condition holds:
\begin{equation} 
\lim_{n\to\infty} \frac{1}{\left(\sum_{i=1}^n \1\{Z_i = z\} \right)^2}  \frac{max_{1 \leq i \leq n} \hat \cG_g(X_i) Y_i^2(z)}{\var(\hat \tau^{(g)})} = 0,
\label{eqn:lindeberg_condition}
\end{equation} 
then the subgroup difference-in-means estimator will converge in distribution to $N(0, 1)$: 
\begin{equation*} 
\frac{\sqrt{n}\left( \hat \tau^{(g)} - \E(\hat \tau^{(g)})\right)}{\sqrt{\var(\hat \tau^{(g)})}} \cid N(0, 1).
\end{equation*} 
\end{theorem} 
The results of Theorem \ref{thm:cid_group_dim} can be viewed as a special case of \citet{schochet2024design}, Theorem 1. The condition in \eqref{eqn:lindeberg_condition} is a Lindeberg-type condition, which restricts the tail behavior of the underlying potential outcome distribution and is relatively weak.

\subsection{Non-parametric test of treatment effect heterogeneity} \label{subsec:nonparam-test}

We propose a nonparametric test of treatment effect heterogeneity. Following \citetappendix{imai2022statistical}, we consider the following null hypothesis of no treatment effect heterogeneity: 
\begin{equation} 
H_0: \tau^{(1)} = \ldots = \tau^{(G)} = \tau, \label{eqn:test}
\end{equation} 
for all subgroups $g \in \{1, \ldots, G\}$, with the following estimator: 
$$\boldsymbol{\hat \tau} = (\hat \tau^{(1)} - \hat \tau, \ldots, \hat \tau^{(g)} - \hat \tau)^\top,$$
where $\hat \tau := \frac{1}{\sum_{i=1}^n Z_i} Y_i Z_i - \frac{1}{\sum_{i=1}^n (1-Z_i)} Y_i (1-Z_i)$ is the usual difference-in-means estimator across the entire sample. Then, following Theorem \ref{thm:cid_group_dim}, we can show that under the null hypothesis, the test statistic can be asymptotically approximated using a $\chi^2$ distribution. The following corollary formalizes. 
\begin{corollary}[Non-Parametric Test of Treatment Effect Heterogeneity] \label{cor:te_test}
Under the null hypothesis in \eqref{eqn:test}, with the alternative $H_\textsc{A}: \exists~\hat \tau^{(g)} \neq \tau$, then: 
$$\boldsymbol{\hat \tau}^\top \boldsymbol{\Sigma}^{-1} \boldsymbol{\hat \tau} \cid \chi^2.$$
\end{corollary}

\subsection{Inference after selection} \label{app:selection-inf}
In the following subsection, we consider the problem of how to perform inference after selecting a particular subgroup (or subgroups). For example, researchers may be interested in performing inference on the subgroup with the highest treatment effect, or the set of subgroups with statistically significant treatment effects. In such settings, researchers must use alternative approaches to construct confidence intervals for valid coverage. We leverage a recent approach proposed in \citet{andrews2024inference} to accomodate settings where researchers are interested in inference of the subgroup ATE after choosing a particular subgroup. For concreteness, we focus on the setting when researchers are interested in inference across a subgroup with the \textit{largest} treatment effect estimate, though the procedure can be generalized for other settings as well. 

To start, define $\hat g^*$ as the subgroup that has the largest treatment effect estimate:
$$\hat g^* = \argmax_{g \in \{1, ..., G\}} \hat \tau^{(g)}.$$ 
Conditional on selecting $\hat g^*$, the distribution of $\hat \tau^{(\hat g^*)}$ will be shifted upwards, resulting in positive median bias. This is commonly referred to as the winner's curse \citep{andrews2024inference, zrnic2024locally, wei2024inference}. Let $\boldsymbol{\hat \tau}^{-(\hat g^*)}$ represent the vector of estimated subgroup average treatment effects, with the $\hat g^*$-th group omitted. Let $\sigma^2_{(\hat g^*)} = \var(\hat \tau^{(\hat g^*)})$. Let $F_{TN}(\hat \tau^{(\hat g^*)}; \tau, \hat g^*, \boldsymbol{\hat \tau}^{-(\hat g^*)})$ represent the cumulative distribution function of a Normal distribution $N(\tau, \sigma^2_{(\hat g^*)})$, truncated to $\left[\max_{g \in [G] \backslash \hat g^*}\boldsymbol{\hat \tau}^{-(\hat g^*)}, \infty \right)$.

Then, let $\hat \tau^{(\hat g^*)}_{1/2}$ be defined as the $\tau$ such that 
$$F_{TN}(\hat \tau^{(\hat g^*)}; \tau, \hat g^*, \boldsymbol{\hat \tau}^{-(\hat g^*)}) = \frac{1}{2}.$$ 
$\hat \tau^{(\hat g^*)}_{1/2}$ is the corrected estimator that accounts for the selection bias. Because $F_{TN}(\hat \tau^{(\hat g^*)}; \tau, \hat g^*, \boldsymbol{\hat \tau}^{-(\hat g^*)})$ is strictly increasing as a function of $\tau$, $\hat \tau^{(\hat g^*)}_{1/2}$ will be the unique solution. 

To construct confidence intervals, let $\hat \tau(\alpha)$ be the $\tau$ such that $F_{TN}(\hat \tau^{(\hat g^*)}; \tau, \hat g^*, \boldsymbol{\hat \tau}^{-(\hat g^*)}) = \alpha.$ Then define the confidence interval $\text{CI}(\alpha) := \left[\hat \tau(\alpha/2), \hat \tau(1- \alpha/2) \right]$. \citet{andrews2024inference}, Proposition 2 shows that 
$$\Pr\left(\tau^{(\hat g^*)} \in \text{CI}(\alpha) \mid \hat g^* \in \tilde g \right) = \alpha, \text{ for all } \tilde g \in \{1,..., G\}.$$
Furthermore, by law of iterated expectations, the proposed confidence intervals will also obtain unconditional coverage. 

The correction procedure is necessary in settings when researchers are interested in performing inference for the estimated treatment effect that corresponds to the subgroup with the largest estimated treatment effect. If researchers are just interested in inference across subgroup ATEs, then they can use the inferential procedures detailed in the main manuscript in Section 3.4. Furthermore, if researchers are interested in evaluating whether or not the estimated treatment effects are statistically significantly different from each other across subgroups, then they should employ the non-parametric test of treatment effect heterogeneity, introduced in Appendix \ref{subsec:nonparam-test}.

\subsection{Detailed CDT Algorithm}

\begin{algorithm}[H]
\small 
\singlespacing
\DontPrintSemicolon
\SetAlgoNoLine
\caption{Causal Distillation Trees}
\label{alg:cdt}
\vspace{-1em}
\KwIn{data $\cD = \{X, Z, Y\}$, teacher (CATE) model $\teacher$, student (decision tree) model $\student$, training proportion $\pi_{train}$, number of cross-fitting replicates $R$ (if necessary)}
Randomly split data $\cD$ into $\cD^{train} = \{X^{train}, Z^{train}, Y^{train}\}$ and $\cD^{est} = \{X^{est}, Z^{est}, Y^{est}\}$ with probability $\pi_{train}$ and $1 - \pi_{train}$, respectively\;
\Comment{Stage 1: estimate heterogeneous treatment effects via teacher model}
\eIf{using model-specific out-of-sample estimation}{
    \Comment{e.g., if teacher model has out-of-bag estimation procedure}
    Fit CATE model $\teacher$ using $\cD^{train}$ $\rightarrow \teacherh$\;
    Use $\teacherh$ to estimate $\tau(X_i)$ for each unit $i$ in $\cD^{train}$ $\rightarrow \hat{\tau}(X_i)$\;
}{
    \Comment{do repeated cross-fitting}
    \For{$r = 1, \ldots, R$}{
        Randomly split $\cD^{train}$ into two equally-sized partitions $\mathcal{D}^{train}_{in}$ and $\mathcal{D}^{train}_{out}$\;
        Fit CATE model $\teacher$ using $\cD^{train}_{in}$ $\rightarrow \teacherh$\;
        Use $\teacherh$ to estimate $\tau(X_i)$ for each unit $i$ in $\cD^{train}_{out}$ $\rightarrow \tauh^{(r)}(X_i)$\;
        Reverse roles of $\mathcal{D}^{train}_{in}$ and $\mathcal{D}^{train}_{out}$ and repeat lines 11-12\;
    }
    Average across cross-fits: $\hat{\tau}(X_i) = \sum_{r = 1}^R \tauh^{(r)}(X_i)$ for each unit $i$ in $\cD^{train}$\;
}
\Comment{Stage 2: estimate subgroups via student model}
Fit decision tree $\student$ on $X^{train}$ to predict $\hat \tau(X_i)$'s $\rightarrow$ estimated subgroups $\left \{ \hat \cG_1(X), \ldots, \hat \cG_G(X)\right\}$\;
\Comment{Honestly estimate subgroup ATEs}
Estimate $\tauh^{(g)}$ via \eqref{eqn:subgroup_dim} for each $g = 1, \ldots, G$ using $\cD^{est}$\;
\end{algorithm}

\begin{figure}[h]
    \centering
    \includegraphics[width=0.99\linewidth]{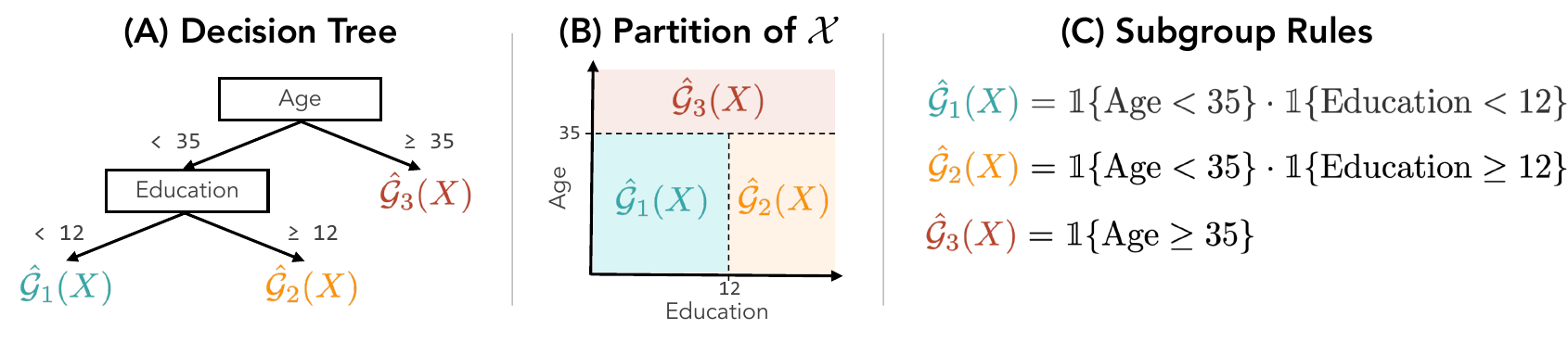}
    \caption{Given a decision tree (A), its decision splits correspond to binary cuts that partition the covariate space $\mathcal{X}$, shown in (B). Furthermore, the decision splits leading to each terminal node map to the collection of rules that define a subgroup, shown in (C).}
    \label{fig:tree_to_subgroup}
\end{figure}

\subsection{ Discussion of Teacher Model Selection Algorithm}\label{app:teacher_model_selection}
We formalize the teacher model selection algorithm in Algorithm \ref{alg:jaccard}. 
In brief, Algorithm~\ref{alg:jaccard} evaluates the Jaccard SSI between estimated subgroups across different bootstrap samples for each candidate teacher model. 
The teacher model that results in the most stable subgroups (i.e., the highest average Jaccard SSI) is selected as the final teacher model for CDT.

{\small 
\begin{algorithm}[t]
\DontPrintSemicolon
\SetAlgoNoLine
\caption{Selecting the teacher model for CDT}\label{alg:jaccard}
\KwIn{training data $\cD^{train} = \{X^{train}, Z^{train}, Y^{train}\}$, set of candidate teacher (CATE) models $\{\teacher^1, \ldots, \teacher^L\}$, student (decision tree) model $\student$, number of bootstraps $B$, depth of tree $d$}
\For{$l = 1, \ldots, L$}{
    Fit $\teacher^l$ on $\cD^{train}$ to obtain estimates of the heterogeneous treatment effects $\tauh^{l}(X_i)$ for each unit in $\cD^{train}$ (via lines 2-16 in Algorithm~\ref{alg:cdt})\;
    \For{$b = 1, \ldots, B$}{
        \For{$t = 1, 2$}{
            Get bootstrap sample of $\{(X_i, \; \tauh^{l}(X_i)) : i \in \text{training unit}\}$\;
            Fit decision tree $\student$ on the bootstrapped $X_i$'s to predict bootstrapped $\tauh^{l}(X_i)$'s $\rightarrow$ estimated decision tree $\studenth^{l}_{b_t}$\;
            Grow/prune decision tree $\studenth^{l}_{b_t}$ to have depth $d$ $\rightarrow$ estimated subgroups $\hat{\cG}^{l}_{b_t}$\;
        }
        Compute $J^{l}_b := SSI(\hat{\cG}^{l}_{b_1}, \hat{\cG}^{l}_{b_2})$\;
    }
    Compute $J^{l} := \frac{1}{B} \sum_{b = 1}^{B} J^{l}_b$\;
} 
Select the teacher model $\teacher^{l^*}$ such that $l^* = \argmax_{l = 1, \ldots, L} \;\; J^l$\;
\end{algorithm}
}

\begin{figure}
    \centering
    \includegraphics[width=.85\linewidth]{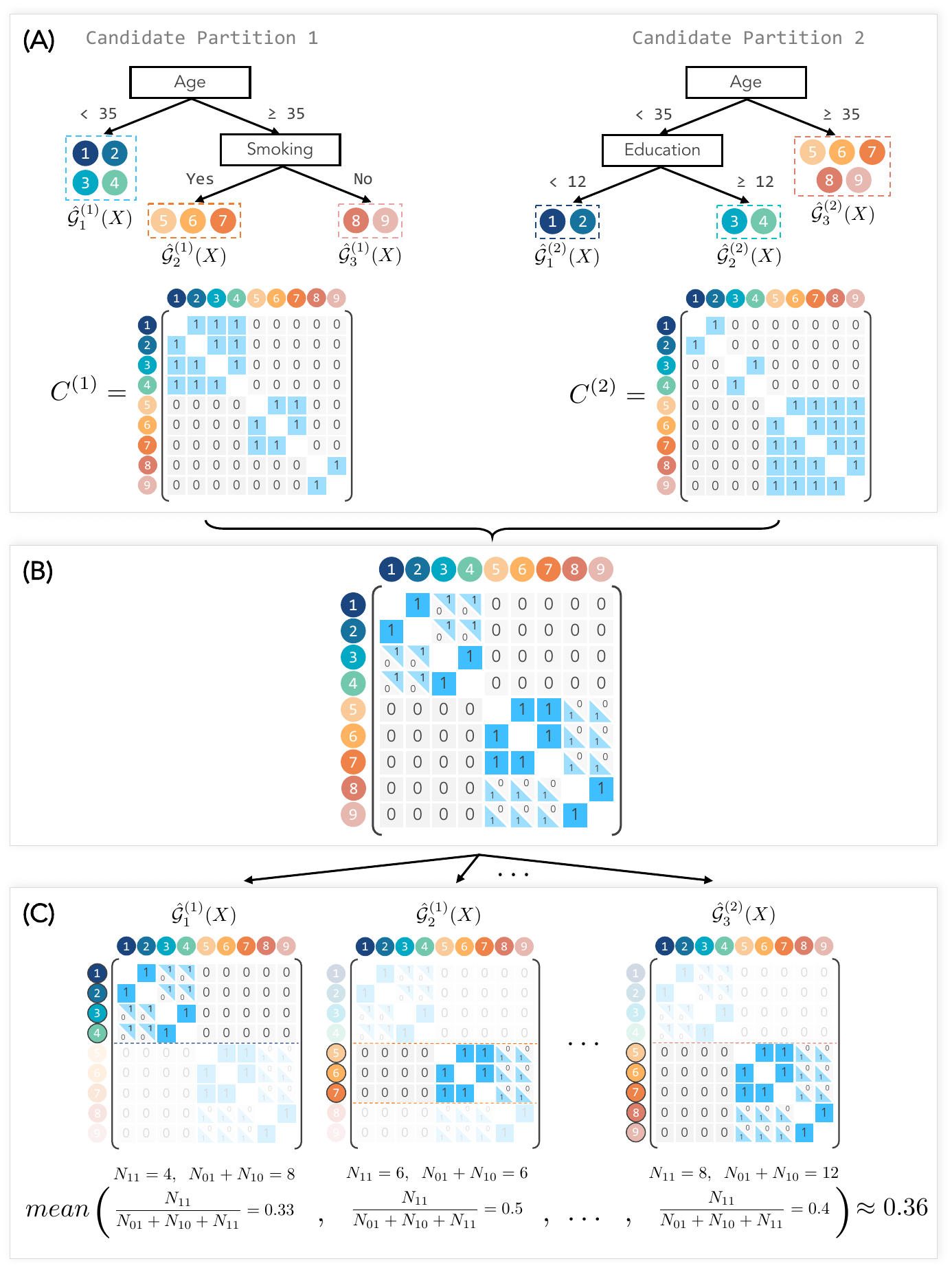}
    \caption{Jaccard subgroup similarity index (SSI). (A) Given two candidate partitionings into subgroups, two binary matrices, denoted $C^{(1)}$ and $C^{(2)}$, are constructed, indicating whether or not two distinct units belong in the same subgroup in $\hat{\cG}^{(1)}$ and $\hat{\cG}^{(2)}$, respectively. (B) $C^{(1)}$ and $C^{(2)}$ are compared to identify the sample pairs which (i) belong to the same subgroup in both partitions (i.e., $N_{11}$ shown as full blue cells), (ii) belong to the same subgroup in one partition and a different subgroup in the other partition (i.e., $N_{01} + N_{10}$ shown as divided blue and gray cells), and (iii) belong to different subgroups in both partitions (i.e., $N_{00}$ shown as full gray cells). (C) Finally, $N_{11}(\cH) / (N_{01}(\cH) + N_{10}(\cH) + N_{11}(\cH))$ is computed per subgroup $(\cH)$ and averaged across all subgroups to obtain the Jaccard SSI.}
    \label{fig:ssi}
\end{figure}

However, to also ensure that the Jaccard SSIs are comparable across both bootstrap resamples and teacher models, we require that the student model outputs the same number of subgroups (i.e., terminal nodes) across all runs.
Otherwise, a teacher model, such as a constant predictor, that results in fewer decision tree splits is inherently going to be more stable than a teacher model that results in more decision tree splits.
One way to meet this requirement, which we incorporate in Algorithm~\ref{alg:jaccard} and our R package, is to grow the student decision tree to a fixed depth $d$ (forcing splits to occur until the tree reaches depth $d$), resulting in $2^d$ terminal nodes.\footnote{In situations where practitioners have a pre-specified number of subgroups in mind, the teacher selection algorithm can be easily modified to grow trees with a fixed number of leaves, as opposed to growing trees with a fixed depth.}

Consequently, the proposed teacher model selection procedure requires choosing the desired tree depth $d$ at which to evaluate the Jaccard SSI.
This is equivalent to choosing the order, or the number of binary decision rules characterizing each subgroup. 
For example, a subgroup in a depth $d = 2$ tree could be $\cG_g(X) = \mathbbm{1}\{X_{\text{gender}} = \text{Female}\} \cdot \mathbbm{1}\{X_{\text{age}} < 35\}$ while $\cG_g(X) = \mathbbm{1}\{X_{\text{gender}} = \text{Female}\} \cdot \mathbbm{1}\{X_{\text{age}} < 35\} \cdot \mathbbm{1}\{X_{\text{height}} \geq 66\}$ is a subgroup in a depth $d = 3$ tree. 
Domain expertise can hence sometimes inform the choice of $d$. 
However, when researchers do not have a strong substantive prior for $d$, we recommend that researchers repeat the above teacher model selection procedure for multiple choices of $d$ and select the teacher model that yields the highest subgroup Jaccard stability scores across a majority of these $d$'s.
We demonstrate the effectiveness of this approach for teacher model selection in Section~\ref{sec:sims}, where we select the teacher model by evaluating their stability across a range of $d$'s, namely, $d = 1, 2, 3, 4$.

It is important to note that the choice of $d$ and the overall teacher model selection procedure are confined to the training data and hence independent of the held-out estimation data used to estimate the final subgroup ATEs.
Moreover, this choice of $d$ does not necessarily correspond to the depth of the final decision tree used to estimate the subgroups and subgroup ATEs in CDT.
The $d$ here is only used to ensure that the Jaccard SSIs are comparable across bootstrap resamples and teacher models, and is hence only used for the teacher model selection procedure.
The depth of the final decision tree in CDT can still be chosen by modifying the tree hyperparameters and/or applying standard pruning approaches (discussed in Appendix~\ref{app:pruning}). 
In this work, we used the default \texttt{rpart} hyperparameters and post-pruning based upon the cross-validated complexity parameter $\alpha$ to obtain the final CDT tree (more discussion in Appendix~\ref{app:sims}).

Figure~\ref{fig:timing} investigates the runtime of our proposed teacher selection procedure with $B = 100$ bootstraps, depths $d = 1, 2, 3, 4$, and sample sizes up to $n = 5000$ based on a single core of a 2023 MacBook Pro with Apple M3 Pro chip.
The two potential bottlenecks in the teacher model selection procedure are the repeated fitting of the student decision tree (line 6 in Algorithm~\ref{alg:jaccard}) and the repeated computation of the Jaccard SSI between two sets of subgroups (line 9 in Algorithm~\ref{alg:jaccard}).
Since the Jaccard SSI metric requires computing pairwise comparisons between all units in the training data, we have implemented a highly efficient version of Jaccard SSI in \texttt{Rcpp}, which we make available in our R package \texttt{causalDT}. 
Using this implementation, we assessed the time taken to fit a single decision tree and the time taken to compute the Jaccard SSI between two sets of subgroups (Figure~\ref{fig:timing}(A)); both are on the order of a fraction of a second even for $n = 5000$.
The teacher selection procedure is hence also reasonably fast under the given setup.
When $n = 5000$, evaluating $J^l$ (without parallelization) for a single teacher model (i.e., causal forest) across $B = 100$ bootstraps completes in under 1.5 minutes on average, and the full CDT fit including this $J^l$ evaluation completes in under 2 minutes on average (Figure~\ref{fig:timing}(B)).

\begin{figure}
    \centering
    \includegraphics[width=\linewidth]{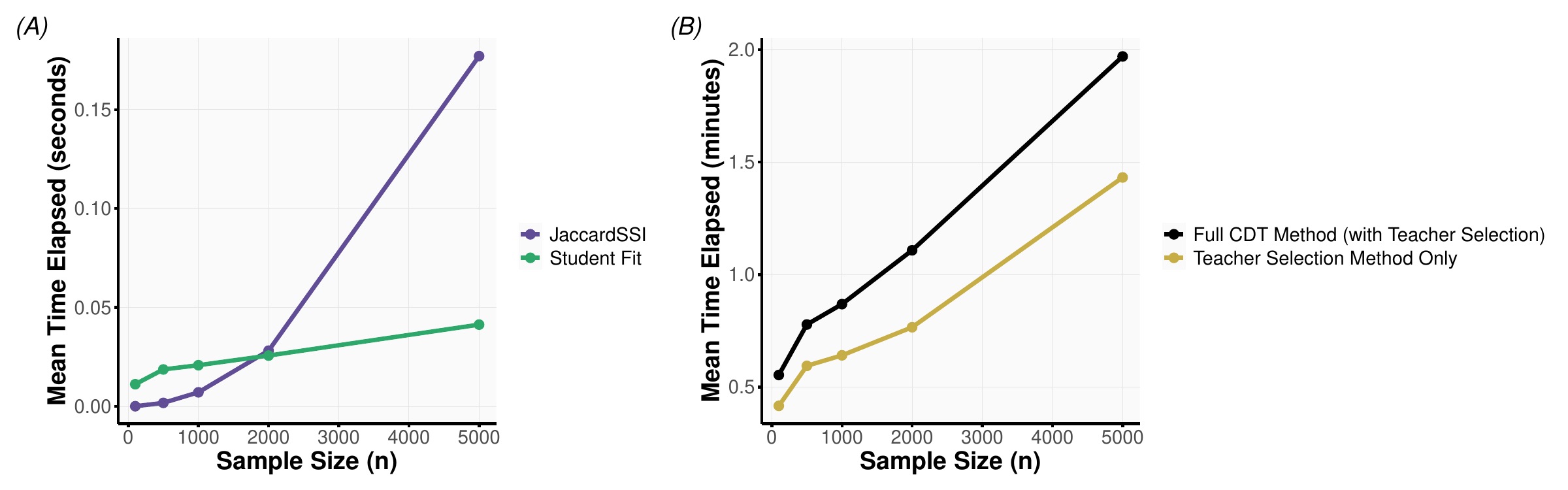}
    \caption{Runtime analysis of teacher model selection procedure across varying samples sizes. (A) Time taken (in seconds) for a single student model fit and a single Jaccard SSI evaluation between two sets of subgroups. (B) Time taken (in minutes) to run only the teacher model selection procedure (i.e., compute $J^l$) versus the full CDT fit (including the teacher model selection procedure). Results are averaged across 500 simulation replicates with ribbons denoting $\pm1SE$.}
    \label{fig:timing}
\end{figure}

\subsection{Diagnostic Tools for CDT}\label{app:diagnostics}

To help researchers evaluate the quality of the estimated subgroups from CDT, we recommend several important diagnostics.

\paragraph{Prediction accuracy of student model.} First, researchers can assess how well the student model fits the teacher model's estimated CATEs. We quantify this via the prediction accuracy (e.g., root mean squared error (RMSE)) between the student model predictions and the CATEs, estimated by the teacher model. High concordance between the student model predictions and the teacher model CATEs is a positive sign while low concordance suggests that the student model is not compatible or a good proxy for the teacher model and should lower our degree of trust in the student model's estimated subgroups.

\paragraph{Distribution of teacher-estimated CATEs.} Secondly, researchers can quantify and visualize the variance or distribution of CATEs, estimated by the teacher model, in each decision tree split in the student model. This is important for understanding both how much heterogeneity occurs within each subgroup (or node) and how much the subgroup CATEs (i.e., the mean CATE per node) change after making each split. A good split (or decision rule) will result in a large change in the mean subgroup CATE and low heterogeneity within the resulting subgroups. In contrast, splits that do not substantively change the mean subgroup CATE nor reduce the amount of heterogeneity within the subgroups can be deemed as uninformative and unnecessary splits.

\paragraph{Number of treated and control units per subgroup.} Thirdly, we recommend reporting the number of treated and control units from the training data that fall in each subgroup. Since the student model only uses $X$ to predict the teacher-estimated CATEs, it is possible that only control or only treated units fall into a student-estimated subgroup. If the training and estimation data splits (i.e., $D^{train}$ and $D^{est}$ from Algorithm~\ref{alg:cdt}) are from similar populations, then the estimation data split may similarly yield only control or only treated units (or some highly imbalanced ratio of treated-to-control units) in that subgroup. In these cases, estimation of the subgroup ATE would be highly variable or not possible using \eqref{eqn:subgroup_dim}. It may also suggest that the tree has been grown too deep and that there is currently not sufficient sample size to estimate such a subgroup.

\paragraph{Stability of selected features.} Finally, in addition to evaluating the subgroup Jaccard stability across the different bootstrap samples in Algorithm~\ref{alg:jaccard}, we can also track the selected features across these bootstraps for each chosen tree depth and plot their distribution (e.g., Figures~\ref{fig:sim-stability-and-0}-\ref{fig:sim-stability-or-cov}). Similar to the subgroup Jaccard stability, a more stable (or homogeneous) distribution of selected features should elucidate greater trust in that model or tree depth. On the flip side, a very heterogeneous distribution of selected features indicates that the estimated decision tree structure is highly volatile, depending heavily on the training data subset.

\section{Proofs and Derivations} \label{app:proofs}

Throughout the theoretical derivations, we assume that $X \sim \text{Unif}([0,1]^p)$. While this assumption imposes independence among the covariates, the restriction to uniform marginal distributions is without loss of generality for continuous covariates: any continuous distribution can be transformed to a uniform distribution via the probability integral transform, which is monotone and thus preserves the ordering of the covariate values. Since tree-based models are invariant to monotone transformations of the covariates, the uniform distribution assumption does not restrict the generality of our results.

\subsection{Proof of Proposition \ref{prop:converge_one}}
We can re-write the difference in $\hat \cG_1(X)$ and $\cG_1(X)$ as:  
\begin{align*} 
\E\left[ \left \lvert \hat \cG_1(X) - \cG_1(X) \right \rvert \right]
&= \E\left[ \left \lvert \1\{ X \leq \hat s_n\} - \1\{ X \leq s\}\right \rvert \right] \\
&= \E \left[ \1\{ X \leq \hat s_n\} - \1\{ X \leq s\} \mid \hat s_n  > s \right] \Pr(\hat s_n > s) \\
& \qquad + \E \left[ \1\{ X \leq s\} -  \1\{ X \leq \hat s_n\} \mid \hat s_n  \leq s \right] \Pr(\hat s_n \leq s) \\
&= \left \lvert F_X(\hat s_n) - F_X(s) \right \rvert \\\
&\leq K \left \lvert \hat s_n - s\right \rvert,
\end{align*} 
where the last inequality follows from $F_X$ being continuously differentiable by Assumption \eqref{assum:min} near split point $s$. As a result, there must exist a positive, real-valued constant $K$ such that $f_X(x) \leq K < \infty$. 

We will now show that $\left \lvert \hat s_n - s \right \rvert$ converges to zero at a rate that depends on the underlying smoothness of the data generating process and the rate of convergence of the teacher model. 

To begin, let $M_n(a, s) := \E_n[\ell(a; X, s)]$ and $M(a,s) = \E[\ell(a; X, s)]$ represent the sample-level loss and population-level loss respectively. $M_n$ is a stochastic process, as it depends on the underlying sample, while $M$ is a deterministic function. Let $v(X) = \hat \tau(X) - \tau(X)$.  Let $f_{\tau}(X; s) := \E[\tau(X) \mid X \leq s] \1\{ X \leq s\} + \E[\tau(X) \mid X > s] \1\{ X > s\}$, $f_{\tau}^d(X; s) := \E[\hat \tau(X) \mid X \leq s] \1\{ X \leq s\} + \E[\hat \tau(X) \mid X > s] \1\{ X > s\}$, and $f_\tau^v(X;s) = \E[v(X) \mid X \leq s] \1\{X \leq s\} + \E[v(X) \mid X > s] \1\{ X > s\}$. Finally, let $\hat f_{\tau}(X; s)$, $\hat f_{\tau}^d(X; s)$, and $\hat f_\tau^v(X;s)$ represent the sample analogs, respectively. 

Throughout, we will leverage the following algebraic expressions: 
\begin{equation}
\begin{aligned}
M_n(\hat \tau(X), s) =& M_n(\tau(X), s) + 2 \E_n \left[ \{ \tau(X) - \hat f_{\tau}(X; s)\} \cdot \{ \hat \tau(X) - \tau(X) \} \right]\\
&+ \E_n \left[ \{ \hat \tau(X) - \tau(X)\}^2\right]  - \E_n[\hat f^v_\tau(X;s)^2 ].
\end{aligned}
\label{eqn:sample_gap}
\end{equation}
\begin{equation} 
\begin{aligned}
M(\hat \tau(X), s) =& M(\tau(X), s) + 2 \E \left[ \{ \tau(X) - f_{\tau}(X; s)\} \cdot \{ \hat \tau(X) - \tau(X) \} \right]\\
&+ \E \left[ \{ \hat \tau(X) - \tau(X)\}^2\right]  - \E[ f^v_\tau(X;s)^2 ].
\end{aligned}
\label{eqn:pop_gap}
\end{equation}
\begin{proof} 
To start, 
\begin{align*} 
 \hat f_{\tau}^d(X; s) =& \E_n(\hat \tau(X) \mid X \leq s) \1\{X \leq s\} + \E_n(\hat \tau(X) \mid X > s) \1\{X > s\} \\
 =& \E_n(\tau(X) \mid X \leq s) \1\{X \leq s\} + \E_n(\tau(X) \mid X > s) \1\{X \geq s\} \\
 &+ \E_n(\hat \tau(X) - \tau(X) \mid X \leq s) \1\{X \leq s\} + \E_n(\hat \tau(X) - \tau(X) \mid X > s) \1\{X > s\}\\
 =&  \E_n(\tau(X) \mid X \leq s) \1\{X \leq s\} + \E_n(\tau(X) \mid X > s) \1\{X > s\} + \E_n(\hat \tau(X) - \tau(X)) \\
 =& \hat f_{\tau}(X; s) + \E_n(\hat \tau(X) - \tau(X))
\end{align*} 
We can re-write $M_n(\hat \tau(X),s)$ as a function of $M_n(\tau(X),s)$ and the gap between $\hat \tau(X)$ and $\tau(X)$: 
\begin{align*} 
M_n(\hat \tau(X), s) =& \E_n \left[ \ell(\hat \tau(X); X, s)\right]\\
=& \E_n \left[ \{\hat \tau(X) - \hat f_{\tau}^d(X; s) \}^2 \right] \\
=& \E_n \left[ \{\tau(X) + v(X) - \hat f_{\tau}(X; s) - \hat f_\tau^v(X;s) \}^2 \right] \\
=& \E_n \left[ \{ \tau(X) - \hat f_\tau(X;s)\}^2 \right]  + \E_n \left[ \{ v(X) - \hat f^v_\tau(X;s)\}^2 \right] \\
&+ 2 \E_n \left[\{ \tau(X) - \hat f_\tau(X;s)\} \cdot \{v(X) - \hat f^v_\tau(X;s) \} \right] \\ 
=&  \E_n \left[ \{ \tau(X) - \hat f_\tau(X;s)\}^2 \right]  + \E_n \left[ v(X)^2 \right] + \E_n[\hat f^v_\tau(X;s)^2 ] - 2\E_n[ v(X) \cdot \hat f_\tau^v(X;s)]\\
&+ 2 \E_n \left[\{ \tau(X) - \hat f_\tau(X;s)\} \cdot v(X) \right] - 2\E_n \left[\{ \tau(X) - \hat f_\tau(X;s)\} \cdot \hat f^v_\tau(X;s) \} \right] \\ 
=&  \E_n \left[ \{ \tau(X) - \hat f_\tau(X;s)\}^2 \right]  + \E_n \left[ v(X)^2 \right] - \E_n[\hat f^v_\tau(X;s)^2 ] \\
&+ 2 \E_n \left[\{ \tau(X) - \hat f_\tau(X;s)\} \cdot v(X) \right] \\
=& M_n(\tau(X), s)  + \E_n[\{\hat \tau(X) - \tau(X)\}^2] \\
&+2 \E_n\left[ \{ \tau(X) - \hat f_\tau(X;s)\} \cdot \{ \hat \tau(X) - \tau(X)\} \right] - \E_n[\hat f^v_\tau(X;s)^2 ],
\end{align*} 
where we have made use of the fact that $\E_n \left[\{ \tau(X) - \hat f_\tau(X;s)\} \cdot \hat f_\tau^v(X) \right] = 0$: 
\begin{align*} 
\E_n &\left[\{ \tau(X) - \hat f_\tau(X;s)\} \cdot \hat f^v_\tau(X) \right]  \\
=& \E_n \left[\{ \tau(X) - \hat f_\tau(X;s)\} \E_n[v(X) \mid X \leq s] \1\{X \leq s\}  \right] + \E_n \left[\{ \tau(X) - \hat f_\tau(X;s)\} \E_n[v(X) \mid X > s] \1\{X > s\}  \right] \\
=& \E_n \left[\{ \tau(X) - \hat f_\tau(X;s)\} \1\{X \leq s\}  \right] \E_n[v(X) \mid X \leq s]  \\
& + \E_n \left[\{ \tau(X) - \hat f_\tau(X;s)\}\1\{X > s\}  \right]  \E_n[v(X) \mid X > s]  \\
=& \underbrace{\E_n \left[\tau(X) - \E_n[\tau(X) \mid X \leq s] \mid X \leq s  \right]}_{=0} \E_n[v(X) \mid X \leq s] \E_n [ \1\{X \leq s\} ] \\
& + \underbrace{\E_n \left[ \tau(X) - \E_n[\tau(X) \mid X > s] \mid X > s  \right] }_{=0} \E_n[v(X) \mid X > s]  \E_n [ \1\{ X > s\}] \\
=& 0,
\end{align*} 
and $\E_n[v(X) \cdot \hat f_\tau^v(X;s)] = \E_n [\hat f_\tau^v(X;s)^2]$.
We can re-write $M(\hat \tau(X),s)$ the same way.
\end{proof} 

\begin{lemma} \label{lem:g_bound} 
Let $s, s'$ be such that $|s-s'| < \kappa$. Then, under Assumption \ref{assum:min}, the gap between $f_{\tau}(X; s)$ and $f_{\tau}(X; s')$ is bounded: 
$$\sup_{|s-s'|< \kappa} \norm{f_{\tau}(X; s) - f_{\tau}(X; s')}_2^2 \lesssim \kappa.$$
\end{lemma} 
\begin{proof} 
Without loss of generality, let $s' > s$. Then, 
\begin{align} 
\sup_{|s-s'|< \kappa} &\norm{f_{\tau}(X; s) - f_{\tau}(X; s')}_2^2 \label{eqn:g_sup} \\
\leq& \sup_{|s-s'|< \kappa} \left(\E[\tau(X)  \mid X \leq s] - \E[\tau(X)  \mid X \leq s'] \right)^2 \Pr(X \leq s) \label{eqn:g_term1} \\
&+ \sup_{|s-s'|< \kappa} \left(\E[\tau(X)  \mid X > s] - \E[\tau(X)  \mid X \leq s']\right)^2 \Pr(s < X \leq s') \label{eqn:g_term2}\\
&+ \sup_{|s-s'|< \kappa} \left( \E[\tau(X)  \mid X > s] - \E[\tau(X)  \mid X > s'] \right)^2 \Pr( X > s') \label{eqn:g_term3}
\end{align} 

We will show that \eqref{eqn:g_sup} is dominated by \eqref{eqn:g_term2}. We will bound each piece separately. Consider first \eqref{eqn:g_term1}. By Assumption \ref{assum:smooth} and positivity, there exists some constant that upper bounds $\partial \E[\tau(X) \mid X< s]/\partial s$. This implies that within a neighborhood of $s$, $\left \lvert \E[ \tau(X) \mid X < s] - \E [ \tau(X) \mid X \leq s'] \right \rvert \lesssim | s- s'|$, and 
\begin{equation} 
\sup_{|s-s'| < \kappa} (\E [ \tau(X) \mid X \leq s] - \E[\tau(X) \mid X \leq s'])^2 \Pr(X \leq s) \lesssim \kappa^2.
\label{eqn:g_term1_bound}
\end{equation} 
We can similarly show for \eqref{eqn:g_term3}:
\begin{equation} 
\sup_{|s-s'| < \kappa} (\E [ \tau(X) \mid X > s] - \E[\tau(X) \mid X > s'])^2 \Pr(X > s') \lesssim \kappa^2.
\label{eqn:g_term3_bound}
\end{equation} 

\noindent For \eqref{eqn:g_term2}, let $c_2 := \E[\tau(X) \mid X > s] - \E[\tau(X) \mid X > s']$, where $c_2 < \infty$ under boundedness. Then, 
\begin{align*} 
(\E&[\tau(X) \mid X > s] - \E[\tau(X) \mid X \leq s'])^2 \\
=& (\E[\tau(X) \mid X > s] - \E[\tau(X) \mid X > s'] + \E[\tau(X) \mid X  > s'] - \E[\tau(X) \mid X \leq s'])^2 \\
=& (\E[\tau(X) \mid X > s] - \E[\tau(X) \mid X > s'])^2 + (\E[\tau(X) \mid X > s'] - \E[\tau(X) \mid X \leq s'])^2\\
&+2 (\E[\tau(X) \mid X > s] - \E[\tau(X) \mid X > s']) \cdot (\E[\tau(X) \mid X > s'] - \E[\tau(X) \mid X \leq s']) \\
\leq& c_1  |s-s'|^2 + 2 c_2 | s-s'| + c_2^2
\end{align*} 
Furthermore, we bound $\Pr(s < X \leq s')$. By definition of continuity, for any $c > 0$, there exists some $b > 0$ such that $|f_X(t) - f_X(s)| < c$ whenever $|t- s| < b$. As a result, when $| s- s'| < \kappa$: 
\begin{align*} 
\Pr(s < X \leq s') &= \int_{s}^{s'} f_X(t) dt \\
&= \int_s^{s'} f_X(s) (s - s') + \int_{s}^{s'} \left\{ f_X(t) - f_X(s) \right\}dt \\
&\leq f_X(s)(s-s') + c | s - s'| \\
&\lesssim \kappa
\end{align*} 
As a result, \eqref{eqn:g_term2} is bounded:
\begin{equation} 
\sup_{|s-s'|<\kappa} \left( \E[\tau(X) \mid X \geq s] - \E[\tau(X) \mid X < s']\right)^2 \Pr( s \leq X < s') \lesssim \kappa.
\label{eqn:g_term2_bound}
\end{equation} 

\noindent Substituting \eqref{eqn:g_term1_bound}-\eqref{eqn:g_term2_bound} into the expression for \eqref{eqn:g_sup}: 
$$\sup_{|s-s'|< \kappa} \norm{f_{\tau}(X; s) - f_{\tau}(X; s')}_2^2 \lesssim \kappa,$$
which concludes the proof. 
\end{proof} 

\begin{lemma}[Consistency of $\hat s_n$]\label{lem:hatsn_consistency}
Under Assumption \ref{assum:min}-\ref{assum:consistent_teacher},
$$\hat s_n \cip s$$
\end{lemma}
\begin{proof} 
To start, we will establish 
\begin{equation} 
\sup_{s} \left\lvert \E_n \left[ \ell(\hat \tau(X); X, s) \right] - \E \left[ \ell(\tau(X); X, s) \right] \right \rvert \cip 0.
\label{eqn:s_cip_bound}
\end{equation}

\begin{align*} 
\sup_{s} & \big\lvert\E_n \left[ \ell(\hat \tau(X); X, s) \right] - \E \left[ \ell(\tau(X); X, s) \right] \big \rvert \\
&= \sup_{s} \big\lvert \E_n \left[ \ell(\hat \tau(X); X, s) \right] -  \E \left[ \ell(\hat \tau(X); X, s) \right] +\E \left[ \ell(\hat \tau(X); X, s) \right]  -  \E \left[ \ell(\tau(X); X, s) \right] \big \rvert  \\
&\leq \underbrace{\sup_{s} \big\lvert \E_n \left[ \ell(\hat \tau(X); X, s) \right] -  \E \left[ \ell(\hat \tau(X); X, s) \right]\big \rvert}_{(a)}  + \underbrace{\sup_{s}\big\lvert \E \left[ \ell(\hat \tau(X); X, s) \right]  -  \E \left[ \ell(\tau(X); X, s) \right] \big \rvert}_{\text{(b) Teacher Model Error}}
\end{align*}

\paragraph{Term (a). Gap between population and sample-level loss.} 
We begin by bounding the first term, which is the gap between the population and sample-level loss with respect to the distilled outcomes $\hat \tau(X)$. Because $\E\left[\hat \tau(X)^2 \right] < \infty$, we can directly apply \citet{banerjee2007confidence} (page 562) and we have invoked cross-fitting,
$$\sup_{s} \left \lvert  \E_n \left[ \ell(\hat \tau(X); X, s)\right] - \E \left[ \ell(\hat \tau(X); X, s) \mid \hat \theta \right] \right \rvert = o_p(1),$$
where conditioned on $\hat \theta$, $\hat \tau(X)$ is no longer data-dependent. Because the bound is uniform in $\hat \theta$, $\sup_{s} \left \lvert  \E_n \left[ \ell(\hat \tau(X); X, s)\right] - \E \left[ \ell(\hat \tau(X); X, s) \right] \right \rvert = o_p(1).$
\paragraph{Term (b). Teacher Model Error.} 
We will show that term (b) is bounded as a function of the error rate of $\hat \tau(X)$. We begin by substituting in \eqref{eqn:pop_gap}:

\begin{align*} 
\E&\left[ \ell(\hat \tau(X); X, s) \right]  -  \E \left[ \ell(\tau(X); X, s) \right] \\
=&~2 \E \left[ \{ \tau(X) - f_{\tau}(X; s)\} \cdot \{ \hat \tau(X) - \tau(X) \} \right] + \E \left[ \{ \hat \tau(X) - \tau(X)\}^2\right]  - \E[f^v_\tau(X;s)^2 ]\\
\leq &~2 \E \left[ \{ \tau(X) - f_{\tau}(X; s)\} \cdot \{ \hat \tau(X) - \tau(X) \} \right] + \E \left[ \{ \hat \tau(X) - \tau(X)\}^2\right] \\
\leq&~ 2\sqrt{\E \left[ \lvert \tau(X) - f_{\tau}(X; s) \rvert^2 \right] }\sqrt{\E \left[ \lvert \hat \tau(X) -\tau(X) \rvert^2 \right] } + \E \left[ \{ \hat \tau(X) - \tau(X)\}^2\right]\\
\leq&~ c_0 \delta_n + \delta_n^2 \\
\lesssim&~ \delta_n, 
\end{align*} 
where $c_0 > 0$ is a constant. This upper bounds the gap between the population-level loss for $\hat \tau(X)$ and the population-level loss for $\tau(X)$ with $\delta_n$. Because the upper bound does not depend on the split choice $s$, 
$$\sup_{s} \E\left[ \ell(\hat \tau(X); X, s) \right]  -  \E \left[ \ell(\tau(X); X, s) \right] \lesssim \delta_n.$$
Under uniqueness of split $s$ and continuity of $\tau(X)$ in $(l,s)$ and $(s,u)$, ${M}(\tau(X), s) > \sup_{s' \neq G} {M}(\tau(X), s')$, for every open interval $G$ that contains $s$ \citep[][Lemma 5.1]{banerjee2007confidence}. Then, we can directly apply Corollary 3.2.3 from \citet{van1996weak}, to establish $\hat s_n \cip s$. 
\end{proof} 

\vspace{3mm} 

To establish the rate of convergence, we extend \citet{song2015asymptotics}, which is a modification of \citet{van1996weak}, Theorem 3.2.5 and \citet{banerjee2007confidence}, Theorem 5.2 to account for settings in which the underlying data generating process is not linear. \citet{song2015asymptotics} considers a similar idea through the lens of misspecification, where they consider the setting in which a change point detection model is fit to an inherently smooth underlying surface. A similar result is also derived in \citet{escanciano2020estimation}, who explicitly considers the decision tree set-up, though the findings are analogous to \citet{song2015asymptotics}. Our derivation extends the theoretical results considered in the aforementioned papers by account for the additional complexity that the decision tree being estimated is using the estimated CATEs and not the true CATEs.

As an overview, we will establish two key bounds. The first (i.e., Lemma~\ref{lem:m_bound}) bounds the difference in the population-level loss around a neighborhood of the optimal split $s$, while the second (i.e., Lemma~\ref{lem:emp_process-bound}) bounds the fluctuations in the empirical process (i.e., the modulus of continuity of the empirical process, with respect to the gap between the optimal split $s$ and $s'$).

We formalize these with the following lemmas.
\begin{lemma}  \label{lem:m_bound} 
Under Assumption \ref{assum:min}-\ref{assum:bounded_moments}, there exists some constant $C>0$ such that for all $|s- s'| \leq \kappa$ where $\kappa > 0$,
\begin{equation} 
M(\tau(X), s) - M(\tau(X), s') \geq - C \kappa^\alpha.
\label{eqn:m_bound} 
\end{equation}
\end{lemma} 
\begin{proof} 
To begin, let $\tau_l(s) = \E[\tau(X) \mid X \leq s]$ and $\tau_r(s) = \E[\tau(X) \mid X> s]$. For a generic split value $k$, 
\begin{align*} 
M(\tau(X), k) =& \underbrace{\E\left[ (\tau(X) - \tau_l(s) \1\{X \leq k\} - \tau_r(s) \1\{ X > k\} )^2 \right] }_{:= A(k)}\\
&+ \underbrace{\E[(\tau(X) - g_\tau(X; k))^2 - \E[(\tau(X) -   \tau_l(s)  \1\{X \leq k\} - \tau_r(s) \1\{ X > k\})^2]}_{:= B(k)},
\end{align*} 
where we can re-write term 1 (i.e., $A(k)$) as: 
\begin{align*} 
A(k) &= \E\left[ (\tau(X) -  \tau_l(s)  \1\{X \leq k\} -  \tau_r(s)  \1\{ X > k\} )^2 \right]\\
&= \E\left[ (\tau(X) - \tau_r(s) - \tau_l(s)  \1\{X \leq k\} +  \tau_r(s)  \1\{ X \leq k\} )^2 \right]\\
&= \E\left[ (\tau(X) - \tau_r(s) - ( \tau_l(s) - \tau_r(s))  \1\{ X \leq k\} )^2 \right]\\
&= \E\left[(\tau(X) - \tau_r(s))^2 \right] - 2 ( \tau_l(s)  -  \tau_r(s) ) \E[\tau(X) - \tau_r(s)] \1\{X \leq k\} + (\tau_l(s) - \tau_r(s))^2 \1\{ X \leq k\} \\ 
&= \E\left[(\tau(X) - \tau_r(s))^2 \right] - 2 ( \tau_l(s)  -  \tau_r(s) ) \E \left[ \left\{ \tau(X) - \tau_r(s) - \frac{\tau_l(s) - \tau_r(s)}{2} \right\} \1\{ X \leq k\}\right] \\ 
&= \E\left[(\tau(X) - \tau_r(s))^2 \right] - 2 ( \tau_l(s)  -  \tau_r(s) ) \E[(\tau(X) - \tau_0) \1\{ X \leq k\}],
\end{align*} 
where $\tau_0 = \frac{1}{2} \left\{\E(\tau(X) \mid X_i < s) + \E(\tau(X) \mid X \geq s) \right\}$, 
and similarly, we can re-write term 2 (i.e., $B(k)$) as: 
\begin{align*} 
B(k) =& \E[(\tau(X) - g_\tau(X; k))^2 - \E[(\tau(X) -   \tau_l(s)  \1\{X \leq k\} - \tau_r(s) \1\{ X > k\})^2] \\
=& \E[(\tau(X) - \tau_l(k) \1\{X \leq k\} - \tau_r(k) \1\{X > k\} )^2 \\
& \quad - \E[(\tau(X) -   \tau_l(s)  \1\{X \leq k\} - \tau_r(s) \1\{ X > k\})^2]\\
=& (\tau_l(k)^2 - \tau_l(s)^2) \E[\1\{X \leq k\}] + (\tau_r(k)^2 - \tau_r(s)^2) \E[\1\{X > k\}] \\
& \quad - 2 \E[\tau(X) \left(  \left\{ \tau_l(k) - \tau_l(s) \right\} \1\{X \leq k\}  +  \left\{ \tau_r(k) - \tau_r(s) \right\} \1\{X > k\} \right) ] \\
=&(\tau_l(k)^2 - \tau_l(s)^2) \E[\1\{X \leq k\}] + (\tau_r(k)^2 - \tau_r(s)^2) \E[\1\{X > k\}] \\
& \quad  - 2 \left\{ \tau_l(k) - \tau_l(s) \right\} \E[\tau(X) \1\{X \leq k\}] - 2 \left\{ \tau_r(k) - \tau_r(s) \right\}  \E[\tau(X) \1\{ X > k\} ]\\
=& (\tau_l(k)^2 - \tau_l(s)^2) \E[\1\{X \leq k\}] + (\tau_r(k)^2 - \tau_r(s)^2) \E[\1\{X > k\}]\\
& \quad - 2 \left\{ \tau_l(k) - \tau_l(s) \right\} \tau_l(k) \E[\1\{X \leq k\}]- 2 \left\{ \tau_r(k) - \tau_r(s) \right\}  \tau_r(k) \E[\1\{X > k\}]  \\
=& - (\tau_l(k) - \tau_l(s))^2 F_X(k) - (\tau_r(k) - \tau_r(s))^2 (1-F_X(k)),
\end{align*}
Then, because $B(s) = 0$, $B(s) - B(s') = - B(s') \geq 0$, 
\begin{align*} 
 M(\tau(X), s) - M(\tau(X), s') &=  A(s) - A(s') + (B(s) - B(s')) \\
 &=  A(s) - A(s') - B(s') \\
 &\geq A(s) - A(s').
\end{align*} 
Because $M(\tau(X), s) - M(\tau(X), s') \leq 0$, $A(s) - A(s') \leq B(s') \leq 0$. There exists some constant $C$ such that 
\begin{align*} 
A(s) - A(s') &=  2(\tau_l(s) - \tau_r(s)) \E \left[( \tau(X) - \tau_0) )(\1\{X \leq s\} - \1\{X \leq s'\})\right]\\
&= -2\lvert \tau_l(s) - \tau_r(s)\rvert \lvert H(s) - H(s')\rvert \\
&\geq - C |s-s'|^\alpha,
\end{align*} 
where the final inequality follows from Assumption \ref{assum:smooth}-(b). As such, we have shown that for $|s-s'| \leq \kappa$ where $\kappa > 0$, there exists some constant $C > 0$ such that $M(\tau(X), s) - M(\tau(X), s') \geq - C \kappa^\alpha$. 
\end{proof} 

\begin{lemma} \label{lem:emp_process-bound}
For $\kappa \in (0,\kappa_1]$, under Assumptions \ref{assum:min}-\ref{assum:consistent_teacher}, there exists a function $\phi_n(\cdot)$ such that: 
\begin{align} 
\E \left[ \sup_{|s - s'| < \kappa} \bigg \lvert M_n(\hat \tau(X), s) - M(\tau(X), s) - \left\{M_n(\hat \tau(X), s') - M(\tau(X), s') \right\}  \bigg\rvert \right] \leq \frac{\phi_n(\kappa)}{\sqrt{n}}.
\label{eqn:emp_process_bound}
\end{align} 
\end{lemma} 

\begin{proof} 
To start, we can re-write the gap between $M_n(\hat \tau(X), s) - M(\tau(X), s)$ and $M_n(\hat \tau(X), s') - M(\tau(X), s')$. Let $\mG_n(\cdot, s) =  \sqrt{n}\{M_n(\cdot, s) - M(\cdot, s)\}$. Then, 
\begin{align*} 
\E &\left[ \sup_{|s - s'| < \kappa} \sqrt{n} \bigg \lvert M_n(\hat \tau(X), s) - M(\tau(X), s) - \left\{M_n(\hat \tau(X), s') - M(\tau(X), s') \right\}  \bigg\rvert \right]  \\
&= \E \bigg[ \sup_{|s - s'| < \kappa}\sqrt{n} \bigg \lvert M_n(\tau(X), s) - M(\tau(X), s) + \big\{ M_n(\hat \tau(X), s) - M_n(\tau(X), s) \big\} \\
&\qquad \qquad \qquad- \big\{ M_n(\tau(X), s') - M(\tau(X), s') + \left( M_n(\hat \tau(X), s')- M_n(\tau(X), s') \right)\big\}  \bigg\rvert \bigg] \\
&= \E \bigg[ \sup_{|s - s'| < \kappa} \bigg \lvert \mG_n(\tau(X), s) - \mG_n(\tau(X), s') + \\
&\qquad \sqrt{n} \big\{M_n(\hat \tau(X), s) - M_n(\tau(X), s) - (M_n(\hat \tau(X), s') - M_n(\tau(X), s')) \big\} \bigg \rvert \bigg] \\
&\leq\underbrace{\E \left[ \sup_{|s - s'| < \kappa} \bigg \lvert \mG_n(\tau(X), s) - \mG_n(\tau(X), s') \bigg \rvert \right] }_{\text{Term 1 (Empirical Process)}}  + \\
&\qquad  + \sqrt{n} \underbrace{\E \left[ \sup_{|s - s'| < \kappa} \bigg \lvert \big\{M_n(\hat \tau(X), s) - M_n(\tau(X), s) - (M_n(\hat \tau(X), s') - M_n(\tau(X), s')) \big\} \bigg \rvert \right]}_{\text{Term 2 (Teacher model error)}}.
\end{align*} 
Usefully, term 1 is directly bounded by Theorem 3.2 from \citet{escanciano2020estimation}: 
 \begin{align*} 
 \E \left[ \sup_{|s - s'| < \epsilon} \bigg \lvert \mG_n(\tau(X), s) - \mG_n(\tau(X), s') \bigg \rvert \right] \lesssim \epsilon^\eta.
\end{align*} 

Bounding the teacher model error (i.e., term 2) is a more involved argument. We rely on cross-fitting to ensure that $\hat \tau(X; \hat \theta)$ is a deterministic quantity, once conditioning on the nuisance parameters $\hat \theta$. More specifically, let the sample data $\mathcal{D}_n$ be partitioned into two random splits $\cD_n^{(1)}$ and $\cD_n^{(2)}$. Without loss of generality, assume $\cD_n^{(1)}$ is used to estimate the nuisance parameters $\hat \theta$, while $\hat \tau(X; \hat \theta)$, $\hat g_{\hat \tau}(s)$, and $M_n(\hat \tau(X), s)$ is computed using $\cD_n^{(2)}$. This ensures that conditional on $\hat \theta$ (or equivalently, $\cD^{(1)}_n$), $\hat \tau(X; \hat \theta)$ is a deterministic function of $X$, and by extension, $\hat \tau(X; \hat \theta) - \tau(X)$ is also a fixed function. 

Substituting in Equation \eqref{eqn:pop_gap}, we have that
\begin{align} 
\E &\left[ \sup_{|s - s'| < \kappa} \bigg \lvert \big\{M_n(\hat \tau(X), s) - M_n(\tau(X), s) - (M_n(\hat \tau(X), s') - M_n(\tau(X), s')) \big\} \bigg \rvert \right] \nonumber \\
=& \E \left[ \sup_{|s - s'| < \kappa} \bigg \lvert 2 \E_n \big[\{\hat f_{\tau}(X;s) - \hat f_{\tau}(X; s') \} \cdot \{\hat \tau(X) - \tau(X) \} \big] - \left\{ \E_n \left[ \hat f_\tau^v(X;s)^2 \right] - \E_n \left[ \hat f_\tau^v(X;s')^2 \right]\right\}  \bigg \rvert \right] \nonumber \\
\leq& \underbrace{2 \E \left[ \sup_{|s - s'| < \kappa} \bigg \lvert  \E[(f_{\tau}(X;s) - f_{\tau}(X; s') ) \cdot \{\hat \tau(X) - \tau(X) \}] \bigg \rvert \right] }_{(a)} \nonumber \\
&  + \underbrace{\frac{2}{\sqrt{n}} \E \left[ \sup_{|s - s'| < \kappa} \bigg \lvert   \mG_n(( f_{\tau}(X;s) - f_{\tau}(X; s')) (\hat \tau(X) - \tau(X))) \bigg \rvert \right] }_{(b)}\label{eqn:teacher_model_error} \\
& + \underbrace{2 \E \left[ \sup_{|s - s'| < \kappa} \bigg \lvert \E_n \bigg[ \left\{ \hat \tau(X) - \tau(X) \right\} \cdot \left\{ \hat f_{\tau}(X; s) - f_{\tau}(X; s) - (\hat f_\tau(X; s') - f_{\tau}(X; s')) \right\}\bigg]\bigg \rvert \right]}_{(c)}\nonumber \\
&+ \underbrace{\E\left[ \sup_{|s-s'| < \kappa} \left \lvert \E_n \left[ \hat f_\tau^v(X;s)^2 \right] - \E_n \left[ \hat f_\tau^v(X;s')^2 \right] \right \rvert \right]}_{(d)}
\end{align}  

We will bound each quantity in \eqref{eqn:teacher_model_error} and show that asymptotically, \eqref{eqn:teacher_model_error}-(a) will dominate. 

\paragraph{Term (a).} To bound this quantity, we invoke cross-fitting and condition on $\hat \theta$:
\begin{align*} 
\E&\left [\left\{ g_{\tau}(X;s) - f_{\tau}(X; s') \right\} \cdot \left\{ \hat \tau(X; \hat \theta) - \tau(X) \right\} \mid \hat \theta \right]\\
&\leq^{(i)} \sqrt{\E\left [ \left\{ f_{\tau}(X;s) - f_{\tau}(X; s') \right\}^2 \mid \hat \theta \right]} \cdot \sqrt{\E \left[ \left\{\hat \tau(X; \hat \theta) - \tau(X) \right\}^2 \mid \hat \theta \right]}\\
&\leq^{(ii)}\norm{f_{\tau}(X;s) - f_{\tau}(X; s')}_2 \delta_n
\end{align*} 
where (i) follows from applying Cauchy-Schwarz, and (ii) follows from noting that $g_\tau$ is a deterministic function that is not dependent on $\hat \theta$ and the assumption that $\E \left[ \left\{\hat \tau(X; \hat \theta) - \tau(X) \right\}^2 \mid \hat \theta \right]^{1/2} \leq \delta_n$. Then, applying Lemma \ref{lem:g_bound}, 
\begin{align*} 
\sup_{|s-s'|<\kappa} \left \lvert \E[(f_{\tau}(X;s) - f_{\tau}(X; s') ) \cdot (\hat \tau(X) - \tau(X))] \right \rvert  \lesssim \delta_n \kappa^{1/2} 
\end{align*} 

\paragraph{Term (b).} To begin, we define the random, $\hat \theta$-dependent function class: 
$$\cF_{\kappa} = \left\{ (\hat \tau(X; \hat \theta) - \tau(X)) (f_{\tau}(X; s) - f_\tau(X; s'): |s-s'| < \kappa \right\}.$$
Conditional on $\hat \theta$, $\cF_{\kappa }$ is a fixed function class. There are two things worth noting about $\cF_{\kappa}$. First, there exists a constant envelope $F$ for $\cF_{\epsilon}$, which follows directly from boundedness. More specifically, $\norm{\hat \tau(X; \hat \theta) - \tau(X)}_{\infty} \leq c_1$, where $c_1$ is a constant, and $\norm{f_{\tau}(X;s) - f_{\tau}(X; s')}_\infty \leq C_0$, where $C_0 < \infty$. Then, $\mathcal{F}_\kappa$ has a constant envelope.

Furthermore, conditioned on $\hat \theta$, the $L_2$ radius of $\cF_{\kappa}$ is bounded. More precisely, denote the $L_2$ radius of $\cF_{\kappa}$, conditioned on the parameter $\hat \theta$ as $\sigma_{\hat \theta}$ (i.e., $\sigma_{\hat \theta} := \sup_{f \in \cF_{\epsilon}} \norm{f}_{2, \hat \theta},$ where we use the notation $\norm{\cdot}_{2, \hat \theta}$ to represent the $L_2$ norm, conditioned on $\hat \theta$). For any $f  \in \mathcal{F}_{\kappa}$:
\begin{align*} 
\E\left(f^2 \mid \hat \theta \right) =& \E \left[ (\hat \tau(X; \hat \theta) - \tau(X))^2 (f_{\tau}(X; s) - f_\tau(X; s') )^2 \mid \hat \theta \right] \\
\leq& \E\left[(\hat \tau(X; \hat \theta) - \tau(X))^2 \mid \hat \theta \right] \norm{(f_{\tau}(X; s) - f_\tau(X; s')}_{\infty} \\
&\quad \wedge \norm{\hat \tau(X; \hat \theta) - \tau(X)}^2_{\infty} \norm{(f_{\tau}(X; s) - f_\tau(X; s')}_{2} \\
\lesssim& \delta^2_n \wedge \kappa^{1/2} \\
=& \kappa^{1/2},
\end{align*} 
which follows from $\delta_n = o(1)$. Then, taking the supremum of $f \in \cF_{\kappa}$: 
\begin{align*} 
\sigma_{\hat \theta} :&= \sup_{f \in \cF_{\kappa}} \norm{f}_{2, \hat \theta} \lesssim \kappa^{1/2}.
\end{align*} 

We can then directly compute the uniform entropy integral: 
\begin{align*} 
\mathcal{J}(\sigma_{\hat \theta}) \lesssim \int_{0}^{\sigma_{\hat \theta}} \sqrt{1 + \log(1/\kappa)} d\kappa \lesssim \sigma_{\hat \theta} \sqrt{\log(1/\sigma_{\hat \theta})},
\end{align*} 
which follows by noting that the function class of one-dimensional indicator functions has a bracketing number that is upper bounded by $\log(1/\kappa)$.

Then, we can apply \citet{van1996weak}, Theorem 2.14.1: 
\begin{align*} 
\E &\left[ \sup_{f\in \cF_{\kappa}} \left \lvert \mG_n(f) \right \rvert \mid \hat \theta  \right] \lesssim \mathcal{J} \left(\sup_{f\in \cF_{\kappa}} \norm{f}^2_{2, \hat \theta} \right) \lesssim \kappa^{1/2}.
\end{align*} 
Because the upper bound does not depend on $\hat \theta$: 
\begin{align*} 
\frac{2}{\sqrt{n}}\E &\left[ \sup_{|s-s'| < \kappa} \bigg \lvert \mG_n(( f_{\tau}(X;s) - f_{\tau}(X; s'))(\hat \tau(X) - \tau(X)) \bigg \rvert  \right]\\
&\leq \frac{2}{\sqrt{n}} \E \left\{ \E \left[ \sup_{|s-s'| < \kappa} \left \lvert \mG_n(( f_{\tau}(X;s) - f_{\tau}(X; s'))(\hat \tau(X) - \tau(X)) \right \rvert \mid \hat \theta \right] \right\} \\
&\lesssim \frac{\kappa^{1/2}}{\sqrt{n}}
\end{align*} 

\paragraph{Term (c).} We now bound the third term. 
\begin{align*} 
\sup_{|s - s'| < \kappa}& \left \lvert \E_n \left[ (\hat \tau(X) - \tau(X)) \cdot (\hat f_{\tau}(X; s) - f_{\tau}(X; s) - (\hat f_\tau(X; s') - f_{\tau}(X; s'))\right] \right \rvert  \\
&\leq^{(i)} \norm{\hat \tau(X) - \tau(X)}_2 \sup_{|s - s'| < \kappa} \norm{\hat f_{\tau}(X; s) - f_{\tau}(X; s) - (\hat f_\tau(X; s') - f_{\tau}(X; s'))}_2 \\
&\leq^{(ii)} \delta_n O_p(1/\sqrt{n}) \\
&= O_p(\delta_n/\sqrt{n})
\end{align*} 
where $(i)$ follows from Cauchy-Schwarz, and $(ii)$ follows from applying Theorem 3.2 in \citet{song2015asymptotics}, where we treat $\tau(X)$ as the outcome and establishes the convergence of $\hat f_\tau(X; s)$ to $f_\tau(X; s)$. 

\paragraph{Term (d).} Finally, we turn to bounding the final term:
$$\E\left[ \sup_{|s-s'| < \kappa} \left \lvert \E_n \left[ \hat f_\tau^v(X;s)^2 \right] - \E_n \left[ \hat f_\tau^v(X;s')^2 \right] \right \rvert \right].$$
To start, we can re-write $\E_n[\hat f_\tau^v(X;s)^2]$ as
\begin{align*} 
\E_n[\hat f_\tau^v(X;s)^2] =& \E_n[v(X) \mid X \leq s]^2 \E_n[\1\{X \leq s\}] + \E_n[v(X) \mid X > s]^2 \E_n[\1\{X > s\}] \\
=& \E_n[v(X) \mid X \leq s]^2 \hat F_X(s) + \E_n[v(X) \mid X > s]^2 \left(1- \hat F_X(s) \right)\\
=& \E_n[v(X)]^2 - \E_n[v(X)]^2 + \E_n[v(X) \mid X \leq s]^2 \hat F_X(s) + \E_n[v(X) \mid X > s]^2 \left(1- \hat F_X(s) \right)\\
=& \E_n[v(X)]^2 - \E_n[v(X)\mid X \leq s]^2 \hat F_X(s)^2 - \E_n[v(X)\mid X > s]^2 (1-\hat F_X(s))^2 \\
&- 2 \E_n[v(X) \mid X \leq s] \E_n[v(X) \mid X > s] \hat F_X(s) (1-\hat F_X(s)) \\
&+ \E_n[v(X) \mid X \leq s]^2 \hat F_X(s) + \E_n[v(X) \mid X > s]^2 \left(1- \hat F_X(s) \right) \\
=& \E_n[v(X)]^2 + \hat F_X(s) \left(1- \hat F_X(s) \right) \left\{ \E_n[v(X) \mid X \leq s] - \E_n[v(X) \mid X > s] \right\}^2 \\ 
=& \E_n[v(X)]^2 + \frac{\E_n\left[ (v(X) - \E_n[v(X)])  \1\{X \leq s\}  \right]^2}{\hat F_X(s) \left(1- \hat F_X(s) \right) }.
\end{align*} 
Let $H_v(s) := \E_n\left[ (v(X) - \E_n[v(X)])  \1\{X \leq s\}  \right]$. Then, $\E_n \left[ \hat f_\tau^v(X;s)^2 \right] - \E_n \left[ \hat f_\tau^v(X;s')^2 \right]$ can be compactly re-written as: 
\begin{align} 
\E_n &\left[ \hat f_\tau^v(X;s)^2 \right] - \E_n \left[ \hat f_\tau^v(X;s')^2 \right] \nonumber \\
&= \frac{H_v(s)^2}{\hat F_X(s) \left(1- \hat F_X(s) \right) } - \frac{H_v(s')^2}{\hat F_X(s') \left(1- \hat F_X(s') \right) } \nonumber \\
&= \frac{H_v(s)^2 - H_v(s')^2}{\hat F_X(s) \left(1- \hat F_X(s) \right) } - H_v(s')^2 \left( \frac{1}{\hat F_X(s') \left(1- \hat F_X(s') \right) } - \frac{1}{\hat F_X(s) \left(1- \hat F_X(s) \right) } \right) \nonumber \\
&= \frac{H_v(s)^2 - H_v(s')^2}{\hat F_X(s) \left(1- \hat F_X(s) \right) } - H_v(s')^2 \cdot  \frac{\hat F_X(s') (1-\hat F_X(s')) - \hat F_X(s) (1-F_X(s))}{\hat F_X(s') \left(1- \hat F_X(s') \right) \cdot \hat F_X(s) \left(1- \hat F_X(s) \right) }
\label{eqn:hv_expression}
\end{align} 
Equation \eqref{eqn:hv_expression} depends on the empirical cdf of $X$. While Assumption \ref{assum:min} guarantees that the \textit{population}-level cdf is bounded away from 0 and 1, the empirical cdf is not necessarily bounded. We consider two scenarios. In the first event, we consider when $\hat F_X(k)$ for all $k \in \mathcal{N}$ is bounded away from 0 and 1. The second event considers when $\hat F_X(k)$ deteriorates towards 0 or 1. We will show that the second event occurs with vanishing probability. 

To formalize, let $\eta_- \leq F_X \leq \eta_+$ represent the range that $F_X$ spans across the neighborhood $\mathcal{N}$, where, by Assumption \ref{assum:min}, $[\eta_-, \eta_+] \in (0,1)$. Let $\eta := \frac{1}{2} \min \{ \eta_-, 1- \eta_+\}$. Define the event $E_n := \{ \eta \leq \hat F_X(k) \leq 1- \eta, \ \text{ for all } k \in \mathcal{N}\}$. We will consider the setting in which $E_n$ holds, and the one in which it does not (i.e., $E_n^c$). 
\begin{itemize} 
\item Conditional on the event $E_n$, $\hat F_X(k) \{ 1-\hat F_X(k)\} \geq \eta (1- \eta)$ for all $k \in \mathcal{N}$. Then: 
\begin{align} 
\E_n &\left[ \hat f_\tau^v(X;s)^2 \right] - \E_n \left[ \hat f_\tau^v(X;s')^2 \right] \nonumber \\
&\leq \frac{H_v(s)^2 - H_v(s')^2}{\eta (1-\eta)} - H_v(s')^2 \cdot  \frac{\hat F_X(s') (1-\hat F_X(s')) - \hat F_X(s) (1-F_X(s))}{\eta^2 (1-\eta)^2} \nonumber \\
&\leq \frac{H_v(s)^2 - H_v(s')^2}{\eta (1-\eta)} - H_v(s')^2 \cdot  \frac{\left \lvert \hat F_X(s') - \hat F_X(s) \right \rvert}{\eta^2 (1-\eta)^2} \nonumber \\
&\leq \frac{\left \lvert H_v(s) - H_v(s') \right \rvert \cdot ( \lvert H_v(s) \rvert + \lvert H_v(s') \rvert )}{\eta (1-\eta)} - H_v(s')^2 \cdot  \frac{\left \lvert \hat F_X(s') - \hat F_X(s) \right \rvert}{\eta^2 (1-\eta)^2} \label{eqn:h_v_gap}
\end{align} 
Without loss of generality, assume $s' \leq s$. Then, we can re-write $H_v(s) - H_v(s')$ as
\begin{align*} 
\lvert H_v(s) - H_v(s') \rvert &= \lvert \E_n[\left\{ v(X) - \E_n(v(X)) \right\} \1\{X \leq s\} ] -  \E_n[\left\{ v(X) - \E_n(v(X)) \right\} \1\{X \leq s'\} ] \rvert\\
&= \lvert \E_n[\left\{ v(X) - \E_n(v(X)) \right\} \1\{s' \leq X \leq s\} ] \rvert \\
&\leq \sqrt{\lvert \E_n [ \{ v(X) - \E_n(v(X))\}^2 ] \rvert \cdot  \lvert \E_n [\1\{ s' \leq X \leq s\}] \rvert } \\
&\leq R_n \cdot \left \lvert \hat F_X(s') - \hat F_X(s) \right \rvert^{1/2},
\end{align*} 
where $R_n := \E_n[v(X)^2]^{1/2}$. Then, substituting into \eqref{eqn:h_v_gap}, $\E_n\left[ \hat f_\tau^v(X;s)^2 \right] - \E_n \left[ \hat f_\tau^v(X;s')^2 \right]$ is upper bounded by the following: 
\begin{align*} 
\E_n&\left[ \hat f_\tau^v(X;s)^2 \right] - \E_n \left[ \hat f_\tau^v(X;s')^2 \right]\\
&\leq \frac{\left \lvert H_v(s) - H_v(s') \right \rvert \cdot ( \lvert H_v(s) \rvert + \lvert H_v(s') \rvert )}{\eta (1-\eta)} - H_v(s')^2 \cdot  \frac{\left \lvert \hat F_X(s') - \hat F_X(s) \right \rvert}{\eta^2 (1-\eta)^2}  \\ 
&\leq \frac{2 R_n^2 \left \lvert \hat F_X(s') - \hat F_X(s) \right \rvert^{1/2}}{\eta(1-\eta)} - R_n^2 \cdot \frac{\left \lvert \hat F_X(s') - \hat F_X(s) \right \rvert}{\eta^2(1-\eta)^2}\\
&\leq R_n^2 \left( \frac{2}{\eta (1-\eta)} - \frac{1}{\eta^2(1-\eta)^2} \right) \cdot \left \lvert \hat F_X(s') - \hat F_X(s) \right \rvert^{1/2}.
\end{align*} 
Under sample splitting, the nuisance parameters $\hat \theta$ are fixed, and 
\begin{align*} 
\E& \left[ \left. \sup_{|s-s'| < \kappa} \left \lvert \E_n \left[ \hat f_\tau^v(X;s)^2 \right] - \E_n \left[ \hat f_\tau^v(X;s')^2 \right] \right \rvert \cdot \1\{E_n\} ~\right | ~ \hat \theta\right]\\
&\leq \E \left[ \left. \sup_{|s-s'| < \kappa} \left \lvert \E_n \left[ \hat f_\tau^v(X;s)^2 \right] - \E_n \left[ \hat f_\tau^v(X;s')^2 \right] \right \rvert  ~\right | ~ \hat \theta, E_n\right]\\
&\lesssim \E \left[  \left. \sup_{|s-s'| < \kappa} R_n^2 \left \lvert \hat F_X(s') - \hat F_X(s)\right \rvert^{1/2} ~\right | ~ \hat \theta, E_n\right]  \\
&\leq \E[R_n^4 \mid  \hat \theta, E_n]^{1/2} \cdot \E \left[  \left. \sup_{|s-s'| < \kappa}  \left \lvert \hat F_X(s') - \hat F_X(s)\right \rvert ~\right | ~ \hat \theta, E_n\right]^{1/2}\\
&\leq \E[R_n^4 \mid  \hat \theta, E_n]^{1/2} \cdot \E \left[  \left. \left \lvert \hat F_X(s+\kappa) - \hat F_X(s-\kappa)\right \rvert ~\right | ~ \hat \theta, E_n\right]^{1/2} \\
&\leq \E[R_n^4 \mid  \hat \theta]^{1/2}  \cdot (2\kappa)^{1/2}.
\end{align*} 
To bound $\E[R^4_n \mid \hat \theta]^{1/2}$, note that $v(X)$ is i.i.d. under cross-fitting: 
\begin{align*} 
\E[R_n^4 \mid  \hat \theta]^{1/2} &= \E[ \E_n[v(X)^2]^{2} \mid \hat \theta]^{1/2} \\
&= \left( \E\left[ \E_n[v(X)]^2 \mid  \hat \theta\right]^2 + \var(\E_n[v(X)] \mid  \hat \theta) \right)^{1/2} \\
&\lesssim \left( \delta_n^4 + \frac{\delta^2_n }{n}\right)^{1/2},
\end{align*}
where the final inequality arises from noting that we can re-write $\var(\E_n[v(X) \mid \hat \theta)$ as
\begin{align*} 
\var(\E_n[v(X)] \mid  \hat \theta) &\leq \frac{1}{n} \var(v(X) \mid \hat \theta) \leq \frac{c_1^2 \E[v(X)^2 \mid \hat \theta ]}{n} \leq \frac{c_1^2 \delta_n^2}{n}.
\end{align*} 
Because the bound does not depend on $\hat \theta$, 
\begin{align*} 
\E& \left[ \left. \sup_{|s-s'| < \kappa} \left \lvert \E_n \left[ \hat f_\tau^v(X;s)^2 \right] - \E_n \left[ \hat f_\tau^v(X;s')^2 \right] \right \rvert ~\right | ~ E_n\right] \lesssim \left( \delta_n^2 + \frac{\delta_n}{\sqrt{n}}\right) \kappa^{1/2} 
\end{align*} 
\item We now consider $E_n^c$. To start, the probability of $E_n$ not occurring is bounded by the Dvoretzky–Kiefer–Wolfowitz inequality: 
\begin{align*} 
\Pr(E_n^c) \leq \Pr\left( \sup_k \left \lvert \hat F_n(k) - F_n(k) \right \rvert > \eta \right) &\leq 2 \exp(-2 n \eta^2).
\end{align*} 
Then, we invoke the fact that $\sup_{|s-s'| < \kappa} \left \lvert \E_n \left[ \hat f_\tau^v(X;s)^2 \right] - \E_n \left[ \hat f_\tau^v(X;s')^2 \right] \right \rvert^2$ is upper bounded by $R_n^4$.
\begin{align*} 
\E& \left[\left.  \sup_{|s-s'| < \kappa} \left \lvert \E_n \left[ \hat f_\tau^v(X;s)^2 \right] - \E_n \left[ \hat f_\tau^v(X;s')^2 \right] \right \rvert  \cdot \1\{E_n^c\}~ \right |~ \hat \theta~\right] \\
&\leq \E \left[\left. \sup_{|s-s'| < \kappa} \left \lvert \E_n \left[ \hat f_\tau^v(X;s)^2 \right] - \E_n \left[ \hat f_\tau^v(X;s')^2 \right] \right \rvert^2 ~\right |~ \hat \theta~\right]^{1/2} \E\left[ \1\{E_n^c\}\right]^{1/2} \\ 
&\leq \E[R_n^4 \mid \hat \theta ]^{1/2} \cdot \sqrt{2}\exp(-n \eta^2) \\
&\lesssim \left(\delta_n^2 + \frac{\delta_n}{\sqrt{n}} \right) \exp(-n \eta^2)
\end{align*} 
\end{itemize} 
Combining both, we have 
\begin{align*} 
\sqrt{n} \E \left[ \sup_{|s-s'| < \kappa} \left \lvert \E_n \left[ \hat f_\tau^v(X;s)^2 \right] - \E_n \left[ \hat f_\tau^v(X;s')^2 \right] \right \rvert \right] \lesssim \left( \sqrt{n} \delta_n^2 + \delta_n \right)\left( \kappa^{1/2} + \exp(- n \eta^2)\right).
\end{align*} 
As a result, we see that this term is dominated by $\delta_n \kappa^{1/2}$.

Plugging each of the bounds into \eqref{eqn:teacher_model_error}, 
\begin{align*} 
\E &\left[ \sup_{|s - s'| < \kappa} \sqrt{n} \bigg \lvert M_n(\hat \tau(X), s) - M_n(\tau(X), s) - \left\{M_n(\hat \tau(X), s') - M_n(\tau(X), s') \right\}  \bigg\rvert \right]
&\lesssim  \sqrt{n} \delta_n \kappa^{1/2} 
\end{align*} 
As such, we have shown that: 
\begin{align*} 
\E \left[ \sup_{|s - s'| < \kappa} \bigg \lvert M_n(\hat \tau(X), s) - M(\tau(X), s) - \left\{M_n(\hat \tau(X), s') - M(\tau(X), s') \right\}  \bigg\rvert \right] \lesssim \frac{\kappa^{\eta} + \sqrt{n} \delta_n \kappa^{1/2} }{\sqrt{n}}.
\end{align*} 
Setting $\phi_n(\kappa) := \kappa^{\eta} + \sqrt{n} \delta_n \kappa^{1/2}$ concludes the proof. 
\end{proof} 

With Lemma \ref{lem:m_bound} and Lemma \ref{lem:emp_process-bound}, we can apply \citet{van1996weak}, Theorem 3.2.5 to establish the rate of convergence, where we set $\kappa = \epsilon^{2/\alpha}$. The final rate then depends on how fast the teacher model converges to the true CATE. In setting 1, $\delta_n \leq \epsilon^{\frac{2\eta - 1}{\alpha}}/\sqrt{n}$. Then, the maximum fluctuation bound is dominated by the first term (i.e., $\epsilon^{2\eta/\alpha}$, and we recover the same rate as \citet{escanciano2020estimation} of $O_p(n^{-\frac{1}{2\alpha - 2\eta}})$. In setting 2, $\delta_n \geq \epsilon^{\frac{2\eta - 1}{\alpha}}/\sqrt{n}$, and $|\hat s_n - s| = O_p(\delta_n^{\frac{2}{2\alpha - 1}})$. The final rate then depends on the maximum between the two rates. Combining both, we arrive at 
$$|\hat s_n - s| = O_p\left(n^{-\frac{1}{2 \alpha - 2 \eta}} + \delta_n^{\frac{2}{2 \alpha - 1}} \right),$$
which concludes the proof.

\subsection{Proof of Theorem \ref{thm:subgroup_consistency}}
To prove Theorem \ref{thm:subgroup_consistency}, we leverage the following two lemmas. Lemma \ref{lem:bounded_loss} decomposes the gap between the sample-level loss, with $\hat \tau(X)$ and the estimated split $\hat s_n^{(k)}$, and the population-level loss,  with the true CATE $\tau(X)$ and optimal split $s^{(k)}$. The gap between these two terms can be written as a function of the gap between the estimated and true split and the difference between $\E_n[\ell(\hat \tau(X); X, s)]$ and $\E[\ell( \tau(X); X, s)]$.

\begin{lemma}[Bounded Loss]\label{lem:bounded_loss} 
Without loss of generality, assume $\hat s^{(k)}_n < s$. Then for $k \in \{1, ..., p\}$: 
\begin{align*} 
&\left| \E_n\left[\ell(\hat \tau(X); X^{(k)}, \hat s^{(k)}_n) \right]  - \E\left[\ell(\tau(X); X^{(k)}, s^{(k)}) \right] \right|\\
&\qquad \leq C^{(k)}_\tau \times \left| \hat s^{(k)}_n - s^{(k)} \right| + \bigg \lvert \E_n\left[ \ell(\hat \tau(X); X, s) \right] - \E\left[\ell(\tau(X); X, s) \right] \bigg \rvert,
\end{align*} 
where $C^{(k)}_\tau = 2\max_{c \in [\hat s^{(k)}_n, s^{(k)}]}\E(\hat \tau(X) \mid X^{(k)} < c) \times \E_n\left|\tau^d(X^{(k)};\hat s^{(k)}_n) - \hat \tau(X) \right|$. 
\end{lemma} 
\begin{proof} 
We can re-write the difference in the within-sample loss and the population loss as: 
\begin{align*} 
&\left \lvert\E_n\left[\ell(\hat \tau(X_i), X; \hat s_n) \right]  - \E\left[\ell( \tau(X_i); X, s) \right] \right \rvert  \\
&= \left \lvert\E_n\left[\ell(\hat \tau(X_i), X; \hat s_n) \right]  - \E_n\left[\ell(\hat \tau(X_i); X, s) \right] + \E_n\left[\ell(\hat \tau(X_i); X, s) \right] - \E\left[\ell(\tau(X_i); X, s) \right] \right \rvert  \\
&\leq \underbrace{\left \lvert\E_n\left[\ell(\hat \tau(X_i), X; \hat s_n) \right]  - \E_n\left[\ell(\hat \tau(X_i); X, s) \right]\right \rvert}_{(*)}  + \left \lvert \E_n\left[ \ell(\hat \tau(X_i); X, s) \right] - \E\left[\ell(\tau(X_i); X, s) \right] \right \rvert
\end{align*}

\noindent We will construct an upper bound on the first term $(*)$: 
\begin{align*} 
& \left| \E_n\left[\ell(\hat \tau(X); X, \hat s_n) \right]  - \E_n\left[\ell(\hat \tau(X); X, s) \right] \right|\\
& =\left| \E_n\left[ (\hat \tau(X) - \hat f_\tau^d(X;\hat s_n))^2 - (\hat \tau(X) - \hat f^d_\tau(X;s))^2 \right]\right| \\
& = \left|\E_n\left[ \hat f_\tau^d(X;\hat s_n)^2 - \hat f^d_\tau(X;s)^2 - 2\hat{\tau}(X)\left(\hat f_\tau^d(X;\hat s_n) - \hat f^d_\tau(X;s)\right) \right] \right|\\
& =\left| \E_n\left[ \left(\hat f_\tau^d(X;\hat s_n) - \hat f^d_\tau(X;s)\right)\left(\hat f_\tau^d(X;\hat s_n) + \hat f^d_\tau(X;s) - 2\hat \tau(X_i) \right) \right]\right| \\
& \leq \E_n\left[ \left|\left(\hat f_\tau^d(X;\hat s_n) - \hat f^d_\tau(X;s)\right)\right| \left| \left(\hat f_\tau^d(X;\hat s_n) + \hat f^d_\tau(X;s) - 2\hat \tau(X_i) \right) \right|\right] \\
& = \E_n\left[ \left|\left(\E(\hat \tau(X) | X < \hat s_n) - \E(\hat \tau(X) | \ X < s)\right)\right| \left| \left(\hat f_\tau^d(X;\hat s_n) +\hat  f^d_\tau(X;s) - 2\hat \tau(X) \right) \right|\right] \\
&= \E_n\left[\left| \hat s_n - s \right| \times \max_{c \in [\hat s_n, s]} \E(\hat \tau(X) \mid X < c) \left|\hat f_\tau^d(X;\hat s_n) + \hat f^d_\tau(X;s) - 2\hat \tau(X)\right|\right]\\
&=\left|\hat s_n - s \right| \E_n\left[\max_{c \in [\hat s_n, s]} \E(\hat \tau(X) \mid X < c) \left| \left(\hat f_\tau^d(X;\hat s_n) - \hat \tau(X) \right)+ \left(\hat f^d_\tau(X;s) - \hat \tau(X) \right) \right|\right] \\
& \leq \left| \hat s_n - s \right| \times \max_{c \in [\hat s_n, s]}\E(\hat \tau(X) \mid X < c) \times 
\left(\E_n\left|f_\tau^d(X;\hat s_n) - \hat \tau(X) \right|+ \E_n\left| \hat f^d_\tau(X;s) - \hat \tau(X) \right| \right) \\
&\leq 2 \left| \hat s_n - s \right| \times \max_{c \in [\hat s_n, s]}\E(\hat \tau(X) \mid X < c) \times \E_n\left[ \left|\hat f_\tau^d(X;\hat s_n) - \hat \tau(X) \right| \right] \\
&= C_\tau \times \left \lvert \hat s_n - s \right \rvert
\end{align*} 

Then: 
$$\left \lvert\E_n\left[\ell(\hat \tau(X), X; \hat s_n) \right]  - \E\left[\ell(\hat \tau(X); X, s) \right] \right \rvert  \leq  C_\tau \times \left \lvert \hat s_n - s \right \rvert + \bigg \lvert \E_n\left[ \ell(\hat \tau(X); X, s) \right] - \E\left[\ell(\tau(X); X, s) \right] \bigg \rvert,$$
which concludes the proof.
\end{proof} 
\vspace{2mm} 

An immediate implication of Lemma \ref{lem:bounded_loss} is that the within-sample loss at the sample-optimal split $\hat s_n$ will converge in probability to the population-level loss at the population-optimal split point $s$.

\begin{lemma}\label{lem:bounded_prob_loss}
Let $X^{(a)}$ be a relevant covariate that has been omitted, whereas $X^{(b)}$ is an \textit{irrelevant} covariate, erroneously chosen. We can upper bound the probability of the within-sample loss for an irrelevant covariate $X^{(b)}$ being less than a relevant covariate $X^{(a)}$:
\begin{align*} 
\Pr&\left[\E_n\left\{\ell (\hat \tau(X); X^{(b)}, \hat s_{n}^{(b)})\right\} < \E_n\left\{\ell(\hat \tau(X); X^{(a)}, \hat s^{(a)}_n)\right\}\right] \\
&\leq \frac{2}{\Delta_{ba}} \E \left[ \left|\E_n\left\{\ell(\hat \tau(X); X^{(a)}, \hat s^{(a)}_n)\right\} - \E\left\{\ell(\tau(X); X^{(a)}, s^{(a)})\right\} \right| \right],
\end{align*} 
where $\Delta_{ba} := \E \left\{\ell(\tau(X); X^{(b)}, s^{(b)}) \right\} - \E\left\{ \ell(\tau(X); X^{(a)}, s^{(a)})\right\} $. 
\end{lemma} 
Intuitively, if the \textit{population-level} losses between an irrelevant and relevant covariate is relatively small (i.e., $\Delta_{ba}$ is small), then it will take longer for the upper bound to approach zero.

\begin{proof}
We have that
\begin{align*} 
\Pr&\left[\E_n\left\{\ell (\hat \tau(X); X^{(b)}, \hat s_{n}^{(b)})\right\} < \E_n\left\{\ell(\hat \tau(X); X^{(a)}, \hat s^{(a)}_n)\right\}\right]\\
&= \Pr\left[\E_n\left\{\ell(\hat \tau(X); X^{(b)}, \hat s^{(b)}_n)\right\}- \E_n\left\{\ell(\hat \tau(X); X^{(a)}, \hat s^{(a)}_n)\right\}< 0\right] \\
&= \Pr \Big[\E_n\left\{\ell(\hat \tau(X); X^{(b)}, \hat s^{(b)}_n)\right\} - \E\left\{\ell(\tau(X); X^{(b)}, s^{(b)})\right\} + \\
&~~~~~~~~~\E\left\{\ell(\tau(X); X^{(b)}, s^{(b)})\right\} - \E\left\{\ell(\tau(X); X^{(a)}, s^{(a)})\right\} \\
&~~~~~~~~~+ \E\left\{\ell( \tau(X); X^{(a)}, s^{(a)})\right\} - \E_n\left\{\ell( \hat \tau(X); X^{(a)}, \hat s^{(a)}_n)\right\} < 0 \Big] 
\intertext{By construction, $ \E\left\{\ell(\tau(X); X^{(b)}, s^{(b)})\right\}>\E\left\{\ell(\tau(X); X^{(a)}, s^{(a)})\right\}$, as $X^{(a)}$ is a relevant covariate. Let $\E\left\{\ell(\tau(X); X^{(b)}, s^{(b)})\right\}-\E\left\{\ell(\tau(X); X^{(a)},  s^{(a)})\right\} = \Delta_{ba} > 0$:}
&= \Pr \left(\E_n \left\{ \ell(\hat \tau(X); X^{(b)}, \hat s^{(b)}_n)\right\} - \E\left\{\ell(\tau(X); X^{(b)}, s^{(b)})\right\} - \right.\\
&~~~~~~~~~~~~~~~ \left.\left[ \E_n\left\{\ell(\hat \tau(X); X^{(a)}, \hat s^{(a)}_n)\right\} -\E\left\{\ell(\tau(X); X^{(a)}, s^{(a)})\right\} \right] + \Delta_{ba}< 0 \right) \\
&= \Pr\left( \left[\E_n \left\{\ell(\hat \tau(X); X^{(a)}, \hat s^{(a)}_n)\right\} - \E \left\{\ell(\tau(X); X^{(a)}, s^{(a)})\right\} \right] \right.\\
&~~~~~~~~~~~~~~~ \left.- \left[ \E_n\left\{\ell(\hat \tau(X); X^{(b)}, \hat s^{(b)}_n)\right\} - \E\left\{\ell( \tau(X); X^{(b)}, s^{(b)})\right\}\right] > \Delta_{ba}\right)\\
&\leq \Pr \left( \left|\left[ \E_n\left\{\ell(\hat \tau(X); X^{(a)}, \hat s^{(a)}_n)\right\} - \E\left\{\ell( \tau(X); X^{(a)}, s^{(a)})\right\} \right]\right.\right.\\ 
&~~~~~~~~~~~~~~~- \left.\left.\left[\E_n\left\{\ell(\hat \tau(X); X^{(b)}, \hat s^{(b)}_n)\right\} - \E\left\{\ell( \tau(X); X^{(b)}, s^{(b)})\right\} \right] \right| > \Delta_{ba} \right)
\intertext{Then, applying Markov's Inequality:}
&\leq \frac{1}{\Delta_{ba}} \E \left( \left|\left[ \E_n\left\{\ell(\hat \tau(X); X^{(a)}, \hat s^{(a)}_n)\right\} - \E\left\{\ell( \tau(X); X^{(a)}, s^{(a)})\right\} \right]\right . \right . \\
&~~~~~~~~~~~~~~~ \left . \left . - \left[ \E_n \left\{\ell(\hat \tau(X); X^{(b)}, s^{(b)}_n)\right\} - \E\left\{\ell(\tau(X); X^{(b)}, s^{(b)})\right\}\right] \right| \right)\\
&\leq \frac{1}{\Delta_{ba}} \left\{ \E \left[ \left|\E_n\left\{\ell(\hat \tau(X); X^{(a)}, \hat s^{(a)}_n)\right\} - \E\left\{\ell(\tau(X); X^{(a)}, s^{(a)})\right\} \right| \right]\right . \\
&~~~~~~~~~~~~~~~ \left . + \E \left[ \left| \E_n\left\{\ell(\hat \tau(X); X^{(b)}, \hat s^{(b)}_n)\right\} - \E\left\{\ell(\tau(X); X^{(b)}, s^{(b)})\right\} \right| \right] \right\} 
\intertext{Without loss of generality, assume $\E \left[ \left|\E_n\left\{\ell(\hat \tau(X); X^{(a)}, \hat s^{(a)}_n)\right\} - \E\left\{\ell(\tau(X); X^{(a)}, s^{(a)})\right\} \right| \right]$ is greater than or equal to $\E \left[ \left| \E_n\left\{\ell(\hat \tau(X); X^{(b)}, \hat s^{(b)}_n)\right\} - \E\left\{\ell(\tau(X); X^{(b)}, s^{(b)})\right\} \right| \right]$. Then: }
&\leq \frac{2}{\Delta_{ba}} \E \left[ \left|\E_n\left\{\ell(\hat \tau(X); X^{(a)}, \hat s^{(a)}_n)\right\} - \E\left\{\ell(\hat \tau(X); X^{(a)}, s^{(a)})\right\} \right| \right]
 \end{align*} 
\end{proof}

We can now prove Theorem \ref{thm:subgroup_consistency}. Let $s^{(k)}$ be the population-level optimal split for covariate $k$ and $\hat{s}^{(k)}_n$ be the estimated optimal split for covariate $k$ using a sample size of $n$.

\paragraph{Simple Case: Single rule.} We consider the simple setting in which there is a single rule. Assume $X^{(a)}$ is the covariate that the estimated subgroup chooses to construct a rule with, while $X^{(b)}$ is the covariate that the true subgroup uses to construct a rule. We can bound the difference between the estimated and optimal subgroup: 
\begin{align} 
\E& \left[ \left \lvert \hat \cG_1(X) - \cG_1(X) \right \rvert \right ] \nonumber \\
&=\E \left[ \left \lvert \1\{ X^{(a)} < \hat s_n^{(a)}\} - \1\{ X^{(b)} < s^{(b)}\} \right \rvert \right ] \nonumber \\
&\leq \underbrace{\E \left[ \left \lvert  \1\{ X^{(a)} < \hat s_n^{(a)}\} -\1\{ X^{(b)} < \hat s_n^{(b)}\} \right \rvert \right]}_{(1)} + \underbrace{\E \left[ \left \lvert \1\{ X^{(b)} < \hat s_n^{(b)}\} - \1\{ X^{(b)} < s^{(b)}\}\right \rvert \right]}_{(2)}
\label{eqn:base_case}
\end{align} 

First, consider Equation \eqref{eqn:base_case}-(1): 
\begin{align*} 
\E&\left[ \left \lvert  \1\{ X^{(a)} < \hat s_n^{(a)}\} -\1\{ X^{(b)} < \hat s_n^{(b)}\}  \right \rvert \right]\\
&= \E \left[ \left \lvert  \left. \1\{ X^{(a)} < \hat s_n^{(a)}\} -\1\{ X^{(b)} < \hat s_n^{(b)}\} \right \rvert   ~\right |~ X^{(a)} \neq X^{(b)}\right] \Pr(X^{(a)} \neq X^{(b)}) \\
&= \E \left[ \left \lvert  \left. \1\{ X^{(a)} < \hat s_n^{(a)}\} -\1\{ X^{(b)} < \hat s_n^{(b)}\} \right \rvert  ~\right |~ X^{(a)} \neq X^{(b)} \right]\\ 
&\qquad \times \Pr\left[\E_n\left\{\ell (\hat \tau(X); X^{(a)}, \hat s_{n}^{(a)})\right\} < \E_n\left\{ \ell(\hat\tau(X); X^{(b)}, \hat s_n^{(b)})\right\}\right]\\
&\leq \Pr\left[\E_n\left\{\ell (\hat \tau(X); X^{(a)}, \hat s_{n}^{(a)})\right\} < \E_n\left\{\ell(\hat \tau(X); X^{(b)}, \hat s_n^{(b)})\right\}\right],
\end{align*} 
where the first equality follows from Law of Total Expectation, and the second equality follows from noting that the covariate chosen for the first split will differ from the population-level split covariate if the within-sample loss for covariate $X^{(a)}$ is lower than the within-sample loss for covariate $X^{(b)}$. The final inequality follows from the fact that $\E \left[ \left \lvert  \left. \1\{ X^{(a)} < \hat s_n^{(a)}\} -\1\{ X^{(b)} < \hat s_n^{(b)}\}  ~\right |~ X^{(a)} \neq X^{(b)} \right \rvert \right]$ must be bounded between 0 and 1. In settings when $X^{(a)}$ and $X^{(b)}$ are very correlated, then it could be the case that in fact, $\E \left[ \left \lvert  \left. \1\{ X^{(a)} < \hat s_n^{(a)}\} -\1\{ X^{(b)} < \hat s_n^{(b)}\}  ~\right |~ X^{(a)} \neq X^{(b)} \right \rvert \right]$ is quite close to zero as the set of units for whom $\1\{X^{(a)} < \hat s_n^{(a)}\}$ and $\1\{X^{(b)} < \hat s_n^{(b)}\}$ may be quite similar. 

Equation \eqref{eqn:base_case}-(2) can be bounded by applying the results from Proposition \ref{prop:converge_one}: 
\begin{align*} 
\E\left[ \left \lvert \1\{ X^{(b)} < \hat s_n^{(b)}\} - \1\{ X^{(b)} < s^{(b)}\}\right \rvert \right] &\leq 2 K \left \lvert \hat s_n^{(b)} - s^{(b)} \right \rvert,
\end{align*} 
for some constant $K > 0$.

\paragraph{Extending to $r$ total rules:} For $k = 1, \ldots, r$, assume $X^{(a_k)}$ is the covariate that the estimated subgroup chooses to construct a rule with at the $k^{th}$ step, while $X^{(b_k)}$ is the covariate that the true subgroup uses to construct a rule at the $k^{th}$ step.
\begin{align*} 
\E&\left[ \left\lvert \prod_{k=1}^r \1\{X^{(a_k)} < \hat s_n^{(a_k)} \} - \prod_{k=1}^r \1\{X^{(b_k)} < \hat s_n^{(b_k)} \} \right \rvert \right] \\
\leq& \Pr(X^{(a_1)} \neq X^{(b_1)}) + \Pr(X^{(a_1)} = X^{(b_1)}) \cdot \Pr(X^{(a_2)} \neq X^{(b_2)} \mid X^{(a_1)} = X^{(b_1)}) + \ldots\\
&+ \Pr(X^{(a_1)} = X^{(b_1)}, \ldots, \Pr(X^{(a_{r-1})} = X^{(b_{r-1})})) \\
&~~~\times \Pr(X^{(a_{r})} \neq X^{(b_{r})} \mid X^{(a_1)} = X^{(b_1)}, \ldots, X^{(a_{r-1})} = X^{(b_{r-1})}) \\
\leq& \sum_{k=1}^r  \Pr\left[\E_n\left\{\ell (\hat \tau(X); X^{(a_k)}, \hat s^{(a_k)}_n)\right\} < \E_n\left\{\ell(\hat \tau(X); X^{(b_k)}, \hat s^{(b_k)}_n)\right\}\right] + K \left \lvert \hat s^{(b_k)}_n - s^{(b_k)} \right \rvert
\intertext{Let $k^* := \argmax_{k'} \left \lvert \hat s_n^{(k')} - s^{(k')} \right \rvert + \E_n \left [ \ell(\hat \tau(X); X^{(k')}, \hat s^{(k')}) \right] - \E\left[\ell(\hat \tau(X); X^{(k')}, s^{(k')} \right] $, and $\Delta = \min \Delta_{ba}$. Applying Lemma \ref{lem:bounded_loss} and \ref{lem:bounded_prob_loss}:}
\leq& \frac{2r}{\Delta} \E \left\{ \left \lvert\E_n\left[\ell(\hat \tau(X), X^{(k^*)}; \hat s_{n}^{(k^*)}) \right]  - \E\left[\ell(\tau(X); X^{(k^*)}, s^{{(k^*)}}) \right] \right \rvert \right\} + Kr \left \lvert \hat s^{(k^*)}_{n} - s^{(k^*)} \right \rvert\\
\leq& \frac{2r C_\tau}{\Delta} \left \lvert \hat s^{(k^*)}_{n} - s^{(k^*)} \right \rvert + \frac{2r}{\Delta} \left(\E_n\left[\ell( \hat \tau(X); X^{(k^*)}, s^{(k^*)})\right] - \E\left[ \ell(\tau(X); X^{(k^*)}, s^{(k^*)})\right] \right) + Kr \left \lvert \hat s^{(k^*)}_{n} - s^{(k^*)} \right \rvert \\
=& \underbrace{\frac{( 2C_\tau + K \Delta)r}{\Delta} \left \lvert \hat s^{(k^*)}_{n} - s^{(k^*)} \right \rvert}_{(1)}+ \underbrace{\frac{2r}{\Delta} \big\{ \E_n\left[ \ell(\hat \tau(X); X^{(k^*)}, s^{(k^*)})\right] - \E\left[ \ell(\tau(X); X^{(k^*)}, s^{(k^*)})\right] \big\} }_{(2)}.
\end{align*} 
Term (1) represents the gap between an estimated split and true optimal split. 
 Term (2) is the gap between the empirical loss and the population-level loss. From Proposition \ref{prop:converge_one}, term (1) converges to zero. Lemma \ref{lem:hatsn_consistency}, \eqref{eqn:s_cip_bound} establishes that term (2) converges to zero as well, which concludes the proof.

\subsection{Proof of Theorem \ref{thm:cip_group_dim_finite}}
\begin{proof} 
To prove Theorem \ref{thm:cip_group_dim_finite}, we will introduce a lemma that will be helpful in the following derivation. We define $n_g$ as the total number of units in a subgroup $g$, and $n_g(1) = \sum_{i=1}^n \hat \cG_g(X_i) Z_i$ -- i.e., $n_g(1)$ is equal to the number of units in subgroup $g$ who are treated. We similarly define $n_g(0)$ as the number of units in subgroup $g$ who are not treated. Throughout, we will denote $\E_\cD(A_i) = \E (A_i \mid \cD)$ as the expectation, given the sample data $\cD$.
\begin{assumption}[Assignment Symmetry \citepappendix{miratrix2013adjusting}] \label{assum:assignment_sym}
 \begin{enumerate} 
 \item Equiprobable treatment assignment patterns. For each subgroup $g \in \{1,\ldots, G\}$, all $ n_g \choose n_g(1)$ combination of ways to treat $n_g(1)$ units are equiprobable, given the subgroup size $n_g$. 
 \item Independent treatment assignment patterns: for all subgroups $g, g'$,  the treatment assignment process in group $g$ is independent of the treatment assignment process in group $g'$, given $n_g(1)$ and $n_{g'}(1)$.
 \end{enumerate} 
\end{assumption} 
We are considering designs that satisfy assignment symmetry. For example, experimental designs such as complete randomization, Bernoulli randomization, or blocking with independent treatment assignment across blocks, would satisfy assignment symmetry. However, cluster randomization would not satisfy assignment symmetry. Furthermore, we do not account for settings in which some units within a subgroup are more likely to receive treatment. 
\begin{lemma}[Implications of Assignment Symmetry] \label{lem:assignment_sym}
Under Assumption \eqref{assum:assignment_sym}, we have assumed that conditional on the number of units within a subgroup $g \in \{1,\ldots, G\}$, the probability of receiving treatment for units in a subgroup is equiprobable and independent: 
\begin{equation} 
\E_\cD\left\{ Z_i \mid n_g(1) \right\} = \frac{n_g(1)}{n_g} \equiv \frac{1}{n_g} \sum_{i=1}^n \hat \cG_g(X_i) Z_i, \text{ and }
\E_\cD\left\{ Z_i Z_j \mid n_g(1) = a \right\} = \frac{a(a-1)}{n_g (n_g - 1)}
\label{eqn:as}
\end{equation} 
We can then derive the distribution of $Z_i/n_g(1)$: 
\begin{itemize} 
\item $\E_\cD\left[ \frac{Z_i}{n_g(1)}\right] = \frac{1}{n_g}$
\item $\E_\cD\left[ \frac{Z_i^2}{n_g(1)^2}\right] = \frac{1}{n_g} \E_\cD\left[ \frac{1}{n_g(1)}\right]$
\item $\E_\cD\left[ \frac{Z_i Z_j}{n_g(1)^2}\right] = \frac{1}{n_g (n_g - 1)} \E_\cD\left[ \frac{n_g(1) - 1}{n_g(1)}\right] = \frac{1}{n_g (n_g - 1)} \left( 1 - \E_\cD \left[ \frac{1}{n_g(1)} \right] \right)$
\item $\E_\cD \left[\frac{Z_i (1- Z_j)}{n_g(1) n_g(0)} \right] = \frac{1}{n_g \cdot (n_g-1)} \cdot \E_\cD \left[ \frac{n_g - n_g(1)}{n_g(0)}\right] = \frac{1}{n_g \cdot (n_g-1)} $
\end{itemize} 
\end{lemma} 
The results of Lemma \ref{lem:assignment_sym} follow directly from first noting that conditional on the number of units in each subgroup $g$, randomization within a subgroup can be viewed as complete randomization, and then applying law of iterated expectations. \\

\noindent \textbf{Part 1. Unbiasedness.} To begin, we will show that $\hat \tau^{(g)}$ is unbiased for the within-sample subgroup ATE:  
\begin{align} 
\E_\cD & \left( \hat \tau^{(g)} \right)= \E \left\{\left. \frac{1}{\sum_{i=1}^n Z_i \hat \cG_g(X_i)} \sum_{i=1}^n Z_i Y_i \hat \cG_g(X_i) - \frac{1}{\sum_{i=1}^n (1-Z_i) \hat \cG_g(X_i)} \sum_{i=1}^n (1-Z_i) Y_i \hat \cG_g(X_i) ~\right | \mathcal{D} \right\} \nonumber \\
&=  \sum_{i=1}^n  \E \left\{ \left. \frac{Z_i}{\sum_{i=1}^n Z_i \hat \cG_g(X_i)} ~\right | \mathcal{D} \right\} Y_i(1) \hat \cG_g(X_i) - \E  \left\{  \left. \frac{1-Z_i}{\sum_{i=1}^n (1-Z_i) \hat \cG_g(X_i)} ~\right | \mathcal{D}~\right\} Y_i(0) \hat \cG_g(X_i) \nonumber 
\intertext{Under Assumption \eqref{assum:assignment_sym} (Assignment Symmetry), we can apply the results from Lemma \ref{lem:assignment_sym}:}
&= \frac{1}{n_g} \sum_{i=1}^n \hat \cG_g(X_i) \left\{Y_i(1) - Y_i(0) \right\} \label{eqn:hat_tau_g_exp}
\end{align} 

\noindent \textbf{Part 2. Variance derivation.} To derive the variance of the subgroup difference-in-means estimator, we apply results from \citetappendix{miratrix2013adjusting}, who derives the variance of a post-stratification estimator. We can extend the results by treating each subgroup as a strata, where the subgroup difference-in-means estimator is analogous to the difference-in-means estimator across a specific strata. We provide the full derivation for completeness. 

\begin{align}
\var (\hat \tau^{(g)}) &= \var \left\{\frac{1}{\sum_{i=1}^n \hat \cG_g(X_i) Z_i} \sum_{i=1}^n \hat \cG_g(X_i) Z_i Y_i - \frac{1}{\sum_{i=1}^n \hat \cG_g(X_i) (1-Z_i)} \sum_{i=1}^n \hat \cG_g(X_i) (1-Z_i) Y_i  \right\} \nonumber \\
&= \var \left\{\frac{1}{n_g(1)} \sum_{i: \hat \cG_g(X_i) = 1} Z_i Y_i(1) - \frac{1}{n_g(0)} \sum_{i:\hat \cG_g(X_i)=1} (1-Z_i) Y_i(0)  \right\} \nonumber \\
\begin{split}
&= \underbrace{\var \left\{\frac{1}{n_g(1)} \sum_{i: \hat \cG_g(X_i) = 1} Z_i Y_i(1) \right\}}_{(1)} + \underbrace{\var \left\{\frac{1}{n_g(0)} \sum_{i:\hat \cG_g(X_i)=1} (1-Z_i) Y_i(0)  \right\}}_{(2)}  \\
&~~~~~~~~~~~~~~~~~- 2 \underbrace{\cov \left\{\frac{1}{n_g(1)} \sum_{i: \hat \cG_g(X_i) = 1} Z_i Y_i(1), \frac{1}{n_g(0)} \sum_{i:\hat \cG_g(X_i)=1} (1-Z_i) Y_i(0)  \right\}}_{(3)}
\end{split}
\label{eqn:var_expr}
\end{align} 

\noindent Beginning with term (1): 
\begin{align} 
\var_\cD&\left\{\frac{1}{n_g(1)} \sum_{i: \hat \cG_g(X_i) = 1} Z_i Y_i(1) \right\} \\
&= \sum_{i,j:\hat \cG_g(X) = 1} \cov_\cD \left(\frac{Z_i}{n_g(1)} Y_i(1), \frac{Z_j}{n_g(1)}Y_j(1) \right) \nonumber \\
&= \sum_{i:\hat \cG_g(X_i) = 1} \left[  \var_\cD\left\{ \frac{Z_i}{n_g(1)}Y_i(1)\right\} + 
 \sum_{i\neq j} \cov_\cD\left\{ \frac{Z_i}{n_g(1)}Y_i(1), \frac{Z_j}{n_g(1)}Y_j(1)\right\} \right] \nonumber \\
&= \sum_{i:\hat \cG_g(X_i) = 1} \left[ \frac{1}{n_g} \E_\cD \left\{ \frac{1}{n_g(1)} \right\}Y_i(1)^2 - \frac{1}{n_g^2} Y_i(1)^2\right] +  \nonumber \\
 &\sum_{i\neq j} \frac{1}{n_g} \cdot \frac{1}{n_g - 1} \cdot \left( 1 - \E_\cD \left\{ \frac{1}{n_g(1)}\right\}\right) Y_i(1) Y_j(1) -\frac{1}{n_g^2} Y_i(1) Y_j(1)  \nonumber \\
 &= \frac{1}{n_g} \sum_{i:\hat \cG_g(X_i) = 1} \left(\E_\cD \left\{\frac{1}{n_g(1)}\right\} - \frac{1}{n_g} \right)Y_i(1)^2 + \nonumber \\
 &~~~~\frac{1}{n_g} \sum_{i \neq j: \hat \cG_g(X) = 1} \left(\frac{1}{n_g - 1} \left( 1 - \E_\cD \left\{\frac{1}{n_g(1)}\right\} \right) - \frac{1}{n_g} \right) Y_i(1) Y_j(1) \nonumber 
 \intertext{Let $\tilde \beta_g(z) := \E_\cD \{1/n_g(1)\}$:}
 &= \frac{1}{n_g} \sum_{i:\hat \cG_g(X_i) = 1} \left[ \left\{ \tilde \beta_g(1) - \frac{1}{n_g} \right\} Y_i(1)^2 - \frac{1}{n_g-1} \sum_{i \neq j} \left\{\beta_g(1) - \frac{1}{n_g} \right\} Y_i(1) Y_j(1) \right]  \nonumber \\
 &= \left\{\tilde \beta_g(1) - \frac{1}{n_g} \right\} \cdot \frac{1}{n_g} \sum_{i: \hat \cG_g(X_i) = 1} \left[ Y_i(1)^2 - \sum_{i \neq j} Y_i(1) Y_j(1) \right] \nonumber \\
 &= \left\{\tilde \beta_g(1) - \frac{1}{n_g} \right\}  \var_g \left\{ Y_i(1)\mid \cD \right\},
 \label{eqn:var_term1}
\end{align} 
where $ \var_g \left\{ Y_i(1)\mid \cD \right\} := \frac{1}{n_g} \sum_{i: \hat \cG_g(X_i) = 1} \left[ Y_i(1)^2 - \sum_{i \neq j} Y_i(1) Y_j(1) \right]$. We can similarly show term (2) can be rewritten as: 
\begin{equation} 
\var_\cD \left\{\frac{1}{n_g(0)} \sum_{i:\hat \cG_g(X_i)=1} (1-Z_i) Y_i(0)  \right\} = \left\{ \beta_g(0) - \frac{1}{n_g} \right\} \var_g\left\{Y_i(0) \mid \cD \right\}.
\label{eqn:var_term2}
\end{equation}

\noindent Finally, to derive an expression for term (3), we begin by noting the following: 
\begin{align*} 
\E_\cD &\left[\left\{ \frac{1}{n_g(1)} \sum_{i:\hat \cG_g(X_i) = 1} Z_i Y_i(1) \right\} \left\{ \frac{1}{n_g(0)} \sum_{i:\hat \cG_g(X_i) = 1} (1-Z_i) Y_i(0)\right\} \right]\\
&= \E_\cD \left\{ \frac{1}{n_g(1) \cdot n_g(0)} \sum_{i:\hat \cG_g(X_i) = 1} Z_i (1-Z_i) Y_i(1) Y_i(0) + \sum_{i \neq j} Z_i (1-Z_j) Y_i(1) Y_j(0) \right\}
\intertext{Because $Z_i \cdot (1-Z_i) = 0$:}
&=  \E_\cD \left\{ \frac{1}{n_g(1) \cdot n_g(0)} \sum_{i:\hat \cG_g(X_i) = 1}\sum_{i \neq j} Z_i (1-Z_j) Y_i(1) Y_j(0) \right\}\\
&= \frac{1}{n_g \cdot (n_g-1)}\sum_{i:\hat \cG_g(X_i) = 1} \sum_{i \neq j}  Y_i(1) Y_j(0)
\end{align*} 

\noindent Then, we can re-write term (3) as: 
\begin{align} 
\cov_\cD& \left\{ \frac{1}{n_g(1)} \sum_{i:\hat \cG_g(X_i) = 1} Z_i Y_i(1), \frac{1}{n_g(0)} \sum_{i:\hat \cG_g(X_i) = 1} (1-Z_i) Y_i(0) \right\}\nonumber \\
&= \frac{1}{n_g} \frac{1}{n_g-1} \sum_{i: \hat \cG_g(X_i) = 1} \sum_{i \neq j} Y_i(1) Y_j(0) -\left\{  \frac{1}{n_g} \sum_{i:\hat \cG_g(X_i) = 1} Y_i(1)\right\} \cdot \left\{  \frac{1}{n_g} \sum_{i:\hat \cG_g(X_i) = 1} Y_i(0)\right\}\nonumber \\
&= \frac{1}{n_g} \frac{1}{n_g-1} \sum_{i: \hat \cG_g(X_i) = 1} \sum_{i \neq j} Y_i(1) Y_j(0) - \left[ \frac{1}{n_g^2} \sum_{i:\hat \cG_g(X_i) = 1} \left\{ Y_i(1) Y_i(0) + \sum_{i \neq j} Y_i(1) Y_j(0) \right\}\right] \nonumber \\
&= \frac{1}{n_g} \left(\sum_{i:\hat \cG_g(X_i)} \frac{1}{n_g} Y_i(1) Y_i(0) + \left(\frac{1}{n_g - 1} - \frac{1}{n_g}\right) \sum_{i \neq j} Y_i(1) Y_j(0)  \right) \nonumber \\
&= \frac{1}{n_g} \left(\sum_{i:\hat \cG_g(X_i)} \frac{1}{n_g} Y_i(1) Y_i(0) - \frac{1}{n_g (n_g - 1)} \sum_{i \neq j} Y_i(1) Y_j(0)  \right) \nonumber \\
&= -\frac{1}{n_g} \cov_g(Y_i(1), Y_i(0) \mid \cD)
\label{eqn:var_term3}
\end{align} 

Substituting the expressions in Equation \eqref{eqn:var_term1}-\eqref{eqn:var_term3} into Equation \eqref{eqn:var_expr}: 
\begin{align*} 
\var(\hat \tau^{(g)} \mid \cD) =& \left\{\tilde \beta_g(1) - \frac{1}{n_g} \right\}\var_g \left\{ Y_i(1)  \mid \cD\right\}+ \left\{\tilde  \beta_g(0) - \frac{1}{n_g} \right\}\var_g \left\{ Y_i(0)  \mid \cD\right\} \\
&\quad + \frac{2}{n_g} \cov_g\left\{ Y_i(1), Y_i(0) \mid \cD \right\}.
\end{align*} 

\end{proof} 
\subsection{Proof of Theorem \ref{thm:cip_group_dim}}
\begin{proof} 
We will prove consistency, and then derive the asymptotic variance. To show consistency, we take the expectation of $\hat \tau^{(g)}$. 
\begin{align*} 
\E(\hat \tau^{(g)}) &= \E \left\{ \E \left(\hat \tau^{(g)} \mid \mathcal{D} \right)\right\} \\
&= \E \left[\frac{1}{n_g} \sum_{i=1}^n \hat \cG_g(X_i) \left\{Y_i(1) - Y_i(0) \right\} \right]\\
&= \E \left[\frac{1}{n_g} \sum_{i=1}^n \cG_g(X_i) \cdot \left\{  Y_i(1) - Y_i(0) \right\} \right]  + \E \left[\frac{1}{n_g} \sum_{i=1}^n \left\{ \hat \cG_g(X_i) - \cG_g(X_i) \right\} \cdot \left\{  Y_i(1) - Y_i(0) \right\} \right] 
\intertext{Let $\overline{K_\tau} = \max_{i: \hat \cG_g(X_i) = 1} \left\{Y_i(1) -Y_i(0) \right\}$:}
&\leq \tau^{(g)} + \frac{1}{n_g} \sum_{i=1}^n \overline{K_\tau} \left \lvert \hat \cG_g(X_i) - \cG_g(X_i) \right \rvert 
\end{align*} 
From Theorem \ref{thm:subgroup_consistency}, $\left \lvert \hat \cG_g(X_i) - \cG_g(X_i) \right \rvert \cip 0$. As such, $\hat \tau^{(g)} \cip \tau^{(g)}$.

\noindent To derive the variance, we apply law of total variance: 
\begin{align*} 
\var(\hat \tau^{(g)}) = \E\left[ \var(\hat \tau^{(g)} \mid \cD)\right] + \var \left[ \E(\hat \tau^{(g)} \mid \cD) \right]
\end{align*} 
We then examine both components of the variance expression. First, 
\begin{align*} 
\E&\left[ \var(\hat \tau^{(g)} \mid \cD)\right] \\
&= \left\{\tilde \beta_g(1) - \frac{1}{n_g} \right\}\E \left[ \var_g \left\{ Y_i(1)  \mid \cD\right\} \right] + \left\{\tilde  \beta_g(0) - \frac{1}{n_g} \right\}\E \left[ \var_g \left\{ Y_i(0)  \mid \cD\right\}\right] + \frac{2}{n_g} \E \left[ \cov_g\left\{ Y_i(1), Y_i(0) \mid \cD \right\}\right] \\ 
&= \left\{\tilde \beta_g(1) - \frac{1}{n_g} \right\} \var_g \left\{ Y_i(1)  \right\} + \left\{\tilde  \beta_g(0) - \frac{1}{n_g} \right\}\var_g \left\{ Y_i(0) \right\} + \frac{2}{n_g} \cov_g\left\{ Y_i(1), Y_i(0)  \right\} 
\end{align*} 
where $\var_g(A_i) := \E \left[ \var_g(A_i \mid \cD)\right]$. \\

\noindent Second, we re-write the variance of $\E(\hat \tau^{(g)} \mid \cD)$ as: 
\begin{align*} 
 \var&\left[ \E(\hat \tau^{(g)} \mid \cD) \right] \\
 =& \var \left( \frac{1}{n_g} \sum_{i=1}^n \hat \cG_g(X_i) \{Y_i(1) - Y_i(0) \}  \right) \\
 =& \var \left( \frac{1}{n_g} \sum_{i=1}^n \cG_g(X_i) \{Y_i(1) - Y_i(0) \}  + \frac{1}{n_g} \sum_{i=1}^n \{ \hat \cG_g(X_i) - \cG_g(X_i)\} \cdot \{Y_i(1) - Y_i(0)\} \right) \\
 =& \underbrace{\var \left( \frac{1}{n_g} \sum_{i=1}^n \cG_g(X_i) \{Y_i(1) - Y_i(0) \} \right)}_{(1)}  +\underbrace{\var\left( \frac{1}{n_g} \sum_{i=1}^n \{ \hat \cG_g(X_i) - \cG_g(X_i)\} \cdot \{Y_i(1) - Y_i(0)\} \right)}_{(2)}\\
 &+ 2 \underbrace{\cov \left(\frac{1}{n_g} \sum_{i=1}^n \cG_g(X_i) \{Y_i(1) - Y_i(0) \},  \frac{1}{n_g} \sum_{i=1}^n \{ \hat \cG_g(X_i) - \cG_g(X_i)\} \cdot \{Y_i(1) - Y_i(0)\} \right)}_{(3)} 
\end{align*} 
Term (1) can be re-written as: 
\begin{align*} 
\var& \left( \frac{1}{n_g} \sum_{i=1}^n \cG_g(X_i) \{Y_i(1) - Y_i(0) \} \right) \\
&= \frac{1}{n_g} \left( \var_g\{Y_i(1)\} + \var_g\{Y_i(0)\} - 2 \cov_g\{Y_i(1), Y_i(0)\} \right)
\end{align*} 
For term (2): 
\begin{align} 
\var&\left( \frac{1}{n_g} \sum_{i=1}^n \{ \hat \cG_g(X_i) - \cG_g(X_i)\} \cdot \{Y_i(1) - Y_i(0)\} \right) \nonumber \\
&= \frac{1}{n_g^2} \var \left( \sum_{i=1}^n \{ \hat \cG_g(X_i) - \cG_g(X_i)\} \cdot \tau_i \right) \nonumber \\
&\leq \frac{1}{n_g} \frac{n}{n_g} \overline{K_\tau}^2 \var(\hat \cG_g(X_i)),\label{eqn:asyvar_2}
\end{align} 
where we have assumed $n_g/n \to \pi^*_g$ where $0 < \pi^*_g <1$ for all $g \in \{1, ..., G\}$. Then, Equation \eqref{eqn:asyvar_2} converges to zero by Theorem \ref{thm:subgroup_consistency}. 
For term (3): 
\begin{align*} 
\E& \left\{ \left(\frac{1}{n_g} \sum_{i=1}^n \cG_g(X_i) \{Y_i(1) - Y_i(0) \}\right) \cdot \left( \frac{1}{n_g} \sum_{i=1}^n \{ \hat \cG_g(X_i) - \cG_g(X_i)\} \cdot \{Y_i(1) - Y_i(0)\} \right)\right\} \\
&=\frac{1}{n_g^2} \E \left[ \left( \sum_{i=1}^n \cG_g(X_i) \tau_i \right)  \left( \sum_{i=1}^n \{\hat \cG_g(X_i) - \cG_g(X_i) \}\tau_i \right)  \right] \\
&= \frac{1}{n_g^2} \left(\sum_{i=1}^n \E \left[ \cG_g(X_i) \cdot \{\hat \cG_g(X_i) - \cG_g(X_i) \} \tau_i \right] + \sum_{i \neq j} \E \left[ \cG_g(X_i) \{ \hat \cG_g(X_j) - \cG_g(X_j) \} \tau_i \tau_j \right] \right) \\
&= \frac{1}{n_g^2} \left(\sum_{i=1}^n \cG_g(X_i) \cdot \E \left[ \{\hat \cG_g(X_i) - \cG_g(X_i) \} \tau_i \right] + \sum_{i \neq j}  \cG_g(X_i) \E \left[\{ \hat \cG_g(X_j) - \cG_g(X_j) \} \tau_i \tau_j \right] \right) \\
&= \frac{1}{n_g^2} \left(\sum_{i=1}^n \cG_g(X_i) \cdot \E \left[ \{\hat \cG_g(X_i) - \cG_g(X_i) \} \tau_i \right] + \sum_{i \neq j}  \cG_g(X_i) \cdot \E \left[\{ \hat \cG_g(X_j) - \cG_g(X_j) \} \tau_i \tau_j \right] \right)\\
&\leq \frac{1}{n_g^2} \left(\sum_{i=1}^n \cG_g(X_i) \cdot \overline{K_\tau} \cdot \E \left[ \{\hat \cG_g(X_i) - \cG_g(X_i) \} \right] + \sum_{i \neq j}  \cG_g(X_i) \cdot \overline{K_\rho} \cdot \E \left[\{ \hat \cG_g(X_j) - \cG_g(X_j) \} \right] \right),
\end{align*} 
where, $\overline{K_\rho} := \max_{i \neq j} \tau_i \cdot \tau_j$. Under Theorem \ref{thm:subgroup_consistency}, this term will converge in probability to zero. We can similarly show that the lower bound of this term will similarly converge in probability to zero: 
\begin{align*} 
\E& \left\{ \left(\frac{1}{n_g} \sum_{i=1}^n \cG_g(X_i) \{Y_i(1) - Y_i(0) \}\right) \cdot \left( \frac{1}{n_g} \sum_{i=1}^n \{ \hat \cG_g(X_i) - \cG_g(X_i)\} \cdot \{Y_i(1) - Y_i(0)\} \right)\right\} \\
&\geq \frac{1}{n_g^2} \left(\sum_{i=1}^n \cG_g(X_i) \cdot \underline{K_\tau} \cdot \E \left[ \{\hat \cG_g(X_i) - \cG_g(X_i) \} \right] + \sum_{i \neq j}  \cG_g(X_i) \cdot \underline{K_\rho} \cdot \E \left[\{ \hat \cG_g(X_j) - \cG_g(X_j) \} \right] \right),
\end{align*} 
which will converge in probability to zero under Theorem \ref{thm:subgroup_consistency}. Then, by Squeeze theorem, we have shown: 
$$\E \left\{ \left(\frac{1}{n_g} \sum_{i=1}^n \cG_g(X_i) \{Y_i(1) - Y_i(0) \}\right) \cdot \left( \frac{1}{n_g} \sum_{i=1}^n \{ \hat \cG_g(X_i) - \cG_g(X_i)\} \cdot \{Y_i(1) - Y_i(0)\} \right)\right\} \cip 0.$$
Similarly, we can show: 
\begin{align*} 
\E& \left(\frac{1}{n_g} \sum_{i=1}^n \cG_g(X_i) \{Y_i(1) - Y_i(0) \}\right) \cdot \E \left( \frac{1}{n_g} \sum_{i=1}^n \{ \hat \cG_g(X_i) - \cG_g(X_i)\} \cdot \{Y_i(1) - Y_i(0)\} \right)\\
&= \tau^{(g)} \cdot \frac{1}{n_g} \sum_{i=1}^n \E\left[ \{ \hat \cG_g(X_i) - \cG_g(X_i)\}\cdot \tau_i \right].
\intertext{Then, again, using the boundedness of $\tau_i$ (i.e., $\underline{K_\tau} \leq \tau_i \leq \overline{K_\tau}$), we can apply Squeeze theorem to show:}
&\cip 0.
\end{align*} 
As such, we have shown: 
\begin{equation} 
\cov \left(\frac{1}{n_g} \sum_{i=1}^n \cG_g(X_i) \{Y_i(1) - Y_i(0) \},  \frac{1}{n_g} \sum_{i=1}^n \{ \hat \cG_g(X_i) - \cG_g(X_i)\} \cdot \{Y_i(1) - Y_i(0)\} \right)\cip 0
\label{eqn:cov_tau}
\end{equation} 

\noindent Combining the terms together, we have shown: 
\begin{align*} 
\asyvar(\hat \tau^{(g)}) &= \tilde \beta_g(1) \var_g\{Y_i(1)\} + \tilde \beta_g(0) \var_g\{Y_i(0)\}.
\end{align*} 

\noindent To derive the asymptotic limiting distribution, because \eqref{eqn:asyvar_2} and \eqref{eqn:cov_tau} both converge in probability to zero, we can directly apply Slutsky's theorem to show that 
$$\sqrt{n} \left( \hat \tau^{(g)} - \tau^{(g)} \right) \cid N(0, \asyvar(\hat \tau^{(g)}).$$

\end{proof}

\subsection{Proof of Theorem \ref{thm:cid_group_dim}}
We formally impose the following regularity assumptions.
\begin{assumption}[Regularity Conditions] \label{assum:dist_regularity}
\begin{enumerate} 
\item As the sample size $n \to \infty$, the total number of subgroups $G$ remain fixed.
\item The proportion of units in each subgroup converges to a proportion $\pi_g^* \in (0,1)$ (i.e., $n_g / n \to \pi_g^*$, where $0 < \pi_g^* < 1$ for all $g \in \{1, \ldots, G\}$ and $\sum_{g=1}^G \pi_g^* = 1$.
\item The proportion of treated units converges to $\pi^*_z$, where $\pi^*_z \in (0,1)$ (i.e., $\lim_{n \to \infty} \frac{1}{n} \sum_{i=1}^n Z_i = \pi_z^*$).
\item Lindeberg Condition: 
\begin{equation*} 
\lim_{n\to\infty} \frac{1}{n^2_z}  \frac{max_{1 \leq i \leq n} \hat \cG_g(X_i) Y_i^2(z)}{\var(\hat \tau^{(g)})} = 0.
\end{equation*} 
\item $\lim_{n \to \infty} \var_g(Y_i(z)) \leq c_1, $ and $\cov_g(Y_i(1), Y_i(0)) \leq c_2$ where $ c_1, c_2 < \infty$.
\end{enumerate} 
\end{assumption} 

We will show that as $n \to \infty$, $\left(1-\frac{n_z}{n} \right) \cdot \frac{1}{n_z} \var(\hat \cG_g(X_i)) \to 0$. 

First, note: 
 $$\var(\hat \cG_g(X_i)) = \frac{n}{n-1} \frac{1-\pi_g}{\pi_g}.$$
Then,
\begin{align*} 
\left(1-\frac{n_z}{n} \right) \cdot \frac{1}{n_z} \var(\hat \cG_g(X_ i)) &= \frac{1}{n} \frac{1-n_z/n}{n_z/n} \cdot \frac{n}{n-1} \frac{1-\pi_g}{\pi_g}.
\intertext{As $n \to \infty$, $n_z/n \to \pi_z^*$ and $\pi_g \to \pi_g^*$, both of which will be bounded away from 0 and 1 (by Assumption \eqref{assum:dist_regularity}(b)-(c)):}
\lim_{n \to \infty}\left(1-\frac{n_z}{n} \right) \cdot \frac{1}{n_z} \var(\hat \cG_g(X_i))  &= \lim_{n \to \infty} \frac{1}{n} \frac{1-p^*_z}{p^*_z} \cdot \frac{1-\pi^*_g}{\pi^*_g}=0
\end{align*} 

Then, we can apply \citetappendix{schochet2024design}, Theorem 1 to show: 
$$\frac{\hat \tau^{(g)} - \E\left(\hat \tau^{(g)} \right)}{\sqrt{\var(\hat \tau^{(g)}})} \cid N(0, 1)$$

\subsection{Proof of Corollary \ref{cor:te_test}}
Because each subgroup is a distinct partition, $\hat \tau^{(g)}$ and $\hat \tau_{g'}$ will be asymptotically independent from one another. As a result, we can straightforwardly apply the Cramer-Wold device and
$$\boldsymbol{\hat \tau} - \boldsymbol{\tau} \cid N(0, \boldsymbol{\Sigma}_G)$$
Thus: 
$$\boldsymbol{\hat \tau}^\top \boldsymbol{\Sigma}^{-1}_G\boldsymbol{\hat \tau} \cid \chi^2_G$$

\subsection{Proof of Example \ref{ex:split_stability}}
\begin{assumption}[Regularity Conditions] \label{assum:regularity}
\begin{enumerate}[label=(\alph*)]
\item Smoothness of $\tau(X)$: $\tau(X)$ is continuous, and $\tau'(X)$, $\tau''(X)$ exist and are uniformly bounded in a neighborhood around $s$, and $\tau'(s) \neq 0$. 
\item Smoothness condition on density of $X$ and $v$: $dF'_X$ exists and is uniformly bounded in a neighborhood around $s$. Furthermore, $dF_X(s) \neq 0$.
\item Moment condition: $\E(v_i) = 0$, $\E(v_i^2) = \sigma^2$.
\item Tail condition: $dF_{\tau} = o(|\tau|^{-4 + \delta})$ as $|\tau| \to \infty$ for some $\delta > 0$. 
\item Signal condition: $\var(\tau(X_i)) > 0$.
\end{enumerate} 
\end{assumption}
These are common assumptions (see e.g., \citealp{buhlmann2002analyzing}). 
The first three conditions are straightforward. The tail condition (i.e., Assumption \eqref{assum:regularity}-(d)) rules out settings in which the density of $\tau_i$ concentrates at an infinitely large value. Informally, this implies that the treatment effect cannot explode to be infinitely large, which is reasonable in many substantive contexts of interest. The final condition (i.e., Assumption \eqref{assum:regularity}-(e)) which we refer to as a signal condition, rules out settings in which there is no amount of variation in $\tau_i$ that can be explained by the covariates $X_i$ (i.e., the underlying data generating process is pure noise). In other words, the covariates we have collected must have \textit{some} degree of explanatory power. This rules out the setting considered in \citetappendix{cattaneo2022pointwise}.

\begin{lemma}[Asymptotic Variance of Splits] \label{lem:var_split} 
Under Assumption \eqref{assum:regularity} and squared loss, the asymptotic variance of the estimated splits is: 
    $$\asyvar(\hat s_n) = \left(dF_X(s) \sigma^2 \right)^2 \cdot \frac{2^{-2/3}}{6\pi i} \left(\frac{1}{2} \frac{V}{dF_X(s) \sigma^2} \right)^{-4/3} \int_{- \infty}^{\infty} \frac{t}{Ai(it)^2} dt,$$
    where $V = - dF_X(s) f'(s) \neq 0$. 
\end{lemma}
\begin{proof} 
Under Assumption \eqref{assum:regularity} and squared loss, we apply \citetappendix{buhlmann2002analyzing}, Theorem 3.1 to show that as $n \to \infty$, 
$$n^{1/3} (\hat s_n - s) \cid W_{\sigma^2, s} := \argmax_t \left[Q(t)  \cdot \text{sign} \left(\E(\tau_i \mid X_i < s) - \E(\tau_i \mid X_i \geq s) \right) \right],$$
and $Q(t)$ is a scaled, two-sided Brownian motion, originating from zero, with a quadratic drift: 
$$Q(t) = dF_X(s) \sigma^2 B(t) - \frac{1}{2} Vt^2,$$
where $B(t)$ is a two-sided Brownian motion, originating from zero, and $V = - dF_X(s) f'(s) \neq 0$. To derive the variance of $W_{\sigma^2,s}$, we can exploit the fact that $W_{\sigma^2,s}$ follows Chernoff's distribution \citepappendix{groeneboom1989brownian}.  

To begin, define the random variable $Z$ as: 
$$Z_\gamma := \argmax_t B(t) - \gamma t^2,$$
where $B(t)$ is a standard Brownian motion, and $c > 0$. Then, from \citetappendix{groeneboom1989brownian} (Corollary 3.3), the density of $Z$ is given as $dF_Z(t) = g_\gamma(t) g_\gamma(-t),$ and $g_\gamma(t)$ is defined as: 
\begin{align} 
g_\gamma(t) &= \left(\frac{2}{\gamma}\right)^{1/3} \frac{1}{2\pi i} \int_{c_1 - i \infty}^{c_1 + i \infty} \frac{\exp(-tu)}{\text{Ai}((2\gamma^2)^{-1/3} u)} du, 
\label{eqn:density_of_max}
\end{align} 
where $c_1 > a_1$, and $a_1$ is the largest zero of the Airy function Ai. We can apply \citetappendix{janson2013moments} (Theorem 1.1) to obtain a closed form solution for the variance of $Z_{\gamma}$: 
$$\var(Z_\gamma) = \frac{2^{-2/3} \gamma^{-4/3}}{6 \pi i} \int_{-\infty}^{\infty} \frac{t}{\text{Ai}(it)^2}dt.$$
Furthermore, in settings when the Brownian motion is scaled (i.e., $B(at)$), we can exploit the fact that for all $a > 0$, $B(at) = a^{1/2} B(t)$, and: 
\begin{align*} 
Z_\gamma =^d a \argmax (a^{1/2} B(t) - a^2 \gamma t^2) = a Z_{a^{3/2} \gamma}.
\end{align*} 

We then solve for the variance of $W_{\sigma^2,s}$ by substituting $a^{1/2} = dF_X(s) \sigma^2$, and $\gamma = \frac{1}{2} V/(dF_X(s)^2 \sigma^4)^2$. Thus, the asymptotic variance of a split $\hat s_n$ can be written as follows:
$$\asyvar(\hat s_n) = \left(dF_X(s) \sigma^2 \right)^2 \cdot \frac{2^{-2/3}}{6\pi i} \left(\frac{1}{2} \frac{V}{dF_X(s) \sigma^2} \right)^{-4/3} \int_{- \infty}^{\infty} \frac{t}{Ai(it)^2} dt.$$
\end{proof}

\subsection{Additional Discussion of Assumption~\eqref{assum:sep}}\label{app:assum5}

We next discuss the connection between our separability assumption (Assumption~\eqref{assum:sep}) and two important assumptions in the existing tree literature: the sufficient impurity decrease (SID) condition \citep{chi2022asymptotic, mazumder2023convergence} and the merged staircase property (MSP) \citep{abbe2022merged, tan2026statistical}. 
At a high-level, these two assumptions (SID and MSP) are designed to ensure the consistency of the tree but do not make any guarantees about the features being selected by the tree. 
Assumption~\eqref{assum:sep}, on the other hand, is designed to ensure that the tree selects the ``relevant'', as opposed to ``irrelevant'', features, which is crucial for our goal of accurate and interpretable subgroup discovery.

Despite this difference however, it can be valuable to understand the similarities between Assumption~\eqref{assum:sep} and the existing assumptions.
In this subsection, we will show that if the irrelevant features do not change the population loss by a large amount, then (a) Assumption~\eqref{assum:sep} is equivalent to the SID assumption, (b.i) Assumption~\eqref{assum:sep} implies MSP under the binary-uniform covariates setting, and (b.ii) MSP together with a genericity assumption implies Assumption~\eqref{assum:sep} under the binary-uniform covariates setting.
By making these connections, we reveal insights into various function classes that satisfy Assumption~\eqref{assum:sep}.

Throughout this subsection, for each subgroup index $g \in \{1,\ldots,G\}$ and rule index $r \in \{1,\ldots,r_g\}$, we write the subgroup defined by the $r - 1$ previously selected binary decision rules as $\cG_g^{(r-1)}(X) = \prod_{k=1}^{r-1} \1\left\{ X^{(j_k)} \lesseqgtr s^{(j_k)} \right\}$, where $\cG^{(0)}_g(X) \equiv 1$.
Define $C_g^{(r-1)}$ be the region where $\cG_g^{(r-1)}(X) = 1$, i.e.,
\begin{align*}
    C_g^{(r-1)} = \{X \in \mathcal{X} \mid \cG_g^{(r-1)}(X) = 1\}.
\end{align*}
For any $u \in \R$, define $C_{g, L}^{(r-1)}(X^{(j)}, u)$ and $C_{g, R}^{(r-1)}(X^{(j)}, u)$ to be the left and right children cells of $C_g^{(r-1)}$ after splitting on $X^{(j)}$ at threshold $u$, i.e.,
\begin{align*}
    C_{g, L}^{(r-1)}(X^{(j)}, u) := C_g^{(r-1)} \cap \{ X^{(j)} \leq u\} \quad\text{and}\quad C_{g, R}^{(r-1)}(X^{(j)}, u) = C_g^{(r-1)} \cap \{X^{(j)} > u\}.
\end{align*}
Additionally, define $\mathcal{C}$ to be the collection of all non-terminal cells $C_g^{(r-1)}$ considered by the greedy tree algorithm, i.e., 
\begin{align*}
    \mathcal{C} := \left\{C_g^{(r-1)} \mid g \in \{1, \ldots, G\}, r \in \{1, \ldots, r_g\}\right\}.
\end{align*}
By the uniqueness of the optimal splits (Assumption~\eqref{assum:min}), $\mathcal{C}$ has finitely many elements given a fixed number of subgroups $G$ and maximum depth of the tree $\max_{g \in [G]} r_g$.
Note also for notational simplicity, we will omit the explicit dependence on $g$ and $r$ and write $C = C_g^{(r-1)}$ when the context is clear. We also reserve the notation $s^{(j)} \in \R$ to denote the population-optimal split for feature $X^{(j)}$ conditioned on $C$. 

Finally, for each cell $C \in \mathcal{C}$, feature index $j \in [p]$, and splitting threshold $u \in \R$, we denote the population loss before splitting (i.e., the total variance in $\tau(X)$ for a cell $C$) by
\begin{align*}
    M_C^0 := \E[(\tau(X) - \E[\tau(X) \mid C])^2 \mid C] \equiv \var(\tau(X) \mid C)
\end{align*}
and the population loss after splitting on $X^{(j)}$ at threshold $u$ by
\begin{align*}
    M_C(\tau(X), X^{(j)}, u) &:= \E[\ell(\tau(X); X^{(j)}, u) \mid C].
\end{align*}

\subsubsection{Connection between Assumption~\eqref{assum:sep} and Sufficient Impurity Decrease}\label{app:sid}

The sufficient impurity decrease (SID) condition, introduced in \citet{chi2022asymptotic} and expanded upon in \citet{mazumder2023convergence}, has played a key role in the theoretical analysis of greedy tree algorithms such as CART. Under the SID assumption, \citet{mazumder2023convergence} derived a tight upper bound on the prediction error of CART. This convergence rate is polynomial under the SID assumption and faster than the logarithmic rate obtained in previous work without the SID assumption \citep{klusowski2024large}.

In brief, the SID condition assumes that the best split at each node of the tree decreases the population variance by a constant factor. 

\begin{definition}[Sufficient impurity decrease (SID)]
For each cell $C \in \mathcal{C}$, feature index $j \in [p]$ and splitting threshold $u \in \R$, let $D_C(\tau(X), X^{(j)}, u)$ denote the impurity decrease after splitting $C$ on feature $X^{(j)}$ at threshold $u$, i.e.,
\begin{align*}
    D_C(\tau(X), X^{(j)}, u) &:= \Pr(X \in C)\cdot\var(\tau(X)\mid C)\\
        &\qquad - \Pr(X \in C_L(X^{(j)}, u))\cdot\var(\tau(X)\mid C_L(X^{(j)}, u))\\
        &\qquad - \Pr(X \in C_R(X^{(j)}, u))\cdot\var(\tau(X)\mid C_R(X^{(j)}, u)),
\end{align*}
$\tau(X)$ satisfies sufficient impurity decrease (SID) over $\mathcal{C}$ if there exists a constant $\lambda \in (0, 1]$ such that for each $C \in \mathcal{C}$,
\begin{align}
    \sup_{j \in [p],\, u \in \R} D_C(\tau(X), X^{(j)}, u) \geq \lambda \cdot \Pr(X \in C) \cdot \var(\tau(X) \mid X \in C).
\end{align}
\end{definition}

We next show that the SID condition is equivalent to Assumption~\eqref{assum:sep}, as long as the irrelevant features satisfy a particular bound on the change in population loss after splitting. 

\begin{assumption}[Population loss bound for irrelevant features]\label{assum:irrelevant_bound}
    Assume that for each subgroup index $g \in \{1, \ldots, G\}$ and rule index $r \in \{1, \ldots, r_g\}$, the change in population loss from splitting on irrelevant features conditioned on $C_g^{(r-1)}$ is bounded by
    \begin{align}
        M_C^0 - M_C(\tau(X), X^{(b)}, s^{(b)}) < \lambda \var(\tau(X) \mid C_g^{(r-1)}) \quad \text{for each } b \in B_r, 
    \end{align}
    where $\lambda > 0$ is the SID constant.
\end{assumption}

\begin{proposition}
\label{prop:sid}
\begin{enumerate}[label=(\roman*)]
    \item If Assumption~\eqref{assum:sep} holds, then the sufficient impurity decrease holds.
    \item If the sufficient impurity decrease condition and Assumption~\ref{assum:irrelevant_bound} hold, then Assumption~\eqref{assum:sep}~holds.
\end{enumerate}
\end{proposition}

\begin{proof}
First, we derive a useful identity, which re-writes the impurity decrease as the difference in population losses before and after making a new split.
For each cell $C \in \mathcal{C}$, feature index $j \in [p]$, and splitting threshold $u \in \R$, note that
\begin{align*}
    M_C(\tau(X), X^{(j)}, u) 
    &= \E\left[\left(\tau(X) - \1\left\{X \in C_L(X^{(j)}, u) \right\} \E\left[\tau(X) \mid  C_L(X^{(j)}, u)\right] \right.\right.\\
    &\qquad\qquad\left.\left. - \1\left\{X \in C_R(X^{(j)}, u)\right\} \E\left[\tau(X) \mid  C_R(X^{(j)}, u)\right]\right)^2 \Bigm|  C\right] \\
    &= \E\left[\left(\tau(X) - \E\left[\tau(X) \mid C_L(X^{(j)}, u)\right]\right)^2 \1 \left\{X \in C_L(X^{(j)}, u)\right\} \Bigm|  C \right] \\
    &\qquad\qquad + \E\left[\left(\tau(X) - \E\left[\tau(X) \mid C_R(X^{(j)}, u)\right]\right)^2 \1 \left\{X \in C_R(X^{(j)}, u)\right\} \Bigm|  C \right] \\
    &= \Pr(X \in C_L(X^{(j)}, u) \mid X \in C) \\
    &\qquad\qquad \cdot \E\left[\left(\tau(X) - \E[\tau(X) \mid  C_L(X^{(j)}, u)]\right)^2 \Bigm|  C_L(X^{(j)}, u)\right] \\
    &\qquad + \Pr(X \in C_R(X^{(j)}, u) \mid X \in C)\\
    &\qquad\qquad \cdot \E\left[\left(\tau(X) - \E[\tau(X) \mid  C_R(X^{(j)}, u)]\right)^2 \Bigm|  C_R(X^{(j)}, u)\right] \\
    &= \frac{\Pr(X \in C_L(X^{(j)}, u))}{\Pr(X \in C)} \var\left(\tau(X) \mid  C_L(X^{(j)}, u)\right)\\
    &\qquad + \frac{\Pr(X \in C_R(X^{(j)}, u))}{\Pr(X \in C)} \var\left(\tau(X) \mid  C_R(X^{(j)}, u)\right).
\end{align*}
Thus,
\begin{align}
    D_C(\tau(X), X^{(j)}, u) &= \Pr(X \in C) \cdot \var(\tau(X) \mid C) \nonumber\\ 
    &\qquad - \Pr(X \in C_L(X^{(j)}, u)) \cdot \var\left(\tau(X) \mid  C_L(X^{(j)}, u) \right) \nonumber\\
    &\qquad - \Pr(X \in C_R(X^{(j)}, u)) \cdot \var\left(\tau(X) \mid C_R(X^{(j)}, u)\right) \nonumber\\
    &= \Pr(X \in C) \cdot \left\{ M_C^0 - M_C(\tau(X), X^{(j)}, u) \right\}.\label{eq:impurity_decrease_loss}
\end{align}

\vspace{0.75em}
\noindent
\textit{Proof of (i).} \hspace{2pt}
Now assume that Assumption~\eqref{assum:sep} holds. 
For each $C \in \mathcal{C}$,
\begin{align*}
    \sup_{j \in [p], u \in \R} D_C(\tau(X), X^{(j)}, u) &\geq
    \max_{a \in A_r} D_C(\tau(X), X^{(a)}, s^{(a)}) \\
    &= \Pr(X \in C) \max_{a \in A_r} \left\{ M_C^0 - M_C(\tau(X), X^{(a)}, s^{(a)}) \right\} \\
    &= \Pr(X \in C) \left\{ M_C^0 - \min_{a \in A_r} M_C(\tau(X), X^{(a)}, s^{(a)}) \right\} \\
    &\geq \Pr(X \in C) \left\{ \min_{b \in B_r} M_C(\tau(X), X^{(b)}, s^{(b)}) - \min_{a \in A_r} M_C(\tau(X), X^{(a)}, s^{(a)}) \right\} \\
    &\geq \Pr(X \in C) \Delta, 
\end{align*}
where the second equality follows from \eqref{eq:impurity_decrease_loss}, the fourth inequality follows from the fact that $M_C^0 \geq M_C(\tau(X), X^{(j)}, u)$ for all $j \in [p]$ and $u \in \R$, and the last inequality follows from Assumption~\eqref{assum:sep}.
Moreover, since $\mathcal{C}$ is a finite collection of non-terminal cells by definition, we can define $V_{\max} := \max_{C \in \mathcal{C}} \var(\tau(X) \mid C)$.
Taking $\lambda = \Delta / V_{\max} \in (0, 1]$, SID holds.

\vspace{0.75em}
\noindent
\textit{Proof of (ii).} \hspace{2pt} Next, assume that SID holds, and let $C \in \mathcal{C}$ be given.
By SID, we have that
\begin{align*}
    \sup_{j \in [p], u \in \R} D_C(\tau(X), X^{(j)}, u) \geq \lambda \Pr(X \in C) \var(\tau(X) \mid C),
\end{align*}
Using \eqref{eq:impurity_decrease_loss} and Assumption~\ref{assum:irrelevant_bound}, this implies
\begin{align*}
    \sup_{j \in [p], u \in \R} \left\{ M_C^0 - M_C(\tau(X), X^{(j)}, u) \right\} &\geq \lambda \var(\tau(X) \mid C) \\
    &> M_C^0 - M_C(\tau(X), X^{(b)}, s^{(b)}) \quad \text{for each } b \in B_r.
\end{align*}
This shows that the supremum on the left-hand-side cannot be attained by an irrelevant feature. Since the set of irrelevant features $B_r$ and relevant features $A_r$ partition the covariate space, then the supremum must be attained by some relevant feature. Thus, 
\begin{align*}
    \max_{a \in A_r} \left\{ M_C^0 - M_C(\tau(X), X^{(a)}, s^{(a)}) \right\} &> M_C^0 - M_C(\tau(X), X^{(b)}, s^{(b)}) \quad \text{ for each } b \in B_r\\
    \implies \max_{a \in A_r} \left\{ M_C^0 - M_C(\tau(X), X^{(a)}, s^{(a)}) \right\} &> \max_{b \in B_r} \left\{ M_C^0 - M_C(\tau(X), X^{(b)}, s^{(b)}) \right\}\\
    \implies \min_{a \in A_r} M_C(\tau(X), X^{(a)}, s^{(a)}) &< \min_{b \in B_r} M_C(\tau(X), X^{(b)}, s^{(b)}).
\end{align*}
Define
\begin{align*}
    \Delta_{C} := \min_{b \in B_r} M_C(\tau(X), X^{(b)}, s^{(b)}) - \min_{a \in A_r} M_C(\tau(X), X^{(a)}, s^{(a)}) > 0.
\end{align*}
Since this strictly positive gap holds for each $C \in \mathcal{C}$ and $\mathcal{C}$ is a finite collection of non-terminal cells, we can define $\Delta = \min_{C \in \mathcal{C}} \Delta_{C} > 0$. Using this $\Delta$, Assumption~\eqref{assum:sep} holds.
\end{proof}

As an important special case of Proposition~\ref{prop:sid}, if we define irrelevant features to be those that do not change the population loss after splitting on them (i.e., $M_C^0 - M_C(\tau(X), X^
{(b)}, s^{(b)}) = 0$ for all $b \in B_r$), then Assumption~\ref{assum:irrelevant_bound} is automatically satisfied, and the SID condition is equivalent to separability (Assumption~\eqref{assum:sep}). For example, if $\tau(X) = X^{(1)} + X^{(2)}$, then letting $X^{(3)}, \ldots, X^{(p)}$ be the irrelevant features would satisfy the above criteria.

Consequently, we can leverage existing results from \citet{mazumder2023convergence} to understand when Assumption~\eqref{assum:sep} is satisfied in practice.
In particular, additive models of the form $\tau(X) = f_1(X^{(1)}) + \ldots + f_p(X^{(p)})$, where the component functions $f_j$ adhere to a ``locally reverse Poincaré (LRP) inequality'', satisfy the SID condition \citep{mazumder2023convergence}. 
Examples of LRP component functions include strongly increasing functions, smooth and strongly convex functions, polynomial functions, and piecewise functions with each piece belonging to one of the aforementioned classes \citep{mazumder2023convergence}.
Thus, if $\tau(X) = f_1(X^{(1)}) + \ldots + f_s(X^{(s)})$ for some $s < p$ with each $f_j$ satisfying the LRP inequality, and if we let features $X^{(s+1)}, \ldots, X^{(p)}$ be the irrelevant features, then Assumption~\eqref{assum:sep} is satisfied.

\subsubsection{Connection between Assumption~\eqref{assum:sep} and Merged Staircase Property}\label{app:msp}

Another related assumption is the merged staircase property (MSP), first introduced by \citet{abbe2022merged} in the deep learning literature and further studied by \citet{tan2026statistical} in the context of decision trees.
While previous work in the greedy tree literature focused on \textit{sufficient} conditions for the fast convergence of CART, \citet{tan2026statistical} investigated the \textit{necessary} conditions and proved that MSP is necessary as well as near-sufficient for greedily-trained decision trees to be high-dimensional consistent under the sparse regression setting with binary-uniform covariates.

Formally, following \citet{abbe2022merged} and \citet{tan2026statistical}, the definition of MSP relies on the Hadamard (or Walsh-Fourier) transform and thus assumes that the covariates $X$ are binary $\{-1, +1\}$.\footnote{Though MSP is typically defined for binary covariates only, we refer interested readers to \citet{tan2026statistical} for a discussion of extensions to continuous covariate settings.}
Under this binary covariates setting, we can write any function $\tau: \{-1, +1\}^p \rightarrow \R$ uniquely using its Hadamard transform as:
\begin{align*}
    \tau(X) = \sum_{k = 1}^{q} \theta_{S_k} \chi_{S_k}(X),
\end{align*}
where $q$ is a nonnegative integer, $S_k \subseteq [p]$ is a subset of feature indices, $\theta_{S_k}$ are nonzero Fourier coefficients, and $\chi_{S_k}(X) = \prod_{j \in S_k} X_j$ are monomials. 

\begin{definition}[Merged staircase property (MSP)]
    $\tau(X)$ satisfies the merged staircase property (MSP) if there exists an ordering of the sets $S_1, \ldots, S_q$ such that $\left|S_k \setminus \bigcup_{l = 1}^{k - 1} S_l\right| \leq 1$ for all $k \in \{1, \ldots, q\}$.
\end{definition}

In other words, the monomials in the Fourier decomposition can be ordered such that the support does not grow by more than one new feature at a time.
For instance, if $\tau(X) = X_1 + X_2 + X_1 X_2 X_3$, we have that $S_1 = \{1\}$, $S_2 = \{2\}$, and $S_3 = \{1, 2, 3\}$, which satisfies MSP.
Meanwhile, $\tau(X) = X_1 X_2 X_3$ does not satisfy MSP.

\citet{tan2026statistical} (Theorem 1.1) then showed that for $X \in \text{Unif}(\{-1, +1\}^p)$, if $\tau(X)$ does not satisfy MSP, then CART is high-dimensional inconsistent; on the other hand, if $\tau(X)$ satisfies MSP and the nonzero Fourier coefficients $\{\theta_{S_k}\}_{k = 1}^{q}$ are generic, then CART is high-dimensional consistent.
Here, genericity means that the nonzero Fourier coefficients avoid exact cancellation of monomial terms when restricting to subcubes of the binary hypercube. 
Note however that genericity is a technical condition that holds with probability 1 if $\{\theta_{S_k}\}_{k = 1}^{q}$ are drawn from an absolutely continuous distribution.\footnote{Consider an example from \citet{tan2026statistical}: $\tau(X) = X^{(1)} + X^{(2)} + X^{(1)}X^{(2)} + X^{(2)}X^{(3)}$. If we restrict $\tau(X)$ to the subcube where $X^{(1)}=-1$, then the restricted $\tau(X)$ becomes $\tau(X \mid X^{(1)} = -1) = -1 + X^{(2)} + (-1) \cdot X^{(2)} + X^{(2)}X^{(3)} = -1 + X^{(2)}X^{(3)}$, which no longer satisfies MSP even though $\tau(X)$ originally satisfied MSP. This exact cancellation is a result of the Fourier coefficients being linearly dependent. Such linear dependence however forms a measure 0 set and hence avoided if the Fourier coefficients are drawn from an absolutely continuous distribution, as discussed in \citet{tan2026statistical}.}

Importantly for this current work, as part of their proof, \citet{tan2026statistical} established the connection between MSP and SID, which we use to understand the connection between MSP and our separability assumption (Assumption~\eqref{assum:sep}) below.

\begin{proposition}\label{prop:msp}
Suppose that $X \sim \text{Unif}(\{-1,+1\}^{p})$.
\begin{enumerate}[label=\textnormal{(\roman*)}]
    \item If $\tau(X)$ satisfies Assumption~\eqref{assum:sep}, then $\tau(X)$ satisfies MSP.
    \item If $\tau(X)$ satisfies MSP, the nonzero Fourier coefficients of $\tau(X)$ are generic, and Assumption~\ref{assum:irrelevant_bound} holds, then $\tau(X)$ satisfies Assumption~\eqref{assum:sep}.
\end{enumerate}
\end{proposition}

\begin{proof}
    \textit{Proof of (i).} \hspace{2pt}
    By Proposition G.2 from \citet{tan2026statistical}, $\tau(X)$ satisfies the SID condition if and only if $\tau(X)$ satisfies a stronger version of MSP, called the stable merged-staircase property (SMSP). Briefly, SMSP requires that the restriction of $\tau(X)$ to any subcube $C \subseteq \{-1, +1\}^p$ of the binary hypercube also satisfies MSP. Since the full hypercube $\{-1, +1\}^p$ is one such subcube, then SMSP implies MSP. Finally, since SID is equivalent to our separability assumption (Assumption~\eqref{assum:sep}) as shown in Proposition~\ref{prop:sid}, we can conclude that Assumption~\eqref{assum:sep} implies MSP under the binary-uniform covariates setting.

    \textit{Proof of (ii).} \hspace{2pt}
    By Proposition G.1 from \citet{tan2026statistical}, if $\tau(X)$ satisfies MSP and the nonzero Fourier coefficients of $\tau(X)$ are generic, then $\tau(X)$ satisfies SMSP.
    Then again, by invoking Proposition G.2 from \citet{tan2026statistical}, $\tau(X)$ satisfies SMSP if and only if $\tau(X)$ satisfies the SID condition. Finally, SID together with Assumption~\ref{assum:irrelevant_bound} implies Assumption~\eqref{assum:sep} by our Proposition~\ref{prop:sid}(ii).
\end{proof}

Thus, under the binary-uniform covariates setting where irrelevant features do not change the population loss by a large amount after splitting on them, Assumption~\eqref{assum:sep} can be equivalently viewed through the MSP lens, up to the small caveat of genericity. 
Importantly, whereas Assumption~\eqref{assum:sep} and the SID condition impose constraints on the decrease in population loss after making splits, MSP directly characterizes the type of hierarchical structure underlying $\tau(X)$ that is both necessary and near-sufficient for greedy trees such as CART to achieve fast convergence rates.
This hierarchical structural characterization goes beyond the simple additive model examples discussed in Section~\ref{app:sid} and helps to provide further intuition as to when Assumption~\eqref{assum:sep} may be satisfied in practice.

\section{Method Implementation Details}\label{app:sims}

In this section, we describe the hyperparameters and implementation details for all methods used in the simulations and case study. All methods and code were implemented using \texttt{R}. For ease of reproducibility and communication of the simulation results, the simulation study was conducted using the \texttt{simChef} R package \citepappendix{duncan2024simchef}.

\paragraph{Causal Distillation Trees (CDT).} We considered causal distillation trees with four different teacher models: causal forest \citepappendix{wager2018estimation}, Bayesian causal forest (BCF) \citepappendix{hahn2020bayesian}, R-learner with boosting (Rboost) and R-learner with natural cubic splines (Rspline) \citepappendix{nie2021quasi}. These teacher models were implemented using \texttt{grf::causal\_forest()} \citepappendix{grf}, \texttt{bcf::bcf()} \citepappendix{hahn2020bayesian}, \texttt{rlearner::rboost()} \citepappendix{rlearner}, and \texttt{rlearner::rlasso()} \citepappendix{rlearner} after B-spline (degrees of freedom = 3) transformations, respectively. All default hyperparameters settings were used for causal forest, BCF, and the R-learners. Additional hyperparameters were required for BCF including propensity score weights, which were set to the oracle value of 0.5, the number of burn-in samples, which was set to 2000, and the number of MCMC iterations to save after burn-in, which was set to 1000.

For the student model, we fit a CART decision tree \citepappendix{breiman1984classification} using \texttt{rpart::rpart()} \citepappendix{rpart}, with the default hyperparameters---namely,
\begin{itemize}
    \item \texttt{minsplit} = 20 (i.e., the minimum number of observations that must exist in a node in order for a split to be attempted)
    \item \texttt{maxdepth} = 30 (i.e., the maximum depth of any node of the final tree)
    \item \texttt{cp} = 0.01 (i.e., the minimum complexity parameter)
    \item \texttt{xval} = 10 (i.e., the number of folds in cross-validation)
\end{itemize}
The CART was then pruned using the standard post-pruning procedure (i.e., choosing the complexity parameter $\alpha$ which minimizes the cross-validation error), as described in Section~\ref{app:pruning}. Note that this post-pruning step does not drastically affect the subgroup estimation results when using causal forest or Rboost as the CDT teacher model, as shown in Figure~\ref{fig:sim-prune}. 
Given the strong performance of CDT using these default CART hyperparameters and the standard post-pruning procedure across all simulations, we recommend beginning with these defaults when no additional prior knowledge is available.

To perform honest estimation of the subgroup ATEs, we set $\pi_{train} = 0.70$ in the simulations and $\pi_{train} = 0.50$ in the case study for splitting the data into a training and hold-out estimation set. To estimate the heterogeneous treatment effects $\tauh(X_i)$, we used the out-of-bag sample estimates in the Distilled Causal Forest and Distilled BCF, and we used cross-fitting with the default $10$ folds in Distilled Rboost and Distilled Rspline. We also examined the possibility of using repeated cross-fitting in these distilled R-learners in Figure~\ref{fig:sim-crossfit}. We generally found that the subgroup estimation results were robust to this choice and hence used only a single round of cross-fitting (with $10$ folds) in the main simulations to minimize the computational burden.

\paragraph{Virtual Twins.} We implemented the virtual twins algorithm \citepappendix{foster2011subgroup} using the \\ \texttt{ranger::ranger()} \citepappendix{wright2017ranger} implementation of random forests. Like in CDT, we set aside a hold-out estimation set to perform honest estimation of the subgroup ATEs and used $\pi_{train} = 0.70$.

\paragraph{Causal Trees.} We implemented causal trees using the \texttt{causalTree} R package \citepappendix{athey2016recursive} and used the default parameters as shown in the example usage on GitHub --- that is, \texttt{split.Rule = "CT"}, \texttt{cv.option = "CT"}, \texttt{split.Honest = TRUE}, \texttt{cv.Honest = "TRUE"}, \texttt{split.Bucket = FALSE}, \texttt{xval = 5}, \texttt{cp = 0}, \texttt{minsize = 20}, and \texttt{propensity = 0.5}. As in CDT, we pruned the causal tree using the standard post-pruning procedure, choosing the complexity parameter $\alpha$ which minimizes the cross-validation error. Hyperparameter tuning of this complexity parameter will also be investigated later in Section~\ref{app:sim_results}.

\paragraph{Linear and Lasso Regression.} In the linear and Lasso regressions, we included all main effects $X$ and their interactions with the treatment variable $Z$ as covariates to predict the response $Y$. In the linear regression, we defined a selected subgroup feature to be any covariate whose interaction with the treatment variable yielded a significant p-value ($p < 0.05$). In the Lasso regression, we defined a selected subgroup feature to be any covariate whose interaction with the treatment variable received a non-zero coefficient. To estimate the CATE for each unit, we computed the difference between the predicted response when $Z_i = 1$ and the predicted response when $Z_i = 0$. The linear regression was implemented using \texttt{lm()}, and the Lasso regression was implemented using \texttt{glmnet::cv.glmnet()} \citepappendix{glmnet} with 5-fold cross-validation and the default hyperparameter grid.

\section{Simulations}\label{app:sim_results}
\subsection{Details on Simulation Evaluation Metrics}

To evaluate the effectiveness of each subgroup estimation method, we consider the accuracy of subgroup identification from three perspectives: 
\begin{enumerate}
    \item \textbf{Selected subgroup features}: Whether the features used to define the estimated subgroups match the features used to define the true subgroups (i.e., $X^{(1)}$ and $X^{(2)}$), as measured by the accuracy of the selected subgroup features, we calculate the number of true positives, false positives, and $F_1$ score.\footnote{The $F_1$ score summarizes the number of true positives (TP), false positives (FP), and false negatives (FN) into a single quantity (i.e., $F_1 = \frac{2 \cdot TP}{2 \cdot TP + FP + FN}$)}
    \item \textbf{Estimated subgroup thresholds}: Whether the estimated subgroup thresholds are close to the true subgroup thresholds, as measured via the root mean squared error (RMSE).
    \item \textbf{Estimated Subgroup ATEs}: Whether the estimated subgroup average treatment effects $\hat \tau^{(g)}$ are close to the true subgroup average treatment effects (i.e., $\tau^{(g)} := \E\left[ \tau_i \mid \cG_g(X_i)\right]$), as measured via the RMSE.
\end{enumerate}

\subsection{Additional Simulation Settings}

We next present additional simulation results to complement the main results, shown in Section~\ref{sec:sims}. These results include additional subgroup data-generating processes, evaluation metrics, and alternative modeling choices in the CDT framework (e.g., different number of repeated cross-fits $R$, student model choices, and pruning choices).

\begin{figure}
    \centering
    \includegraphics[width=0.98\linewidth]{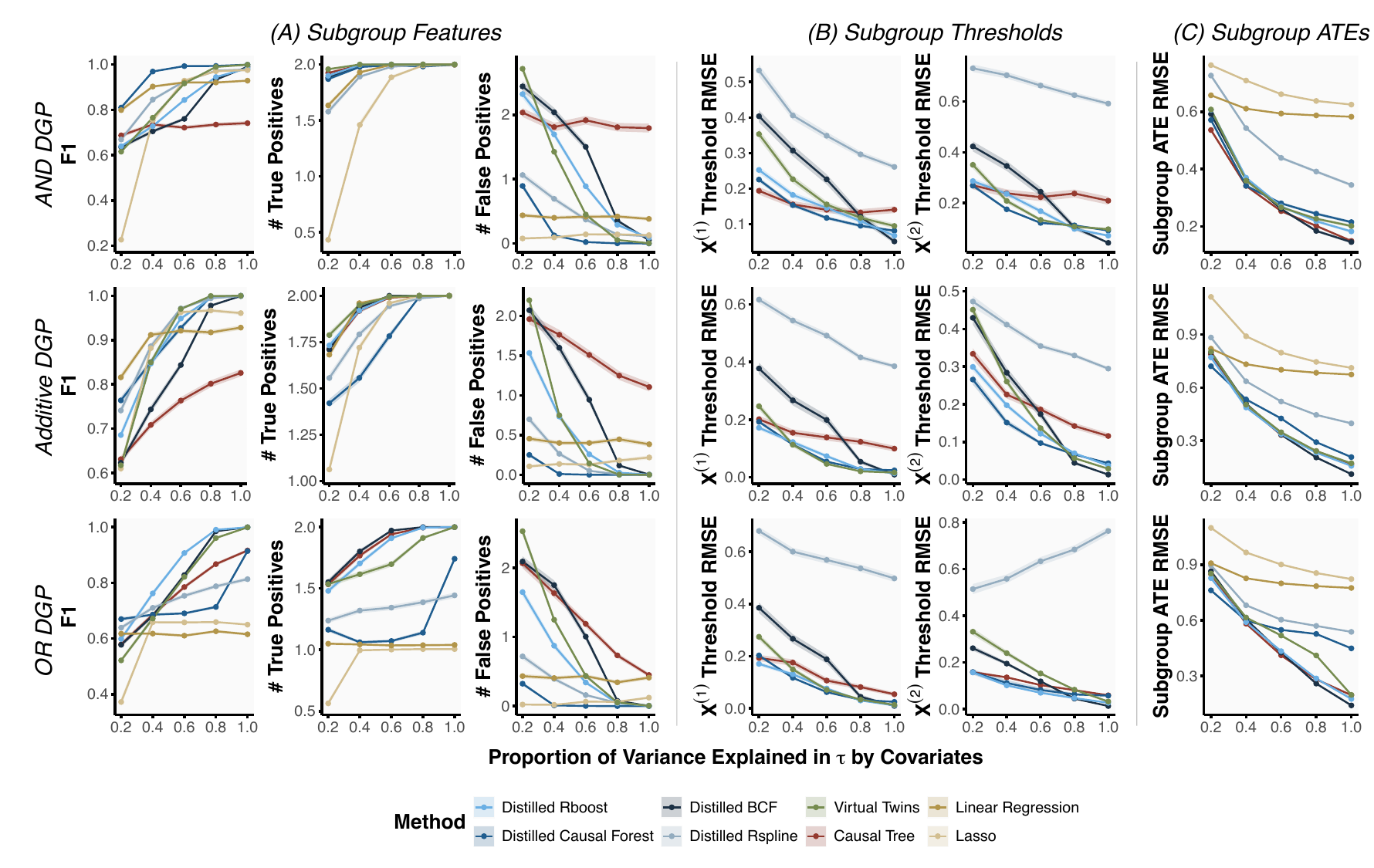}
    \caption{Performance of subgroup estimation methods for (A) identifying the true subgroup features, measured via $F_1$ score, number of true positives, and number of false positives, (B) estimating the true subgroup thresholds, measured via root mean squared error (RMSE) for each true subgroup feature, and (C) estimating the true subgroup ATE, measured via RMSE, across increasing treatment effect heterogeneity strengths (x-axis) and different subgroup data-generating processes in the \textbf{CATE-only outcome model} scenario (rows). Results are averaged across 500 simulation replicates with ribbons denoting $\pm1SE$.}
    \label{fig:sim-subgroup-errors-0}
\end{figure}

\paragraph{CATE-only Outcome Model.} To complement Figure~\ref{fig:sim-subgroup-errors-cov} which examined the subgroup estimation performance under an outcome model including linear covariate effects (i.e., $Y_i = Z_i \cdot \tau_i + X_i^{(3)} + X_i^{(4)} + \nu_i$), we also examined the subgroup estimation performance under the following outcome model, which depends only on the CATE:
\begin{align*}
    Y_i = Z_i \cdot \tau_i + \nu_i.
\end{align*}
Besides this modification in the outcome model, all other simulation settings were kept the same. The results are shown in Figure~\ref{fig:sim-subgroup-errors-0}. As seen under the previous outcome model, the distilled methods with tree-based teacher models generally outperformed existing methods in estimating the subgroup structure and the subgroup ATEs.
Moreover, the performance of the linear models for estimating the subgroup ATEs significantly deteriorated under this CATE-only outcome model. This highlights the sensitivity of the linear model approaches to the form of the outcome model. In comparison, CDT tends to work well in both the setting considered here and in Section~\ref{sec:sims}.

\begin{figure}
    \centering
    \includegraphics[width=1\linewidth]{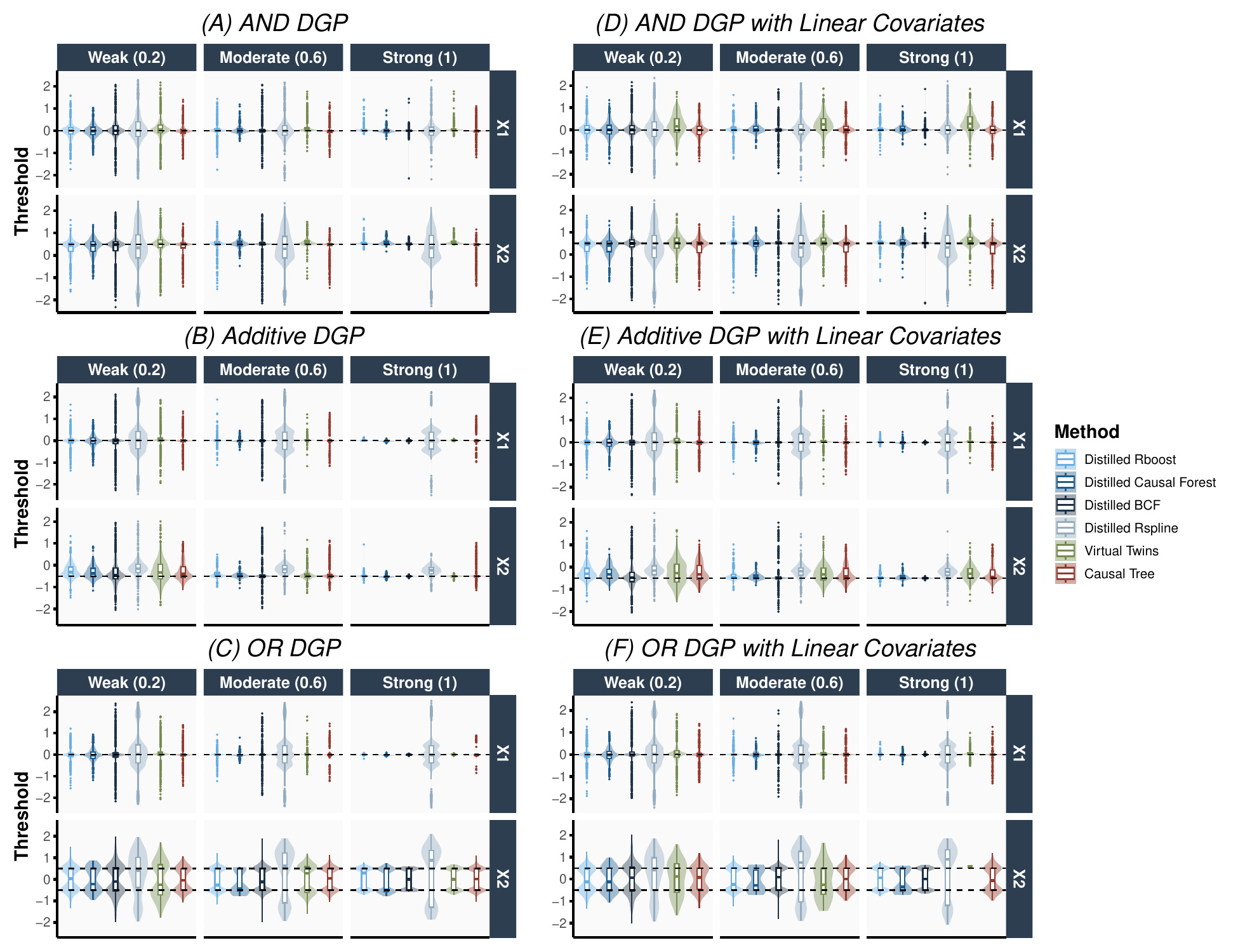}
    \caption{Distribution of estimated thresholds for each true subgroup feature ($X^{(1)}$ and $X^{(2)}$) using different subgroup estimation methods under various treatment effect heterogeneity strengths (columns) and subgroup data-generating processes (subplots). Results are shown for 500 simulation replicates.}
    \label{fig:sim-thr}
\end{figure}

\paragraph{Threshold Distributions.} When examining the distribution of the estimated subgroup thresholds from the tree-based methods in Figure~\ref{fig:sim-thr}, we uncover another benefit of distillation --- the distribution of the CDT-estimated subgroup thresholds (excluding the misspecified Rspline) is much tighter (i.e., has smaller variance) than other tree-based subgroup detection methods without distillation (i.e., causal trees and virtual twins). This pattern becomes more apparent as the treatment effect signal increases. This again reinforces our theoretical understanding of CDT, where we have seen in Example~\ref{ex:split_stability} that an appropriate first-stage learner in CDT acts as a de-noising step, leading to more stable splits (i.e., thresholds) in CDT compared to non-distilled tree-based methods. 

Note that while the thresholds distribution from Distilled Rspline is more variable than the non-distilled methods, Distilled Rspline would have been easily ruled out according to our teacher model selection procedure using Jaccard SSI (see Appendix~\ref{app:jaccard}). 

\begin{figure}
    \centering
    \includegraphics[width=0.9\linewidth]{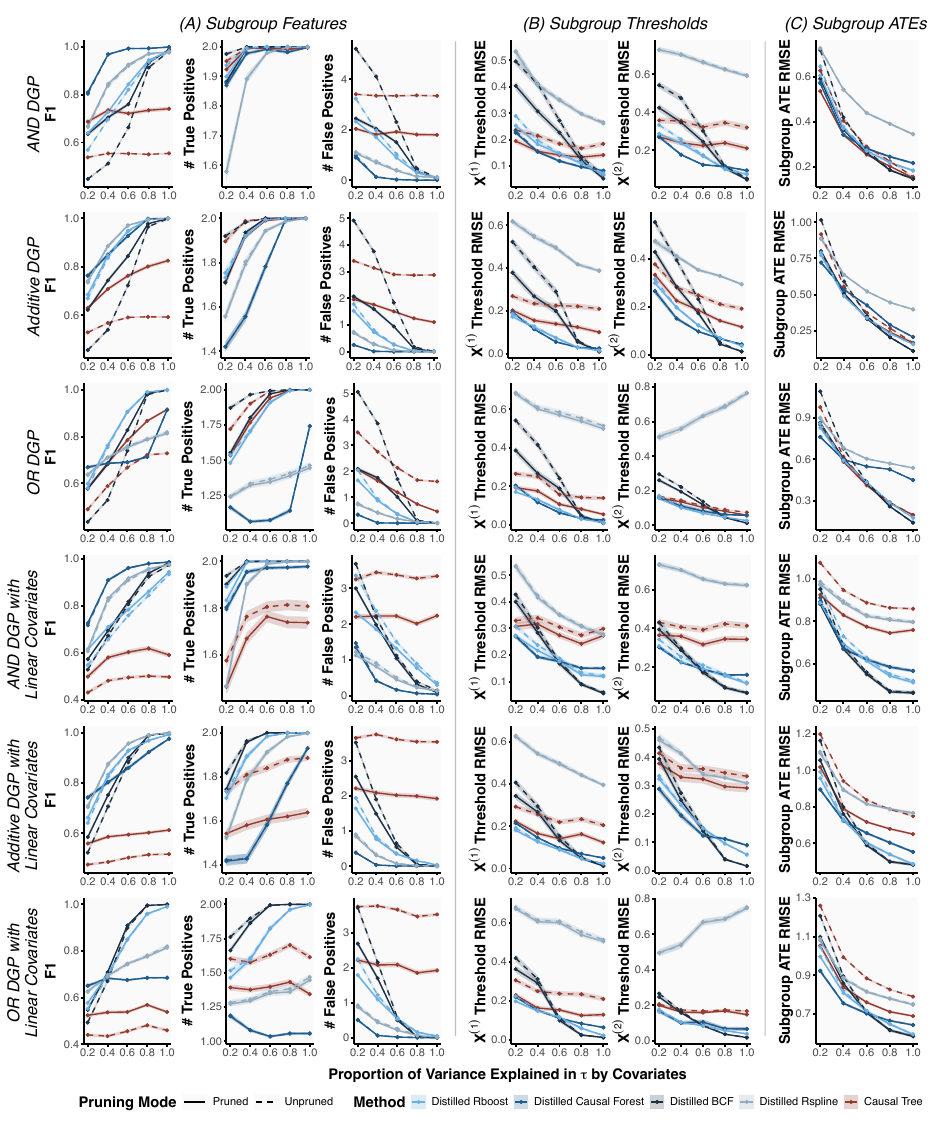}
    \caption{
    Performance of \textbf{pruned and unpruned} versions of CDT and causal trees for (A) identifying the true subgroup features, measured via $F_1$ score, number of true positives, and number of false positives, (B) estimating the true subgroup thresholds, measured via root mean squared error (RMSE) for each true subgroup feature, and (C) estimating the true subgroup ATE, measured via RMSE, across increasing treatment effect heterogeneity strengths (x-axis) and different subgroup data-generating processes (rows). Results are averaged across 500 simulation replicates with ribbons denoting $\pm1SE$.}
    \label{fig:sim-prune}
\end{figure}

\paragraph{Pruned versus Unpruned.} Recall thus far that we have been implementing CDT and causal trees with the standard post-pruning procedure using cross-validation to tune the complexity parameter $\alpha$. To investigate the impact of this post-pruning step, we compare the subgroup estimation performance of CDT and causal trees with and without post-pruning in Figure~\ref{fig:sim-prune}. While the performance of causal trees declines drastically without pruning, the performance of CDT using causal forest or Rboost is similarly strong with and without pruning, demonstrating their robustness with respect to this pruning choice. For Distilled BCF, pruning improves the subgroup estimation performance, particularly in the weak-to-moderate treatment effect heterogeneity regimes.

\begin{figure}
    \centering
    \includegraphics[width=0.9\linewidth]{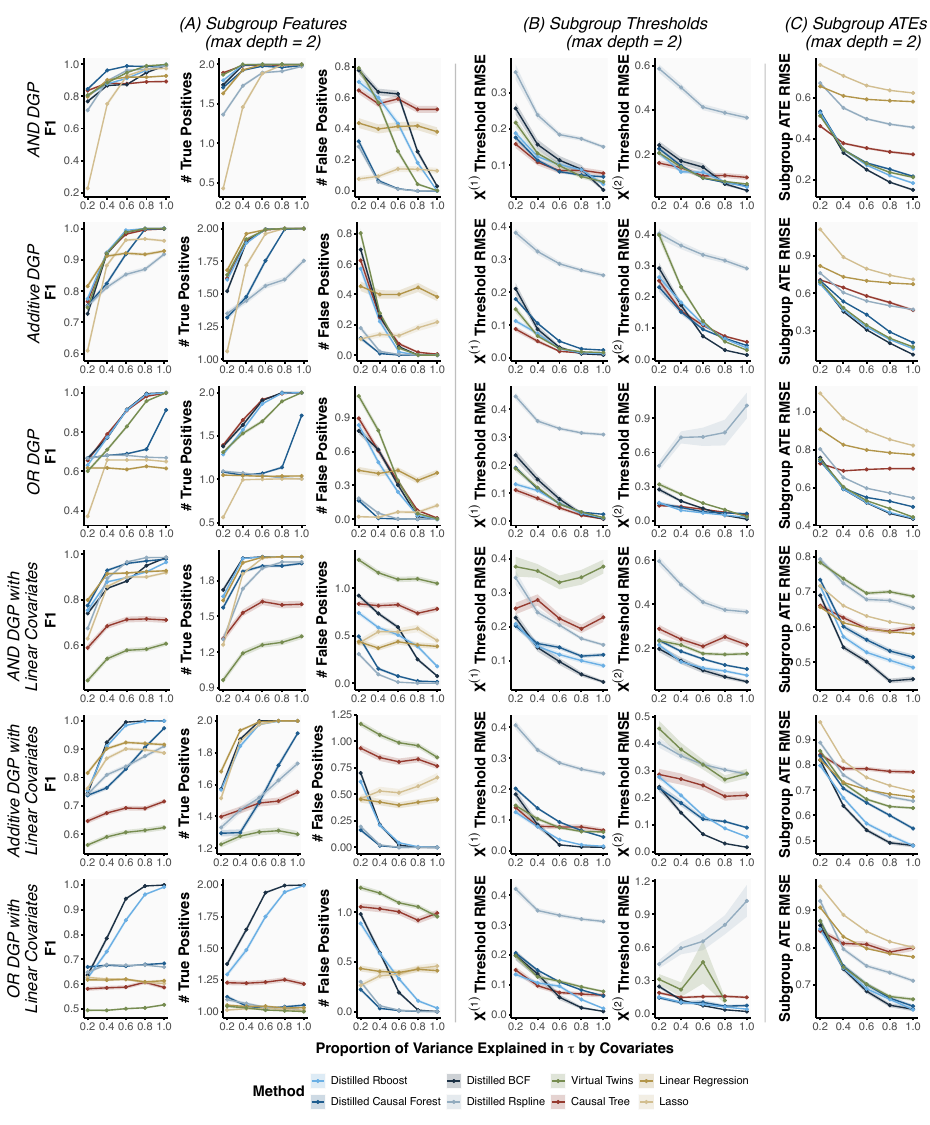}
    \caption{Performance of subgroup estimation methods \textbf{with oracle pruning} for (A) identifying the true subgroup features, measured via $F_1$ score, number of true positives, and number of false positives, (B) estimating the true subgroup thresholds, measured via root mean squared error (RMSE) for each true subgroup feature, and (C) estimating the true subgroup ATE, measured via RMSE, across increasing treatment effect heterogeneity strengths (x-axis) and different subgroup data-generating processes (rows). Results are averaged across 500 simulation replicates with ribbons denoting $\pm1SE$.}
    \label{fig:sim-depth2}
\end{figure}

\paragraph{Evaluating using Oracle Tree Depth.} In general, optimally pruning decision trees is a challenging problem and an active area of research \citepappendix{zhou2023trees}. To investigate whether the observed difference in subgroup estimation performance between tree-based methods is due to suboptimal pruning or not, we show in Figure~\ref{fig:sim-depth2} the subgroup estimation performance of the tree-based methods, pruned to have a fixed depth of 2 (which is the oracle tree depth given our subgroup DGPs, described in Section~\ref{sec:sims}). Even with this oracle pruning, we still observe improvements due to distillation. In particular, compared to causal tree and virtual twins, the CDT methods with the Rboost, Causal Forest, or BCF teacher models yield fewer false positives and more accurate estimation of the subgroup ATEs.

\begin{figure}
    \centering
    \includegraphics[width=0.88\linewidth]{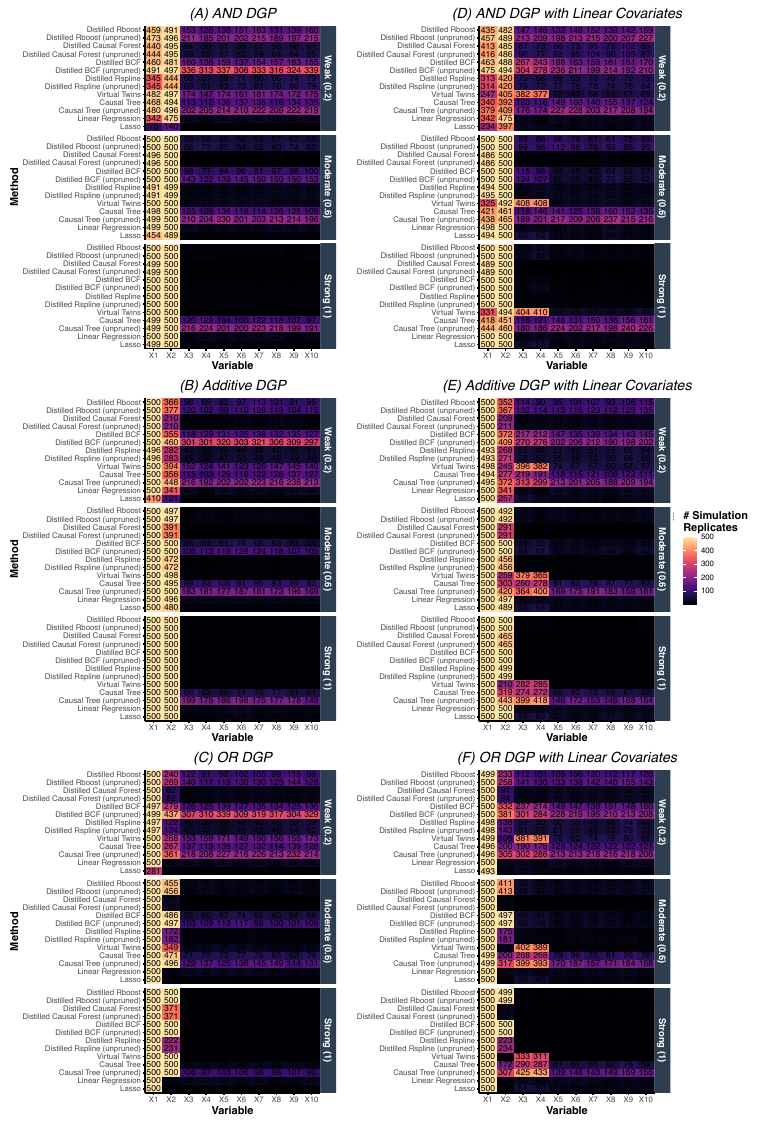}
    \caption{Number of simulation replicates out of 500, for which each variable was selected (at least once) in the estimated subgroups. Results are shown for different variables (x-axis), subgroup estimation methods (y-axis), treatment effect heterogeneity strengths (rows), and data-generating processes (subplots). Large values for variables $X^{(1)}$ and $X^{(2)}$ and small values for all other variables indicate better performance.}
    \label{fig:sim-freq}
\end{figure}

\paragraph{Selection Frequency of Subgroup Features.} In Figure~\ref{fig:sim-freq}, we show the number of simulation replicates (out of 500) that each variable was selected as a subgroup feature. Notably, when the outcome model includes linear covariate effects involving the features $X^{(3)}$ and $X^{(4)}$, causal tree and virtual twins frequently split on these irrelevant features, $X^{(3)}$ and $X^{(4)}$, whereas the distilled methods avoid this pitfall.

\begin{figure}
    \centering
    \includegraphics[width=1\linewidth]{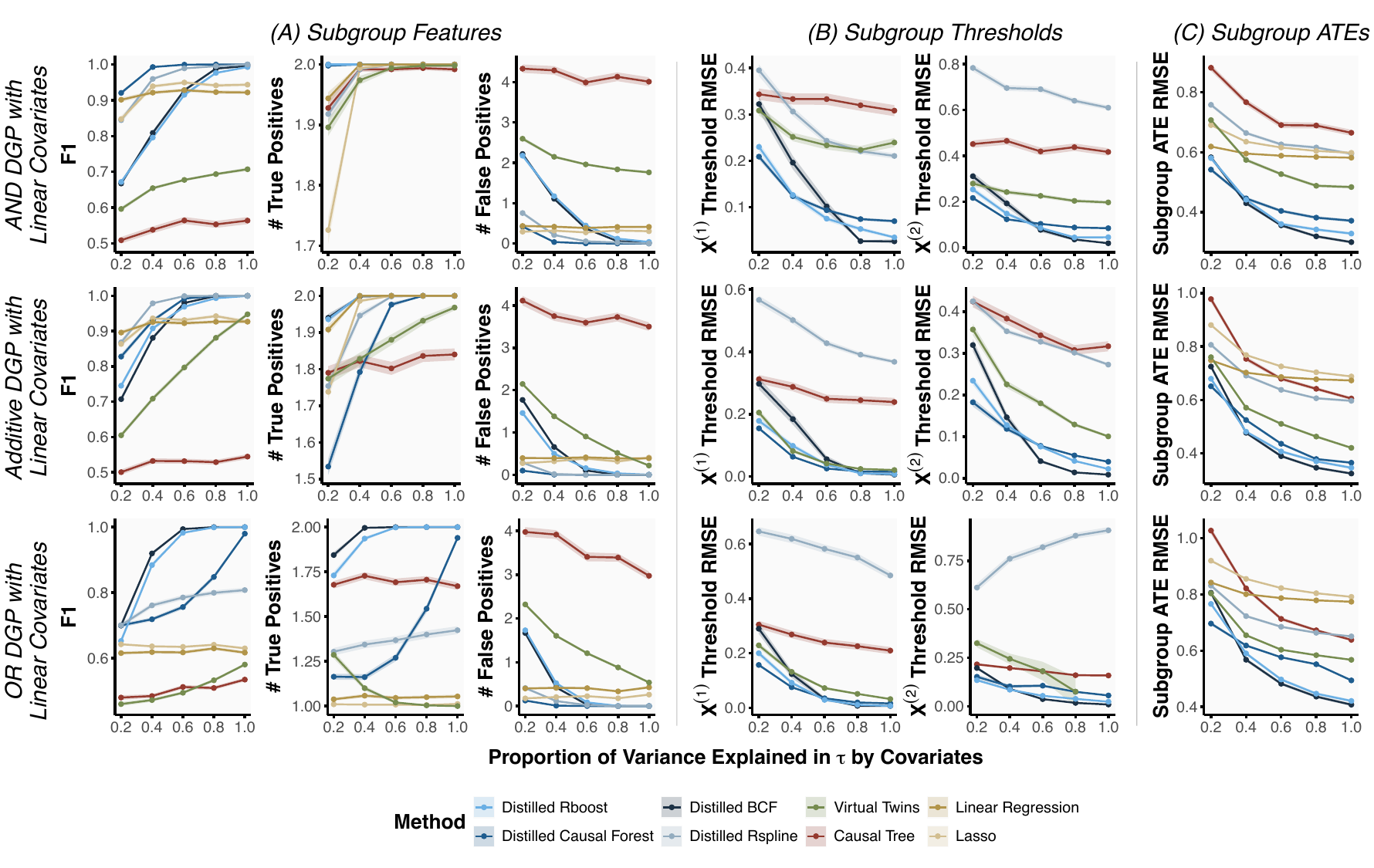}
    \caption{Performance of subgroup estimation methods for (A) identifying the true subgroup features, measured via $F_1$ score, number of true positives, and number of false positives, (B) estimating the true subgroup thresholds, measured via root mean squared error (RMSE) for each true subgroup feature, and (C) estimating the true subgroup ATE, measured via RMSE, across increasing treatment effect heterogeneity strengths (x-axis) and different subgroup data-generating processes under the main simulation settings (Section~\ref{sec:sims}) with a \textbf{different sample size} ($n = 1000$). Results are averaged across 500 simulation replicates with ribbons denoting $\pm1SE$.}
    \label{fig:sim-subgroup-errors-large-n}
\end{figure}

\paragraph{Different Sample Size.} 
To demonstrate the robustness of our findings to sample size, we repeated the simulations in Section~\ref{sec:sims} with a larger sample size of $n = 1000$. 
As seen in Figure~\ref{fig:sim-subgroup-errors-large-n}, the relative performance of methods remains similar to that observed in Section~\ref{sec:sims}, with CDT methods generally outperforming existing methods in estimating the subgroup structure and subgroup~ATEs.

\begin{figure}
    \centering
    \includegraphics[width=1\linewidth]{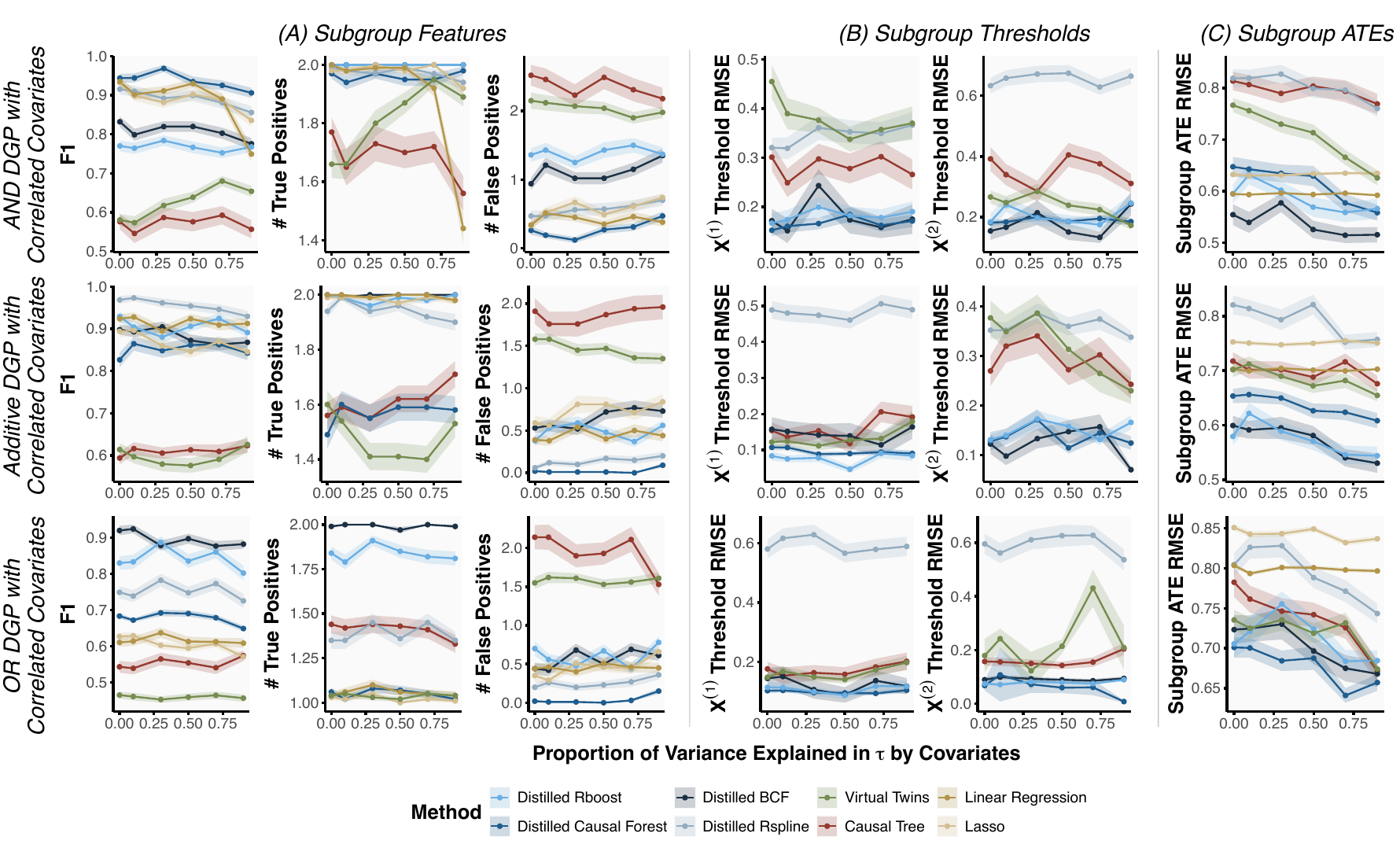}
    \caption{Performance of subgroup estimation methods for (A) identifying the true subgroup features, measured via $F_1$ score, number of true positives, and number of false positives, (B) estimating the true subgroup thresholds, measured via root mean squared error (RMSE) for each true subgroup feature, and (C) estimating the true subgroup ATE, measured via RMSE, across increasing treatment effect heterogeneity strengths (x-axis) and different subgroup data-generating processes in the \textbf{correlated moderators} scenario (rows). Results are averaged across 100 simulation replicates with ribbons denoting $\pm1SE$.}
    \label{fig:sim-subgroup-errors-cor}
\end{figure}

\paragraph{Correlated Moderators.} 
Up to this point, all features in the simulated $X$ have been generated independently of each other. To understand the impact of correlated moderators, we next examined the subgroup estimation performance under the outcome models described in Section~\ref{sec:sims} (i.e., $Y_i = Z_i \cdot \tau_i + X_i^{(3)} + X_i^{(4)} + \nu_i$), where all features $X^{(j)}$ are generated from standard normal distributions but $X^{(1)}$ and $X^{(3)}$ are correlated with a correlation coefficient $\rho \in \{0, 0.1, 0.3, 0.5, 0.7, 0.9\}$. All other features are independent of one another, and all other simulation settings were kept the same. Under this correlated moderator setting, Figure~\ref{fig:sim-subgroup-errors-cor} reveals that the performance of the subgroup estimation methods considered remains fairly consistent across varying correlation strengths.

\begin{figure}
    \centering
    \includegraphics[width=1\linewidth]{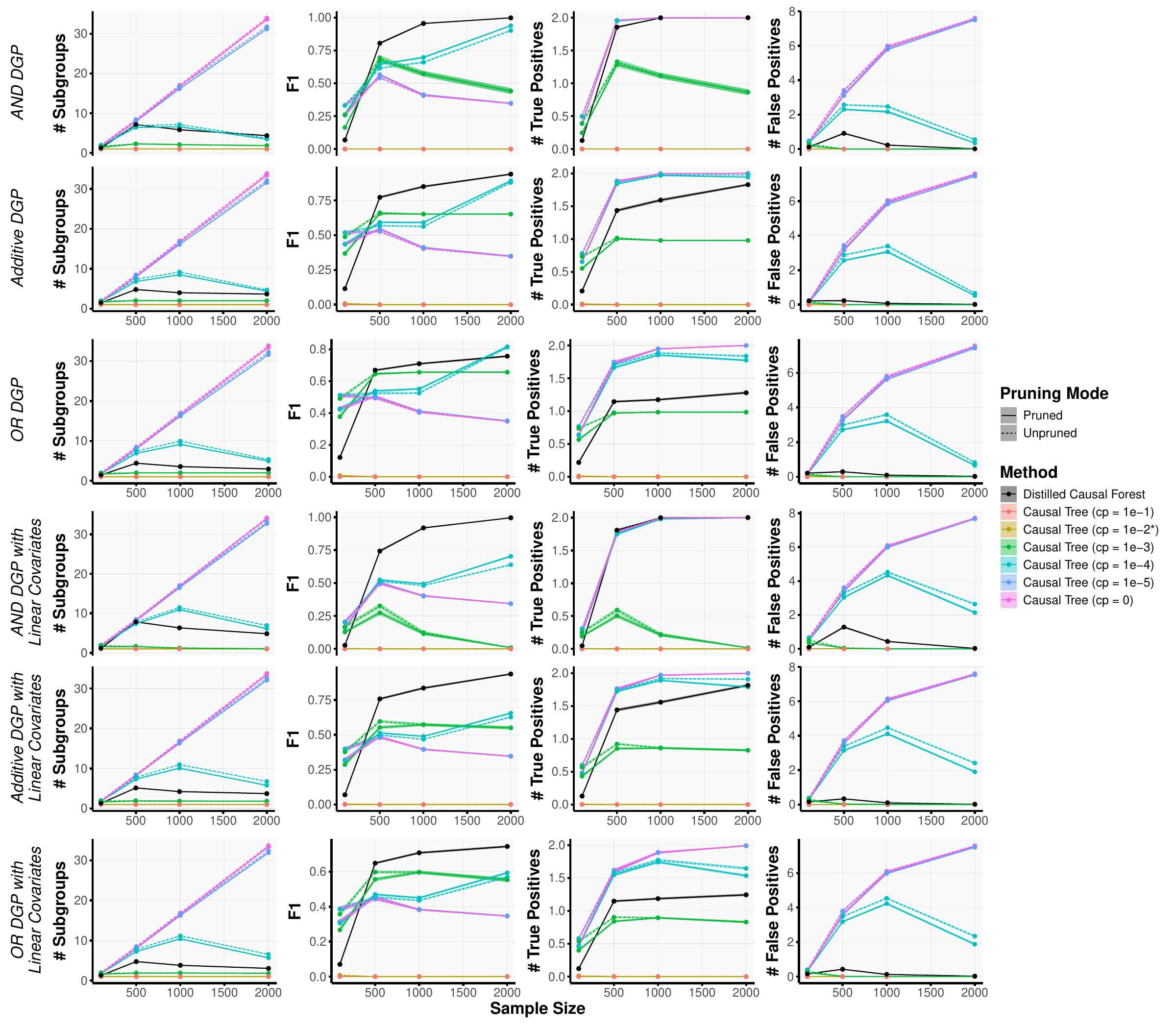}
    \caption{Comparing the number of estimated subgroups and accuracy of the selected subgroup features (measured via $F_1$ score, number of true positives, and number of true negatives) between Distilled Causal Forest (with the default \texttt{cp = 1e-2}) and causal trees with varying complexity (\texttt{cp}) parameters. Results are averaged across 500 simulation replicates with ribbons denoting $\pm 1 SE$.}
    \label{fig:causal-tree-tuning}
\end{figure}

\paragraph{Complexity (\texttt{cp}) Parameter Tuning in Causal Tree.} For both CDT and causal trees, we use their default hyperparameter settings (detailed in Section~\ref{app:sims}) throughout the simulations. One of these hyperparameters is the complexity (\texttt{cp}) parameter used for pruning the tree (i.e., either the causal tree or the CART model within CDT). By default, \texttt{rpart} (for CART) sets \texttt{cp = 0.01} while \texttt{causalTree} sets \texttt{cp = 0}. To ensure that the simulation results are not due to this difference or suboptimal complexity parameter tuning, we investigated the subgroup estimation performance of causal trees across varying choices of \texttt{cp} $\in \{10^{-1}, 10^{-2}, \ldots, 10^{-5}, 0\}$ in Figure~\ref{fig:causal-tree-tuning}. We include the Distilled Causal Forest results for reference. Note also that in Figure~\ref{fig:causal-tree-tuning}, we use the same number of cross-validation folds (\texttt{xval = 10}) for tuning in both causal tree and the CART student model in CDT, to ensure a fair comparison.

\begin{figure}
    \centering
    \includegraphics[width=1\linewidth]{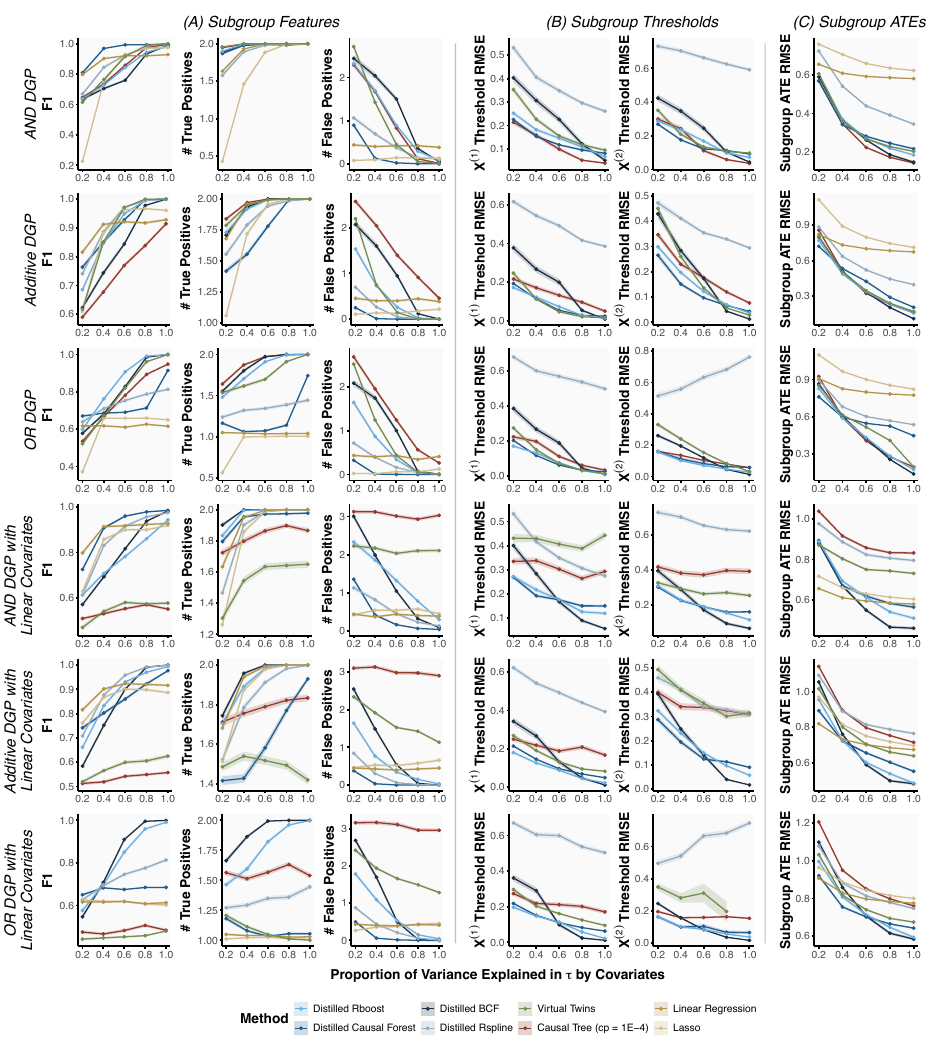}
    \caption{\textbf{Comparing CDT and causal trees with oracle complexity parameter.} Performance of subgroup estimation methods for (A) identifying the true subgroup features, measured via $F_1$ score, number of true positives, and number of false positives, (B) estimating the true subgroup thresholds, measured via root mean squared error (RMSE) for each true subgroup feature, and (C) estimating the true subgroup ATE, measured via RMSE, across increasing treatment effect heterogeneity strengths (x-axis) and different subgroup data-generating processes. Results are averaged across 500 simulation replicates with ribbons denoting $\pm1SE$.}
    \label{fig:sim-subgroup-errors-oracle}
\end{figure}

From Figure~\ref{fig:causal-tree-tuning}, we observe that the performance of causal trees is highly sensitive to the choice of \texttt{cp}. If \texttt{cp} is too small (e.g., 0 or $10^{-5}$), causal tree suffers from a high false positive rate, and if \texttt{cp} is too big (e.g., $10^{-1}$ or $10^{-2}$, which is the default used in CDT), causal tree selects nothing and does not identify any subgroups. While a \texttt{cp} of $10^{-4}$ appears to strike the best balance between the true positives and false positives under the simulation scenarios examined here, note that this ``oracle'' \texttt{cp} can never be known in practice since the ground truth subgroups are unknown. Regardless, for completeness, we repeat the subgroup estimation results from Figures~\ref{fig:sim-subgroup-errors-cov} and \ref{fig:sim-subgroup-errors-0}, but with the causal tree method using this oracle \texttt{cp} of $10^{-4}$. We report this new comparison in Figure~\ref{fig:sim-subgroup-errors-oracle}. Even with this oracle \texttt{cp}, the CDT methods with tree-based teacher models still outperform the oracle causal tree in terms of subgroup identification and estimation of subgroup ATEs, highlighting that these improvements stem from distillation, as opposed to poor hyperparameter tuning.

\begin{figure}
    \centering
    \includegraphics[width=1\linewidth]{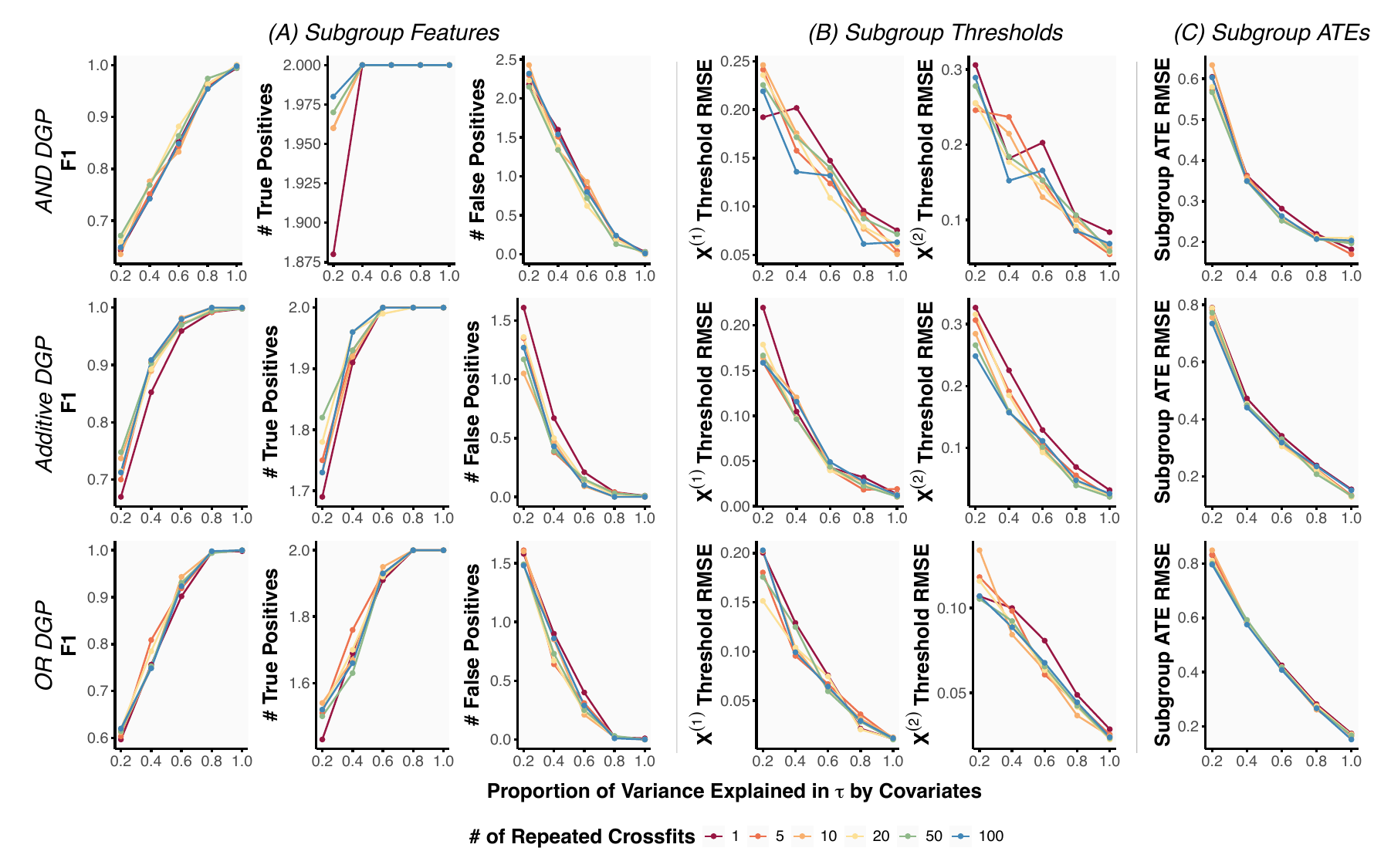}
    \caption{Performance of Distilled Rboost with \textbf{varying number of repeated crossfits $R$} (color) for (A) identifying the true subgroup features, measured via $F_1$ score, number of true positives, and number of false positives, (B) estimating the true subgroup thresholds, measured via root mean squared error (RMSE) for each true subgroup feature, and (C) estimating the true subgroup ATE, measured via RMSE, across increasing treatment effect heterogeneity strengths (x-axis) and different subgroup data-generating processes (rows). Results are averaged across 100 simulation replicates.}
    \label{fig:sim-crossfit}
\end{figure}

\paragraph{Choice of Repeated Cross-Fits $R$.} In CDT, we by default use a single round ($R = 1$) of cross-fitting to estimate the heterogeneous treatment effects in the teacher model. To investigate whether performing repeated cross-fitting yields additional benefit, we provide in Figure~\ref{fig:sim-crossfit} the subgroup estimation performance of Distilled Rboost for varying choices of $R$ between $1$ and $100$. In general, the subgroup estimation performance is fairly robust, regardless of the number of repeats.

\begin{figure}
    \centering
    \includegraphics[width=0.9\linewidth]{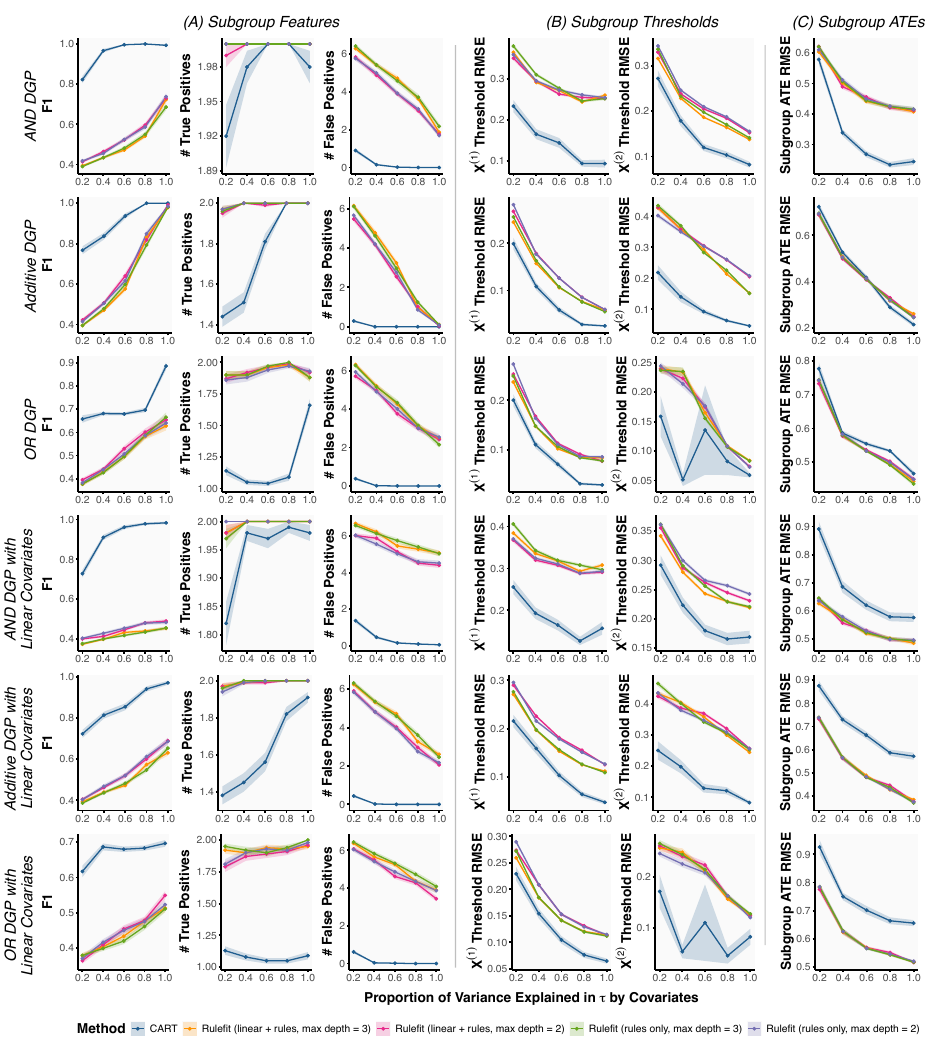}
    \caption{Performance of CDT with \textbf{CART versus Rulefit student models} for (A) identifying the true subgroup features, measured via $F_1$ score, number of true positives, and number of false positives, (B) estimating the true subgroup thresholds, measured via root mean squared error (RMSE) for each true subgroup feature, and (C) estimating the true subgroup ATE, measured via RMSE, across increasing treatment effect heterogeneity strengths (x-axis) and different subgroup data-generating processes (rows). Results are averaged across 100 simulation replicates with ribbons denoting $\pm1SE$.}
    \label{fig:sim-rulefit}
\end{figure}

\paragraph{Rulefit Student Model.} Another choice in the CDT framework is the student model, for which we recommend and have been using a CART decision tree. In previous work \citepappendix{bargagli2020causal, wan2023rule}, Rulefit \citepappendix{friedman2008predictive} has been proposed to either generate subgroup rules or estimate subgroups with heterogeneous treatment effects. In Figure~\ref{fig:sim-rulefit}, we compare the performance of Distilled Causal Forest using CART as the student model to Distilled Causal Forest using various instantiations of rulefit as the student model. Namely, we include four different versions of rulefit, implemented using \texttt{pre::pre()}, with the following hyperparameter settings: 

\begin{enumerate}
    \item Rulefit (rules only, max depth = 2): \texttt{type = "rules"} and \texttt{maxdepth = 2}
    \item Rulefit (rules only, max depth = 3): \texttt{type = "rules"} and \texttt{maxdepth = 3}
    \item Rulefit (linear + rules, max depth = 2): \texttt{type = "both"} and \texttt{maxdepth = 2}
    \item Rulefit (linear + rules, max depth = 3): \texttt{type = "both"} and \texttt{maxdepth = 3} (i.e., the default settings in \texttt{pre::pre()})
\end{enumerate}

Although the rulefit student model sometimes results in more accurate estimation of the subgroup ATE, this is at the cost of a more complex model, illustrated by the substantially higher number of false positive features in rulefit compared to CART. CART thus appears to be a simpler and more interpretable student model choice for CDT.

\newpage 
\section{Additional Teacher Model Selection Simulation Results}\label{app:jaccard}

In this section, we present additional simulations and results to provide intuition into the proposed teacher model selection procedure from Section~\ref{sec:model_selection}.

\begin{figure}[h!]
    \centering
    \includegraphics[width=1\linewidth]{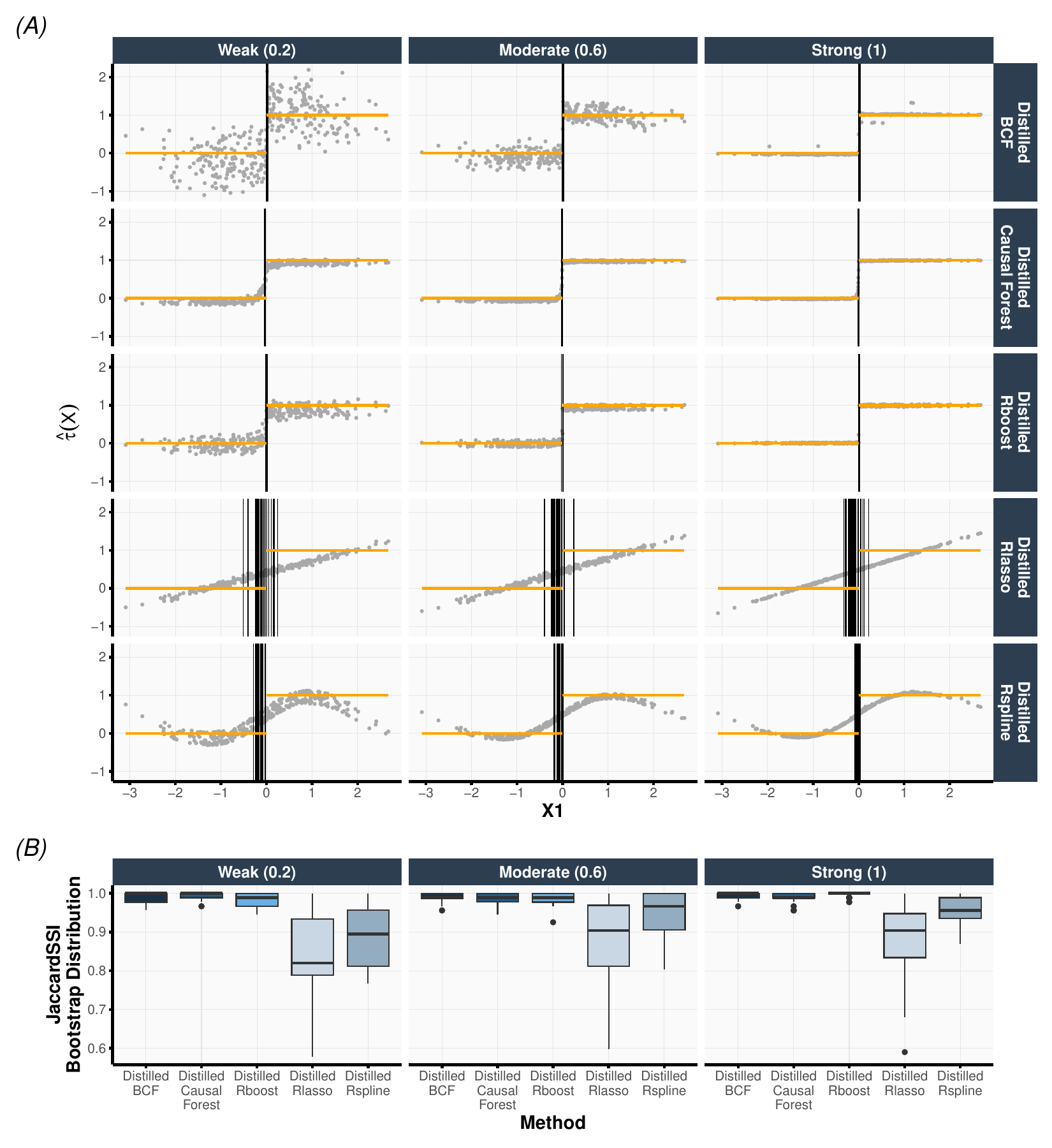}
    \caption{Behavior of teacher models under the simulation setting with two true subgroups defined by a single feature $X^{(1)}$ and a single threshold at $0$, evaluated across different signal-to-noise settings (columns). (A) We plot the teacher model predictions $\hat{\tau}(X_i)$ (gray points), the true treatment effect heterogeneity function (orange), and the partitions obtained from the 100 bootstrap fits in the teacher model selection procedure (vertical black lines). (B) We show the distribution of the resulting Jaccard SSIs from the 100 bootstrap fits for each teacher model.}
    \label{fig:jaccard-example}
\end{figure}

To begin, we first consider the proposed teacher selection procedure, applied to a single simulation replicate generated as in Section~\ref{sec:sims}, but with $Y_i = Z_i \cdot \tau(X_i) + \nu_i$ where $\tau(X_i) = \mathbbm{1}\{X_i^{(1)} > 0\}$. 
Here, there are two true subgroups defined by a single feature $X^{(1)}$ and a single threshold at $0$, and of the teacher models considered (BCF, Causal Forest, Rboost, Rspline, and Rlasso), Rlasso and Rspline would be misspecified teacher models since their inherently smooth nature is a mismatch for the underlying discrete subgroup structure.
Under this simple setting, Figure~\ref{fig:jaccard-example} reveals that the teacher model selection procedure can help to successfully rule out both Rlasso and Rspline from consideration due to their instability. In particular, Figure~\ref{fig:jaccard-example}(A) plots the predictions $\hat{\tau}(X_i)$ from various teacher models (BCF, Causal Forest, Rboost, Rlasso, and Rspline) along with the true treatment effect heterogeneity function (in orange).
We also overlay the partitions (i.e., the black vertical lines) obtained from all 100 bootstrap fits, performed in the teacher model selection procedure, and we display the distribution of the resulting Jaccard SSIs in Figure~\ref{fig:jaccard-example}(B).
Across all three signal-to-noise settings (weak, moderate, and strong), the Rlasso and Rspline teachers result in the most unstable subgroup partitions and hence would not be selected as the teacher model.

\begin{figure}[h!]
    \centering
    \includegraphics[width=1\linewidth]{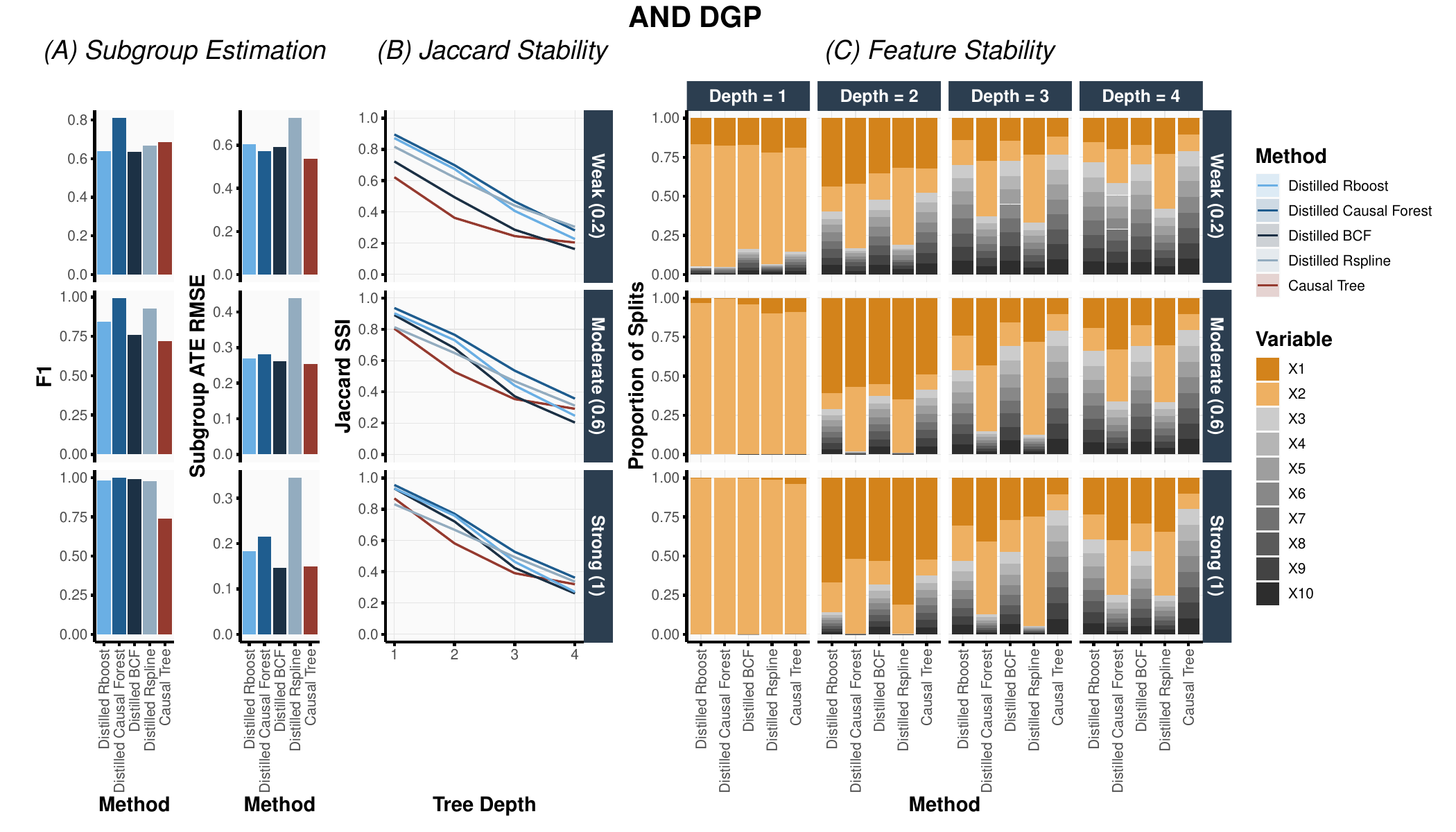}
    \caption{Under the `AND' subgroup data-generating process (CATE-only), we examine (A) the subgroup estimation accuracy alongside (B) the Jaccard SSI for a range of tree depths and (C) the distribution of features selected at each tree depth across the 100 bootstraps. Results are shown for different subgroup estimation methods (colors) and treatment effect heterogeneity strengths (rows).}
    \label{fig:sim-stability-and-0}
\end{figure}

\begin{figure}[h!]
    \centering
    \includegraphics[width=1\linewidth]{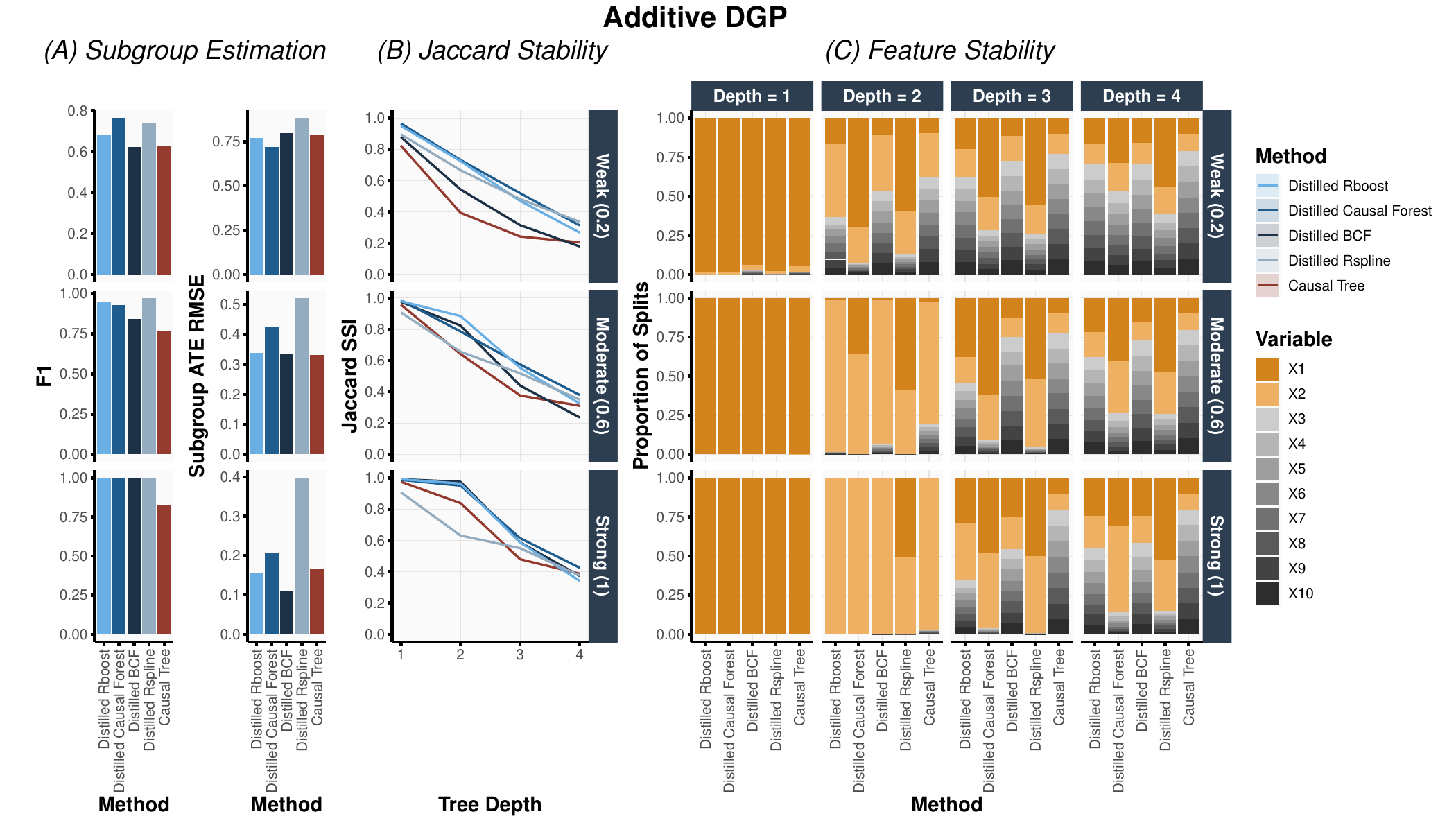}
    \caption{Under the `Additive' subgroup data-generating process (CATE-only), we examine (A) the subgroup estimation accuracy alongside (B) the Jaccard SSI for a range of tree depths and (C) the distribution of features selected at each tree depth across the 100 bootstraps. Results are shown for different subgroup estimation methods (colors) and treatment effect heterogeneity strengths (rows).}
    \label{fig:sim-stability-additive-0}
\end{figure}

\begin{figure}[h!]
    \centering
    \includegraphics[width=1\linewidth]{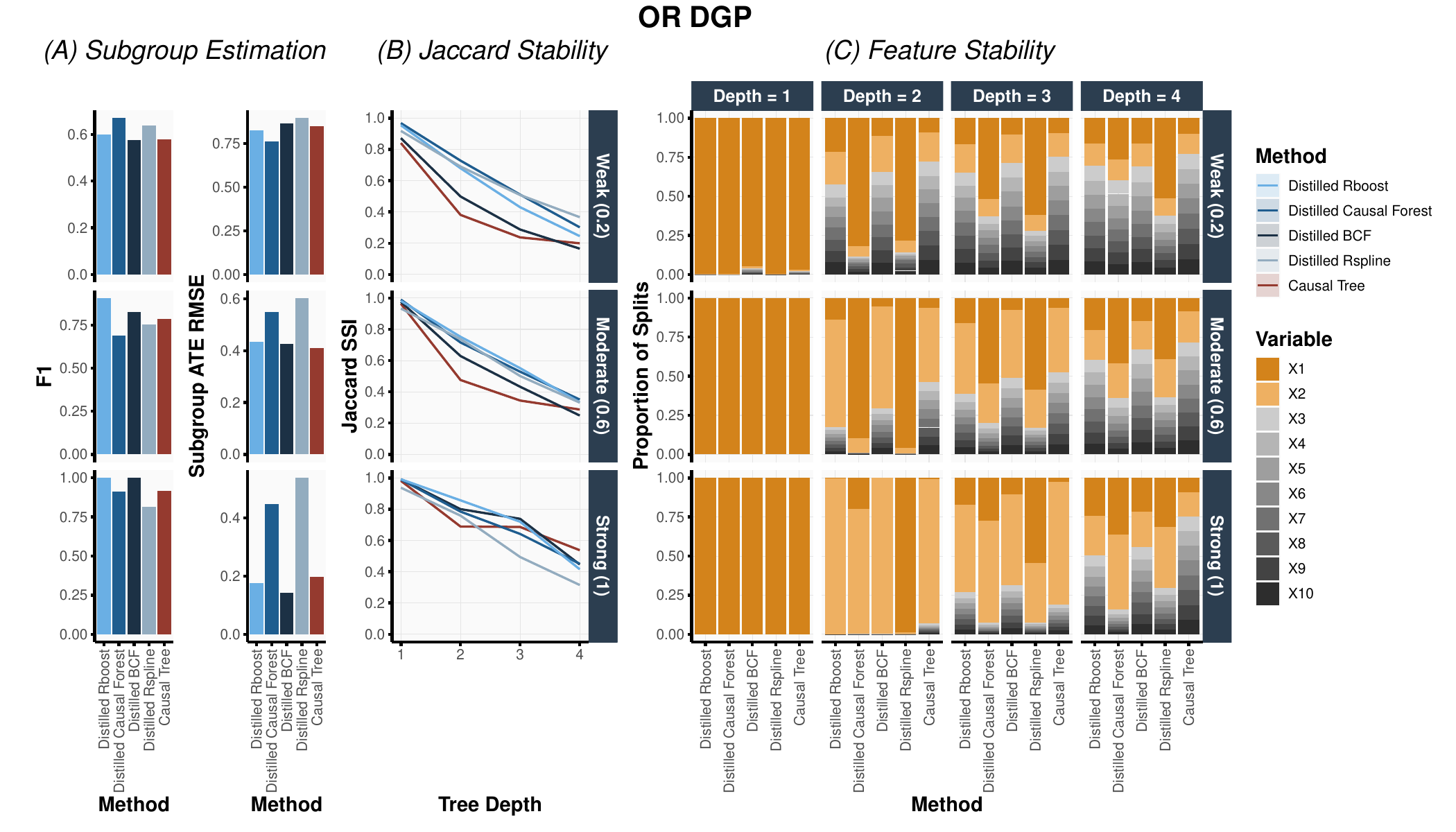}
    \caption{Under the `OR' subgroup data-generating process (CATE-only), we examine (A) the subgroup estimation accuracy alongside (B) the Jaccard SSI for a range of tree depths and (C) the distribution of features selected at each tree depth across the 100 bootstraps. Results are shown for different subgroup estimation methods (colors) and treatment effect heterogeneity strengths (rows).}
    \label{fig:sim-stability-or-0}
\end{figure}

\begin{figure}[h!]
    \centering
    \includegraphics[width=1\linewidth]{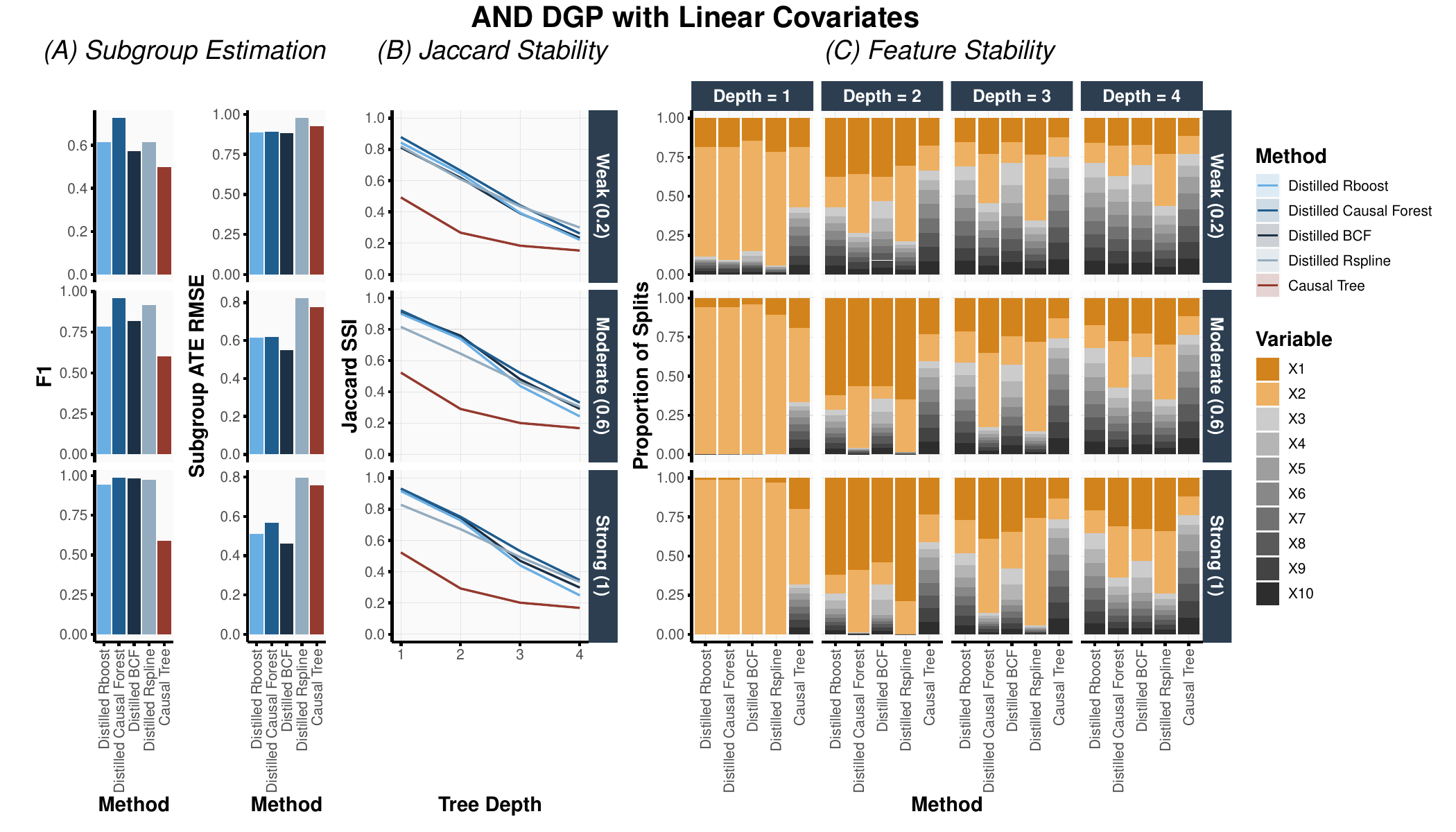}
    \caption{Under the `AND' subgroup data-generating process with linear covariate effects, we examine (A) the subgroup estimation accuracy alongside (B) the Jaccard SSI for a range of tree depths and (C) the distribution of features selected at each tree depth across the 100 bootstraps. Results are shown for different subgroup estimation methods (colors) and treatment effect heterogeneity strengths (rows).}
    \label{fig:sim-stability-and-cov}
\end{figure}

\begin{figure}[h!]
    \centering
    \includegraphics[width=1\linewidth]{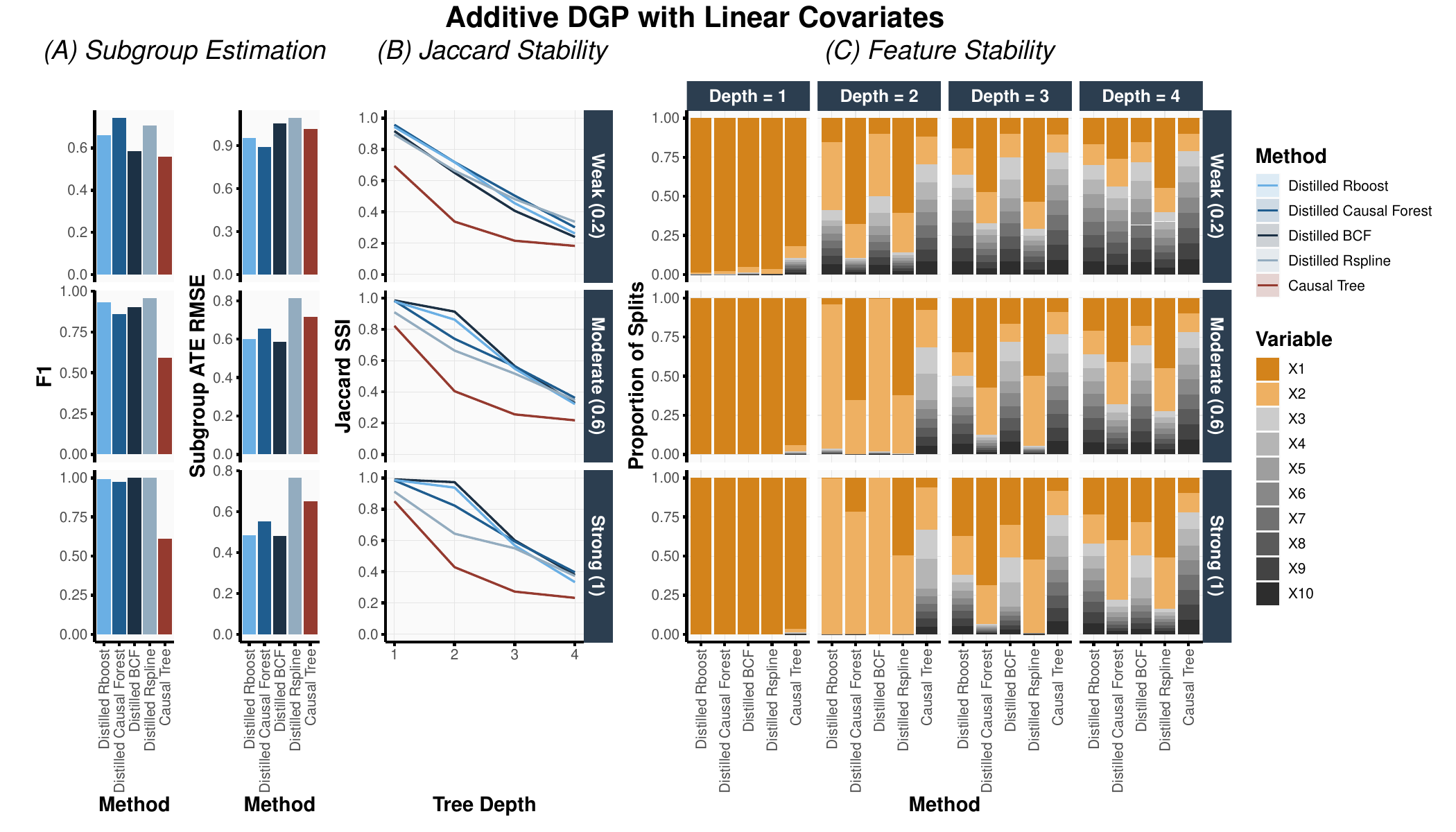}
    \caption{Under the `Additive' subgroup data-generating process with linear covariate effects, we examine (A) the subgroup estimation accuracy alongside (B) the Jaccard SSI for a range of tree depths and (C) the distribution of features selected at each tree depth across the 100 bootstraps. Results are shown for different subgroup estimation methods (colors) and treatment effect heterogeneity strengths (rows).}
    \label{fig:sim-stability-additive-cov}
\end{figure}

\begin{figure}[h!]
    \centering
    \includegraphics[width=1\linewidth]{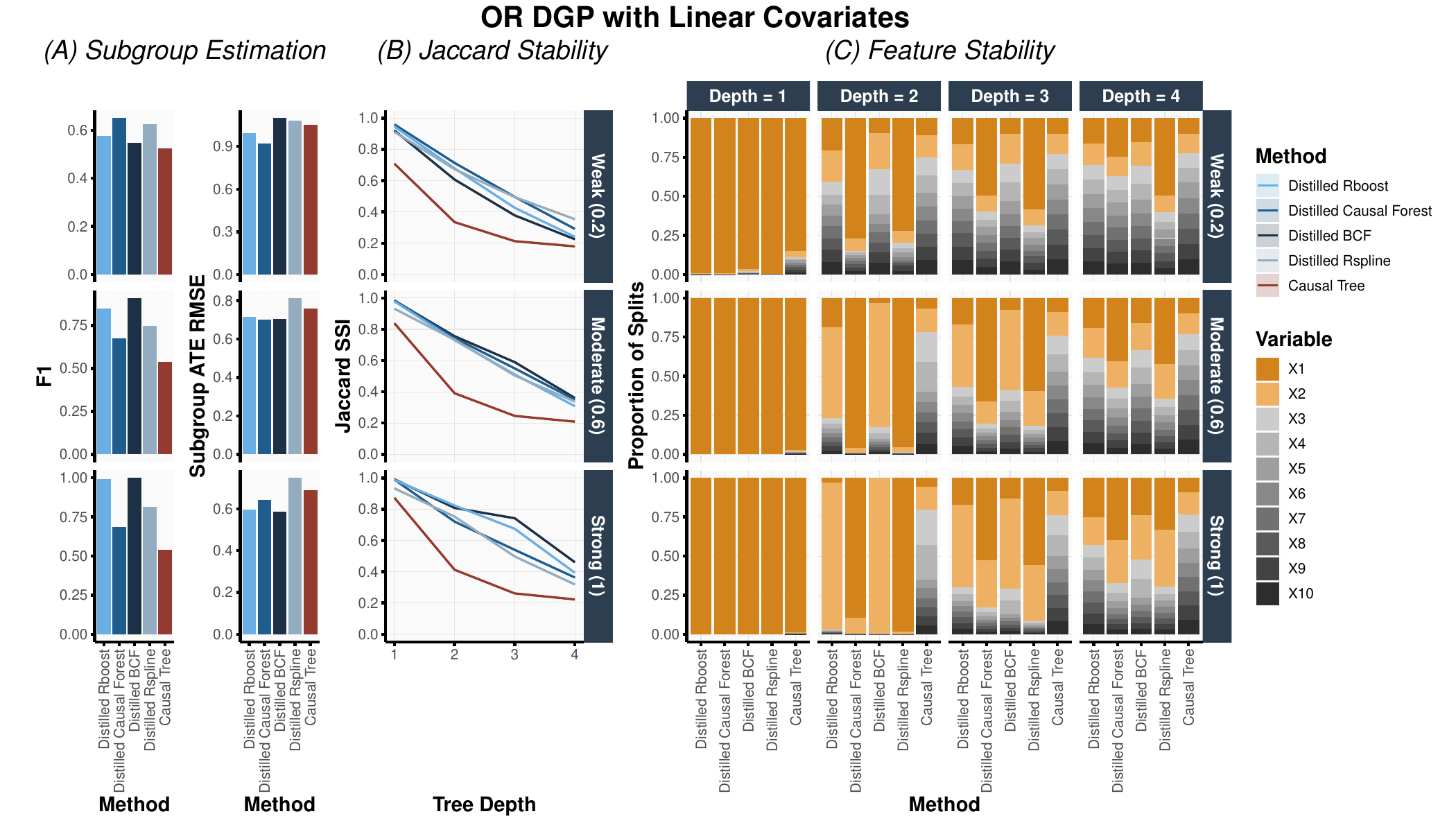}
    \caption{Under the `OR' subgroup data-generating process with linear covariate effects, we examine (A) the subgroup estimation accuracy alongside (B) the Jaccard SSI for a range of tree depths and (C) the distribution of features selected at each tree depth across the 100 bootstraps. Results are shown for different subgroup estimation methods (colors) and treatment effect heterogeneity strengths (rows).}
    \label{fig:sim-stability-or-cov}
\end{figure}

We next present results from a larger-scale simulation study with numerous simulation replicates. 
Specifically, we provide the Jaccard SSI simulation results (Figures~\ref{fig:sim-stability-and-0}-\ref{fig:sim-stability-or-cov}), used to select the teacher model, for each of the subgroup DGPs studied in Section~\ref{sec:sims} and Appendix~\ref{app:sim_results}. As discussed in Section~\ref{sec:sims}, higher Jaccard SSI generally corresponds to more accurate subgroup estimation regardless of the choice of subgroup DGP.

In practice, we reiterate that substantive researchers should leverage their domain knowledge when choosing the tree depth(s) $d$ to consider in the teacher model selection procedure. In cases where such prior knowledge is limited, we recommend that researchers view the Jaccard SSI results alongside the feature stability distributions and other diagnostic tools (see Appendix~\ref{app:diagnostics}) to gain a more holistic perspective. This holistic view of the distilled method and its stability can often shed light on an appropriate choice of tree depth. For example, the feature stability distributions in Figures~\ref{fig:sim-stability-and-0}-\ref{fig:sim-stability-or-cov} clearly illuminate that the trees grown to depths 3 and 4 are substantially more unstable than the depth-2 trees. That is, the distribution of features selected at depths 3 and 4 are far more heterogeneous than that at depth 2. As with the Jaccard SSI, a more stable or homogeneous feature distribution is generally a positive sign and may serve as heuristic to help choose the tree depth.

\clearpage 
\section{Extended Details of AIDS Case Study}\label{app:case-study}

Our analysis of the AIDS case study data consisted of two parts.

\paragraph{Primary Analysis.} 
The first and primary analysis was to identify subgroups of patients with heterogeneous treatment effects using the proposed CDT framework.
This analysis resembles how researchers might use CDT in practice to estimate their own interpretable subgroups with heterogeneous treatment effects.

\begin{figure}[ht]
    \centering
    \includegraphics[width=0.6\linewidth]{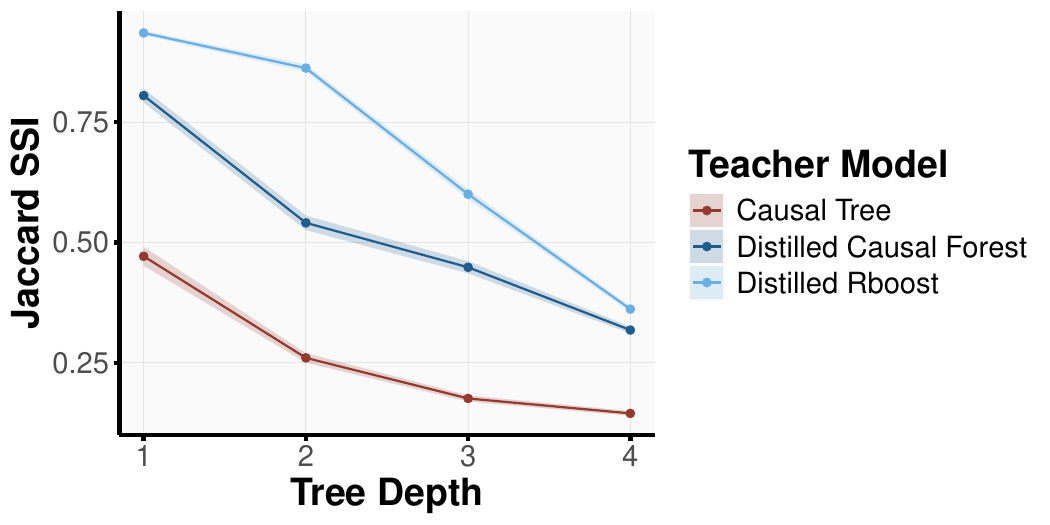}
    \caption{Teacher model selection results from AIDS case study across a range of tree depths $d$ with $B = 100$ bootstraps. Higher Jaccard SSI indicates a more stable teacher model.}
    \label{fig:case-study-jaccard}
\end{figure}

In this analysis, we performed a 50-50\% split of the data into a training and a held-out estimation set. 
Then using the training data, we applied the proposed teacher model selection procedure (with $B = 100$ bootstraps) to choose between the causal forest and the Rboost teacher models.
The results of this model selection are provided in Figure~\ref{fig:case-study-jaccard}, indicating that Distilled Rboost is more stable than Distilled Causal Forest.
We thus selected the Rboost teacher model and refitted Distilled Rboost on the training split to obtain the estimated subgroup structure $\{\hat{\cG}_1(X), \ldots, \hat{\cG}_G(X)\}$, which we depicted in Figure~\ref{fig:case-study-tree}.
Finally, for each of the estimated subgroups $\hat{\cG}_g(X)$, we estimated the subgroup ATE and its corresponding standard errors using the held-out estimation set as described in Section~\ref{subsec:subgroup-ate}.

\paragraph{Secondary Analysis.} 
The secondary analysis was to quantify the stability of CDT and other subgroup estimation methods when trained on different data splits.
While this analysis may also be of interest to practitioners in certain applications, we use it here primarily to provide additional evidence of the increased stability of CDT compared to other subgroup estimation methods.

\begin{figure}
    \centering
    \includegraphics[width=0.9\textwidth]{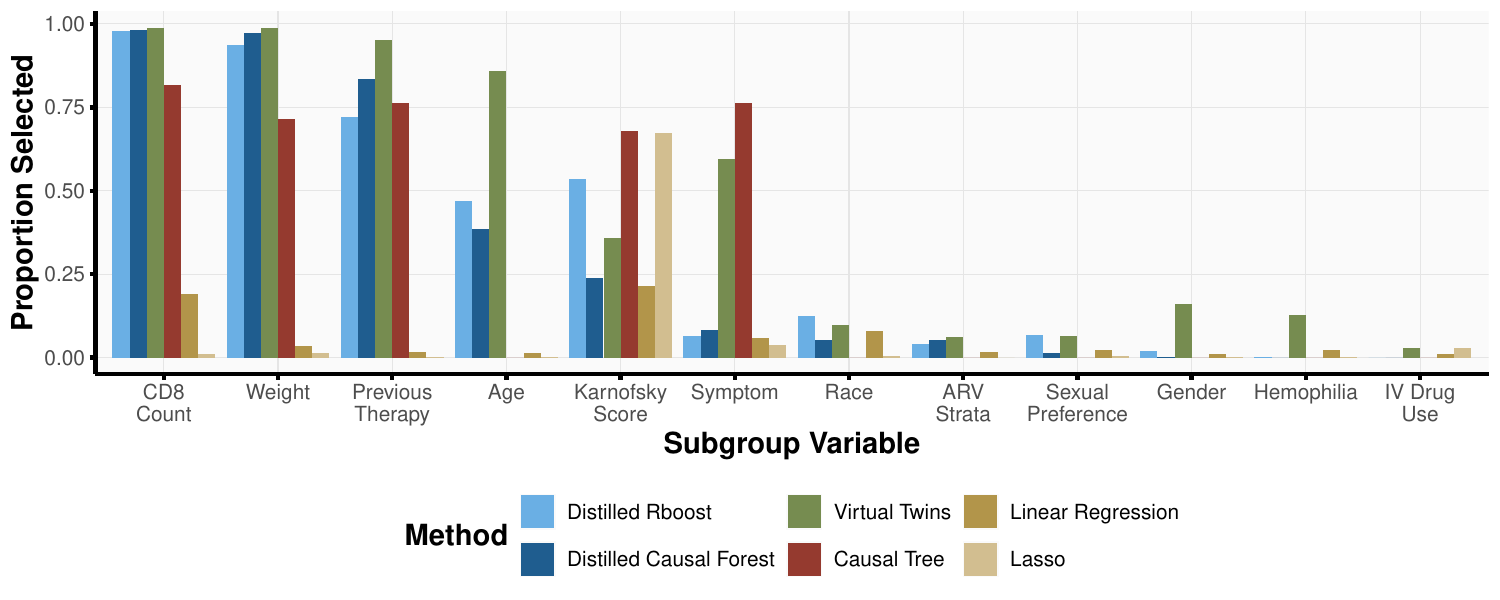}
    \caption{Proportion of splits in which each feature is involved in constructing subgroups across 100 different 50-50\% training/held-out estimation data splits. Proportions closer to 1 indicate that the feature is consistently selected across different data splits while proportions closer to 0 indicate that the feature is rarely selected.}
    \label{fig:case-study-tree-stability}
\end{figure}

In this analysis, we fitted CDT and other competing subgroup estimation methods on 100 different 50-50\% training/held-out estimation data splits and computed the proportion of data splits in which each feature was used to construct subgroups.
In the ideal case, a stable method would construct subgroups involving the same subset of features across all 100 data splits, while never selecting any of the remaining features.
We summarize the results of this stability analysis in Figure~\ref{fig:case-study-tree-stability} and provide brief descriptions of each pre-treatment covariate in Table~\ref{tab:features}.

We highlight four main takeaways from Figure~\ref{fig:case-study-tree-stability}. 
First, we observe that the CDT methods generally construct subgroups with the same set of features, demonstrating high stability, and that the estimated subgroups were overwhelmingly characterized by features with direct clinical relevance. 
Consistent with our primary analysis, we observe that the three main features chosen by the CDT methods were CD8 counts, starting weight, and number of days of antiretroviral therapy. 
Second, even among features that are well-known to be clinically relevant (e.g., CD8 count, starting weight, and previous antiretroviral therapy), causal tree does not select any single feature in more than 82\% of the data splits.
This is less stable in comparison to CDT, which selects CD8 and starting weight in over 94\% of the data splits.
Third, while virtual twins stably selects CD8 count, starting weight, and previous antiretroviral therapy, it constructs substantially more subgroups using other features, resulting in an overwhelming number of complex subgroups that are challenging to interpret. 
Lastly, the linear-based approaches (i.e., linear regression and Lasso) almost never split on CD8 count, starting weight, or previous antiretroviral therapy. 
Instead, they identify subgroups most commonly based upon the Karnofsky score, an assessment tool intended to gauge a patient's functional status and ability to carry out activities of daily living \citepappendix{schag1984karnofsky}. 
While the Karnofsky score would normally be assumed to be an informative feature, the study eligibility criteria required that all participants have a performance score of at least 70; in fact, the median Karnofsky score in the study was 100 (out of 100), translating to a normal status with no evidence of disease. 
Given the limited variability in patients' Karnofsky scores, the Karnofsky score is unlikely to be the primary clinical determinant of treatment effect heterogeneity in this study.

\begin{table}
\centering
\small 
\renewcommand{\arraystretch}{1.1}
\begin{tabular}[t]{lp{13cm}}
\toprule
\textbf{Feature} & \textbf{Description}\\
\midrule
CD8 Count & CD8 T cell count at baseline\\
Weight & Weight in kilograms at baseline\\
Previous Therapy & Number of days of previously received antiretroviral therapy\\
Age & Age in years at baseline\\
Karnofsky Score & Karnofsky Performance Status score on a scale of 0-100\\
Symptom & Symptomatic indicator (0 = asymptomatic, 1 = symptomatic)\\
Race & Race indicator (0 = white, 1 = non-white)\\
ARV Strata & Antiretroviral history stratification (1 = antiretroviral naive, 2 = 1--52 weeks of prior therapy, 3 = 52+ weeks of prior therapy)\\
Sexual Preference & Homosexual activity indicator (0 = no, 1 = yes)\\
Gender & Gender indicator (0 = female, 1 = male)\\
Hemophilia & Hemophilia indicator (0 = no, 1 = yes)\\
IV Drug Use & Intravenous drug use indicator (0 = no, 1 = yes)\\
\bottomrule
\end{tabular}
\caption{Description of pre-treatment covariates included in AIDS case study analysis.}
\label{tab:features}
\end{table}

\newpage
\bibliographystyleappendix{agsm}
\bibliographyappendix{bibliography_appendix}

\end{document}

%% file: preamble.tex
\usepackage[utf8]{inputenc}
\usepackage{amsmath}
\usepackage{amssymb}
\input{epsf}
\usepackage{natbib}
\usepackage[margin=1in]{geometry}

\usepackage[raggedright]{titlesec}
\usepackage[hang,flushmargin]{footmisc} 
\usepackage{graphicx}
\usepackage{framed}
\usepackage{color}
\usepackage{yfonts}
\usepackage[dvipsnames]{xcolor}
\usepackage{pifont}
\usepackage{amsmath,amssymb}
\usepackage{multirow}
\usepackage{amsfonts}
\usepackage{subfigure}
\usepackage{caption}
\usepackage{threeparttable}
\usepackage{esvect}
\usepackage{pgf}
\usepackage{tikz}
\usepackage{tikz-cd}
\usepackage{fancyvrb}
\usetikzlibrary{positioning}
\usepackage{bbm}
\usepackage{colortbl}
\usepackage{pdflscape}
\usepackage{booktabs}
\usepackage{amsmath}
\usepackage{mathdots}
\usepackage{yhmath}
\usepackage{cancel}
\usepackage{siunitx}
\usepackage{array}
\usepackage{multirow}
\usepackage{gensymb}
\usepackage{tabularx}
\usepackage{booktabs}
\usepackage{tabstackengine}
\usetikzlibrary{fadings}
\usetikzlibrary{patterns}
\usetikzlibrary{shadows.blur}
\usetikzlibrary{shapes}
\usepackage{placeins}
\usepackage{comment}
\usepackage[ruled,linesnumbered,commentsnumbered]{algorithm2e}
\SetKwComment{Comment}{// }{}
\usepackage{enumerate}

\usepackage[font=small,labelfont=bf]{caption}

\usepackage{hyperref}
\hypersetup{
    colorlinks=true,
    linkcolor=NavyBlue,
    filecolor=magenta,      
    urlcolor=cyan,
    citecolor=MidnightBlue
}
 
\usepackage{rotating}
\usepackage{ntheorem}
\makeatletter
\renewtheoremstyle{plain} 
{\item{\theorem@headerfont ##1\ ##2\theorem@separator}~}
{\item{\theorem@headerfont ##1\ ##2\ (##3)\theorem@separator}~}
\makeatother

\theoremheaderfont{\normalfont\bfseries}

\newtheorem{theorem}{Theorem}[section]
\newtheorem{definition}{Definition}[section]
\newtheorem{lemma}{Lemma}[section]

\newtheorem{proposition}{Proposition}[section]
\newtheorem{corollary}{Corollary}[section]
\newtheorem{example}{Example}[section]
\newtheorem{assumption}{Assumption}

\newenvironment{proof}{\paragraph{Proof:}}{\hfill$\square$}

\newcommand{\iid}{\stackrel{iid}{\sim}}
\newcommand{\cip}{\stackrel{p}{\rightarrow}}
\newcommand{\cid}{\stackrel{d}{\rightarrow}}
\newcommand{\E}{\mathbb{E}}

\newcommand{\var}{\text{var}}
\newcommand{\asyvar}{\text{asyvar}}
\newcommand{\cov}{\text{cov}}

\newcommand{\cF}{\mathcal{F}}

\newcommand{\cG}{\mathcal{G}}
\newcommand{\mG}{\mathbbm{G}}

\newcommand{\R}{\mathbbm{R}}

\newcommand{\cD}{\mathcal{D}}
\newcommand{\cH}{\mathcal{H}}

\newcommand{\tauh}{\hat{\tau}}

\newcommand{\1}{\mathbbm{1}}
\DeclareMathOperator*{\argmax}{arg\,max}
\DeclareMathOperator*{\argmin}{arg\,min}

\newcommand{\indep}{\raisebox{0.05em}{\rotatebox[origin=c]{90}{$\models$}}}
\definecolor{shadecolor}{gray}{0.9}

\newcommand{\norm}[1]{\left\lVert#1\right\rVert}

\newcommand{\teacher}{\mathcal{M}}
\newcommand{\teacherh}{\widehat{\mathcal{M}}}
\newcommand{\student}{m}
\newcommand{\studenth}{\hat{m}}

\tikzset{every picture/.style={line width=0.75pt}} 

\usepackage{enumitem}
\newlist{Step}{enumerate}{2}
\setlist[Step]{label={{Step \arabic*.}}, leftmargin=*}
\usepackage{setspace}

\makeatletter
\newcommand\circled[1]{%
  \mathpalette\@circled{#1}%
}
\newcommand\@circled[2]{%
  \tikz[baseline=(math.base)] \node[draw,circle,inner sep=2pt] (math) {$\m@th#1#2$};%
}
\makeatother

\makeatletter
\newcommand\circledblue[1]{%
  \mathpalette\@circledblue{#1}%
}
\newcommand\@circledblue[2]{%
  \tikz[baseline=(math.base)] \node[draw,circle, fill=blue!20, inner sep=2pt] (math) {$\m@th#1#2$};%
 }

\renewenvironment{abstract}
 {\begin{center}\normalsize\textsc{Abstract}%
 \end{center}\begin{quote}\normalsize}
 {\end{quote}}

\newcommand{\method}{CDT}

\usepackage{apptools}
\AtAppendix{\counterwithin{assumption}{section}}

\makeatletter
\newcommand{\myfnsymbol}[1]{%
  \expandafter\@myfnsymbol\csname c@#1\endcsname
}
\newcommand{\@myfnsymbol}[1]{%
  \ifcase #1
  \or $^\dagger$
  \or $^*$
  \or 1
  \or 2
  \or 3
  \fi
}
\newcommand{\acknowledgements}{\@myfnsymbol{1}}
\newcommand{\equalcontributor}{\@myfnsymbol{2}}
\newcommand{\affiliationA}{\@myfnsymbol{3}}
\newcommand{\affiliationB}{\@myfnsymbol{4}}
\newcommand{\affiliationC}{\@myfnsymbol{5}}
\makeatother

\definecolor{byzantine}{rgb}{0.74, 0.2, 0.64}